\documentclass[12pt]{article}

\usepackage{csvsimple}
\usepackage{comment}

\usepackage[english]{babel}
\usepackage[T1]{fontenc}
\usepackage[utf8]{inputenc}

\usepackage{beton}
\usepackage{euler}

\usepackage{amsmath, amssymb, amsfonts, amsthm}
\usepackage{mathtools}
\usepackage{bbm}
\usepackage{mathrsfs}
\usepackage{dsfont}           
\DeclareMathOperator*{\argmin}{arg\,min}
\DeclareMathOperator*{\diag}{diag}

\newcommand{\tr}{^\top}
\newcommand{\ev}{\mathrm{E}}
\newcommand{\eps}{\varepsilon}
\newtheoremstyle{boldexample}
  {\topsep}{\topsep}
  {\itshape}{}{\bfseries}{.}{ }
  {\thmname{#1}\thmnumber{ #2}\thmnote{ (#3)}}

\theoremstyle{boldexample}
\newtheorem{theorem}{Theorem}[section]

\newtheorem{lemma}[theorem]{Lemma}
\newtheorem{proposition}[theorem]{Proposition}
\newtheorem{example}[theorem]{Example}

\theoremstyle{definition}

\newtheorem{assumption}[theorem]{Assumption}
\newtheorem*{restriction*}{Restrictions}

\theoremstyle{remark}
\newtheorem{remark}[theorem]{Remark}

\numberwithin{equation}{section}

\usepackage{titlesec}

\titleclass{\subsubsubsection}{straight}[\subsubsection]

\newcounter{subsubsubsection}[subsubsection]
\renewcommand\thesubsubsubsection{%
  \thesubsubsection.\arabic{subsubsubsection}}

\titleformat{\subsubsubsection}
  {\normalfont\normalsize\bfseries}
  {\thesubsubsubsection}{1em}{}

\titlespacing*{\subsubsubsection}
  {0pt}{3.25ex plus 1ex minus .2ex}{1.5ex plus .2ex}

\usepackage[margin=1in]{geometry}
\usepackage{setspace}
\usepackage{indentfirst}

\usepackage{graphicx}
\graphicspath{{images/}}
\usepackage{caption}
\usepackage{subcaption}
\usepackage{threeparttable}
\usepackage{adjustbox}
\usepackage{booktabs}
\usepackage{longtable}
\usepackage{rotating}
\usepackage{array}
\newcolumntype{H}{>{\setbox0=\hbox\bgroup}c<{\egroup}@{}}
\usepackage{nicematrix}

\usepackage{authblk}
\usepackage{hyperref}
\hypersetup{
    colorlinks=true,
    linkcolor=blue,
    citecolor=blue,
    urlcolor=blue
}
\usepackage{enumerate}
\usepackage{textcomp}
\usepackage{xcolor}
\usepackage{soul}

\usepackage[longnamesfirst]{natbib}%

\usepackage[babel]{csquotes}

\begin{document}
\title{Inference on Linear Programs via Feasibility Testing
}
\date{}
\author[1]{Sofya Budanova}
\author[2]{Eric D. Mbakop}
\affil[1]{International College of Economics and Finance, Higher School of Economics}
\affil[2]{Department of Economics, The Ohio State University}
\maketitle

\begin{abstract}
This paper studies inference on the value of a linear program (LP) when the objective and constraints are estimated. Using LP duality, we construct one- and two-sided confidence intervals by inverting tests of the feasibility of systems of linear inequalities. Existing tests yield asymptotically valid confidence intervals but can have poor power. We propose a modification that augments these systems with estimated ``box constraints.'' The resulting procedure has power against $N^{-1/2}$-local alternatives and yields confidence intervals of order $N^{-1/2}$. These results require only that the LP value be finite, and do not require compactness of the feasible region nor standard constraint qualifications. We obtain stronger guarantees under various constraint qualifications and establish uniform validity over large classes of DGPs when known deterministic constraints bound the primal and/or dual variables. Simulations demonstrate good finite-sample performance.

\end{abstract}
\section{Introduction}

This paper studies one- and two-sided inference on the value $v_0$ of a
linear program (LP)
\begin{equation}
\label{eqnLP}
v_0
=
\max\left\{
\langle \rho,x\rangle
\mid
Mx\leq\beta
\right\},
\end{equation}
where $M\in\mathbb{R}^{p\times d}$ and some or all components of $\mu:=\left(\rho\tr,\beta\tr,\operatorname{vec}(M)\tr\right)\tr$ must be estimated from the data. This problem arises naturally in
partially identified models, where the identified set for a scalar
parameter is often an interval whose endpoints are the values of linear
programs. It also arises in point-identified settings, for example when
the object of interest is the value of an optimal matching/coupling or assignment.
Section~\ref{SecExemp} presents several examples.

Inference on the value of an LP is nonstandard for several reasons. First,
even when the population LP is feasible, its sample analogue may be
infeasible with probability that does not converge to zero, making a plug-in estimator infeasible. Second, even
when the sample LP is feasible, its value need not be continuous in the
inputs $\mu$, so the plug-in estimator need not be
consistent.\footnote{See \cite{SR1} and \cite{SR2} for general results on the
stability of the feasible region of an LP and the continuity properties of
its value as a function of the inputs $\mu$. See also \cite{AV}.} Finally,
even in regions of the parameter space where the plug-in estimator is well
behaved, the LP value function need not be Hadamard differentiable
\cite{AS,YG}, so the standard bootstrap may fail; see
\cite{ZFAS,HHJL1,HHJL2}.

Our starting point is a simple application of LP duality. There is a
classical computational equivalence between solving systems of linear
inequalities and solving linear programs: a polynomial-time algorithm for
either problem can be used to construct a polynomial-time algorithm for
the other; see, for example, \citet[Section~10.3]{ALSH}. Our results
establish a statistical analogue of this connection. In particular, we
show that one- and two-sided inference on the value $v_0$ of an LP can be
reduced to testing the feasibility of appropriately constructed systems of
linear inequalities. Thus, statistical procedures for testing linear
feasibility can be transformed into procedures for inference on LP values. More precisely, confidence intervals for $v_0$ can be constructed by
inverting tests, indexed by $v\in\mathbb{R}$, of hypotheses of the form
\begin{equation}
\label{eqnHyp}
H_0:
\ \exists\,\eta
\text{ such that }
b(v)\leq A\eta
\qquad\text{vs.}\qquad
H_1:
\ \nexists\,\eta
\text{ such that }
b(v)\leq A\eta.
\end{equation}
The matrix $A$ is constructed from $\mu$ and does not vary with $v$,
whereas the vector $b(v)$ does. The precise forms of $A$ and $b(v)$ depend
on whether the objective is to construct an upper, lower, or two-sided
confidence interval.

This reduction connects inference on LP values to the recent literature
on testing the feasibility of systems of linear inequalities, including
\cite{BPSSMT}, \cite{CSS}, and \cite{GM25}. Direct application of these
methods already yields an important result: valid confidence intervals
for $v_0$ can be constructed under weaker conditions than those typically
imposed in the existing literature on inference for LP values. In
particular, our baseline validity result requires neither compactness of
the feasible region nor constraint qualifications such as the linear independence constraint qualification (LICQ), the Mangasarian-Fromovitz constraint qualification (MFCQ), or
Slater-type conditions. Apart from regularity conditions governing the
estimation of $\mu$, the essential requirement on the population LP is
simply that $v_0$ be finite.

\begin{figure}[htbp]
    \centering
    \begin{subfigure}[t]{0.495\textwidth}
        \centering
        \includegraphics[width=\textwidth]{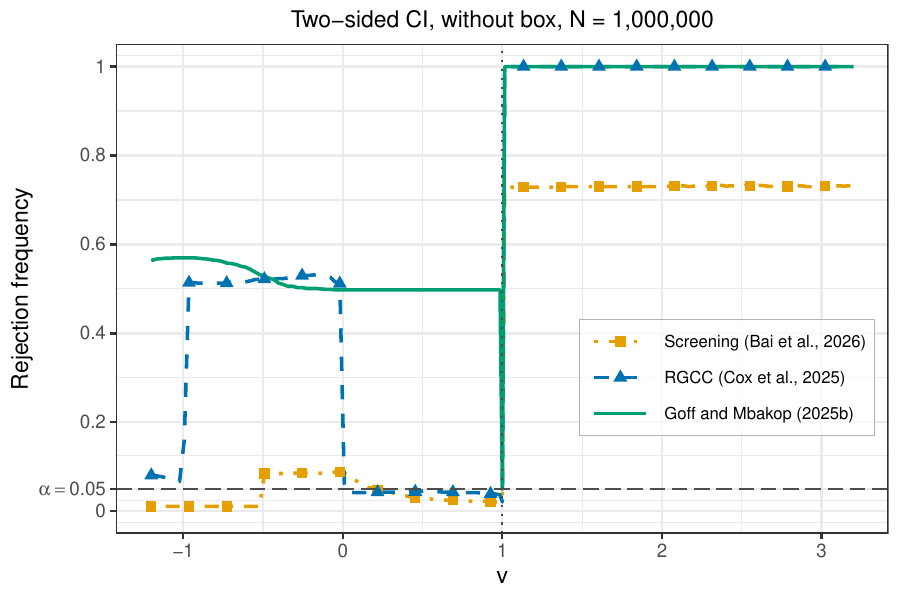}
        \caption{Methods performance \textit{without} the box.}
        \label{fig:intro_nobox}
    \end{subfigure}
    \hfill
    \begin{subfigure}[t]{0.495\textwidth}
        \centering
        \includegraphics[width=\textwidth]{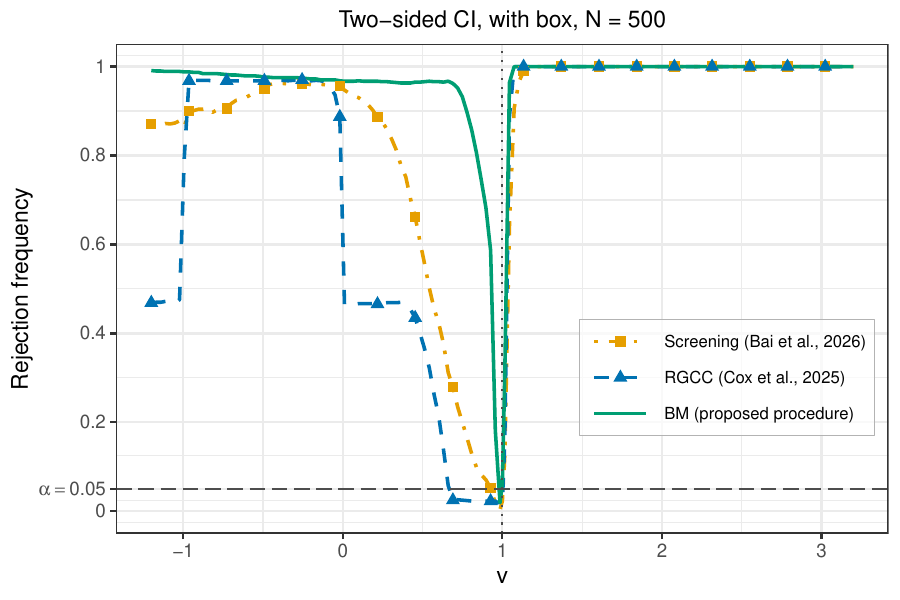}
          \caption{Methods performance \textit{with} the box.}
        \label{fig:intro_box}
    \end{subfigure}
    \caption{\small Power curves for the two-sided confidence interval for $v_0$ under the design in Section \ref{SecMC_AV}, with $b_0=0$, based on $1{,}000$ simulations. \citet{GM25} and BM are implemented with $B=1,000$ bootstrap replications. The significance level is $\alpha=0.05$. The \citet{CSS} procedure is constructed as the intersection of the two one-sided procedures, each conducted at level $\alpha/2$; see Section \ref{SecComp} for details.}
    \label{fig:intro}
\end{figure}

Validity, however, is only part of the problem. Directly inverting
existing tests of \eqref{eqnHyp} can produce confidence intervals with
poor power properties and, consequently, excessive width; see
Figure \ref{fig:intro_nobox}. In some cases, the resulting tests can even be
inconsistent against fixed alternatives, so that the width of a
two-sided confidence interval need not converge to zero as the sample
size increases.

We relate this phenomenon to the impossibility results of \cite{BPSSMT}
and \cite{GM25}. These results show, in particular, that testing procedures that
are uniformly valid over sufficiently large classes of DGPs can have
trivial power against alternatives for which the homogeneous system $0\leq A\eta$ admits a nonzero solution. This condition has a useful geometric
interpretation. Whenever $\mathcal P(v)
:=
\left\{
\eta:b(v)\leq A\eta
\right\}$ is nonempty, the existence of a nonzero solution to $0\leq A\eta$
implies that $\mathcal P(v)$ is unbounded.\footnote{More generally, a
nonempty polyhedron
$\mathcal P=\{\eta:b\leq A\eta\}$ is unbounded if and only if its
recession cone
$\{\eta:0\leq A\eta\}$ contains a nonzero element.}
This observation suggests both an explanation for the poor power of
direct feasibility tests and a way to improve it: augment
\eqref{eqnHyp} with additional constraints that bound $\eta$ without
altering feasibility.

This idea leads to the main methodological contribution of the paper. We
construct data-dependent constraints of the form
\[
\|\eta\|_\infty\leq\hat a_N,
\]
which we refer to as \emph{box constraints}. The bound $\hat a_N$ is
estimated so that, with probability approaching one, the augmented system
remains feasible whenever the original system is feasible. Thus, the box
constraints bound the possible values of the nuisance parameter without asymptotically excluding
the values of $\eta$ needed under the null.

We then modify the procedure of \cite{GM25} by appending these estimated
box constraints to the systems in \eqref{eqnHyp}. The modification
preserves the validity of the original procedure under essentially the
same assumptions, while substantially improving its power properties.
In particular, the resulting tests have nontrivial power against
$N^{-1/2}$-local alternatives and are consistent against fixed
alternatives. Consequently, the associated confidence intervals have
length of order $N^{-1/2}$. The box constraints therefore provide the
localization needed for power without requiring compactness of the
original LP feasible region or additional constraint qualifications. We
also find that appending analogous box constraints to the procedures of
\cite{CSS} and \cite{BPSSMT}, and treating them as deterministic, can substantially improve their power (see Figure \ref{fig:intro_box}),
although we do not formally establish the validity of these modified
procedures.\footnote{For the procedure of \cite{BPSSMT}, a natural
modification would be to estimate the box constraints using the first
sample fold and then treat them as fixed when conducting inference on the
second fold. We conjecture that the resulting procedure remains valid,
although uniform validity would generally hold over a more restricted
class of DGPs, as there cannot exist a uniformly consistent estimator of the box constraints under standard uniformity conditions (see the discussion following Theorem \ref{thm1}). Adapting the procedure of \cite{CSS} to accommodate
estimated box constraints appears less immediate and may require a more
substantial modification of their inference procedure.}

A second contribution is to characterize how constraint qualifications strengthen our results. They are unnecessary for baseline validity but appropriate LICQ conditions yield asymptotically exact coverage, while compactness, MFCQ, or related conditions can make the box constraint unnecessary for the corresponding one- or two-sided problems. 

A third contribution concerns uniform inference. The impossibility
results discussed above imply that uniform validity together with
nontrivial power cannot generally be obtained over large classes of DGPs
without additional structure. We show, however, that such inference is
possible when the systems in \eqref{eqnHyp} contain known deterministic
constraints that bound the relevant nuisance parameters. 
This situation arises naturally, for example, when the nuisance vector
$\eta$ is a probability vector and is therefore known to lie in a bounded
set. In this case, we propose a further modification of our procedure that
is uniformly valid over classes of DGPs satisfying “standard” uniformity
conditions, including a uniform central limit theorem for $\hat{\mu}$,
uniform consistency of its estimated asymptotic covariance matrix, and
uniform validity of a bootstrap estimator of its distribution.\footnote{Conditions of this type are
standard benchmarks for uniform validity in the moment inequality
literature; see, for example, \cite{DAGS}, \cite{RWS}, \cite{CS},
\cite{FSST}, and \cite{ARP}.}
At each fixed DGP, the procedure retains nontrivial power against
$N^{-1/2}$-local alternatives. The price for this stronger uniformity
result is computational: the construction of the critical value requires
solving a nonconvex quadratic program in each bootstrap iteration, which
may become costly in high-dimensional applications. We therefore view this result primarily as establishing
the possibility of nontrivial uniform inference under standard uniformity
conditions, rather than as proposing a computationally attractive procedure.
Developing alternative procedures that retain the same uniform validity
while being computationally tractable and less conservative is an
interesting direction for future research.

\paragraph*{Related literature.}
Our paper is closely related to \citet{JHTR}, who study inference for
linear functionals and scalar subvectors of parameters identified by linear
moment inequalities. Their procedure bootstraps suitably perturbed LPs,
with the perturbations generically producing regular configurations at
which the LP value functions are Hadamard differentiable. The resulting inference
procedure is computationally simple, but is asymptotically conservative
for the original identified set, providing exact coverage instead for an
outer approximation generated by the perturbations. By contrast, our
procedure directly targets the values of the original LPs and produces confidence regions whose excess length shrinks at the
$N^{-1/2}$ rate.

\citet{BG} takes a different regularization approach to the same broad
class of problems. The regularization smooths the
nonregular LP-value map and leads to valid (Gaussian) inference, while retaining consistency for the sharp identified set. An attractive
feature of that approach is computational tractability: the confidence sets
can be obtained from a small number of linear and convex quadratic programs.
The results, however, rely on constraint qualifications that place strong restrictions on the geometry of the LP constraint region. By contrast, our baseline
validity result does not require such constraint qualifications. Instead, we explicitly study what additional statistical conclusions follow when
these constraint qualifications are imposed. In
this sense, our results separate the assumptions needed for validity from those needed for exactness, local power, or the elimination of the box constraints.

Most closely related are \citet{AV} and \citet{chen2026}, who also study inference directly on the value of an LP with estimated objective and constraints. Our work and \citet{AV} are complementary in emphasis. In addition to inference, \citet{AV} studies estimation of the LP value, constructing an estimator that is pointwise $N^{-1/2}$-consistent whenever the population LP is feasible and has a finite value, and characterizing conditions under which such estimators can be uniformly consistent. He also obtains asymptotically normal inference with exact pointwise coverage. Our paper, by contrast, is primarily concerned with inference and with the power and uniform validity of procedures for inference on the LP value.

\citet{chen2026} consider a related penalized estimator, but estimate a finite penalty parameter at which the penalized and original LPs have the same population value, rather than allowing the penalty to diverge with the sample size as in \citet{AV}. This localization is closely related to our estimated box constraints: in both cases, LP duality provides a way to estimate the localization needed for inference.\footnote{Proposition~3.5 of version 2 of \cite{WPGM25} makes explicit the connection between the penalty parameter in \cite{AV} and the box constraints considered below.} \citet{chen2026} address the nonregularity of the resulting estimator using subsampling and obtain asymptotically exact pointwise inference.

Our approach uses localization differently. Rather than regularizing or directly estimating the LP value, we append estimated box constraints to the systems of linear inequalities used for inference and construct confidence intervals by inverting feasibility tests. This perspective connects inference on LP values to impossibility results for testing systems of linear inequalities and provides a geometric interpretation of the role of localization: the box constraints eliminate the recession directions responsible for the poor power of direct feasibility tests. Unlike \citet{AV}, our baseline inference procedure does not require deterministic localization of the LP variables,\footnote{Assumption A0 in \cite{AV} restricts the primal variables to a known compact set, while Assumption B3 imposes a known bound on an optimal dual solution. We discuss their connection with deterministic box constraints in Section~\ref{SecDetBox}.} nor does it rely on sample splitting, making it more directly applicable to dependent data. Finally, we characterize conditions under which our inference procedure is uniformly valid. In particular, when deterministic bounds are known, we show that nontrivial uniform inference is possible over large classes of DGPs satisfying standard uniformity conditions. Comparable uniform validity results are not established in \cite{chen2026}.

Our paper is also related to \citet{bhattacharya} and
\citet{Semenova}. \citet{bhattacharya} is an early example of direct
inference on an optimized value when the primitives of the optimization
problem are estimated, while \citet{Semenova} develops inference for
averages of intersection bounds and extends her approach to functionals
involving LP values. More broadly, through our reduction of inference on
LP values to testing the feasibility of systems of linear inequalities,
our paper is related to the large literature on inference in moment
inequality models; see, among others, \citet{IM04}, \citet{CHT}, \citet{BM08}, \citet{RS},
\citet{DAGS}, \citet{CBS}, \citet{ARP}, \citet{CS}, and \citet{FSST}.

The remainder of the paper is organized as follows. Section~\ref{SecExemp}
presents several motivating applications in which inference on the value
of an LP arises. Section~\ref{SecAC} establishes the connection between
inference on LP values and testing the feasibility of systems of linear
inequalities. Section~\ref{SecInference} introduces our inference
procedures and discusses their implementation. Section~\ref{SecSP} states
the assumptions and main theoretical results, with proofs deferred to the
Appendix. Section~\ref{SecMC} presents simulation evidence.


\subsection{Notation}  
Let \( x \) be a vector in Euclidean space and \( A \) a matrix. Unless stated otherwise, we use \( \|x\| \) to denote the Euclidean norm of \( x \), \(\|x\|_p\) ($1\leq p \leq \infty$) for its general $l_p$-norm, and  \( \|A\| \) to denote the spectral norm of \( A \). We use $ \mathbb{R}_{+}^{p}$ to denote the set of elements of $ \mathbb{R}^{p}$ with non-negative entries. We use $\mathbbm{1}$ to denote a vector with all entries equal to one. For a given vector \( x \), we define \( (x)_+ \) and \( (x)_- \) as the positive and negative parts of \( x \), respectively, where the \( i^{\text{th}} \) entry of \( (x)_+ \) is given by \( \max\{x_i, 0\} \) and the \( i^{\text{th}} \) entry of \( (x)_- \) is given by \( \max\{-x_i, 0\} \). Given a point \( x \) and a subset \( A \) of some Euclidean space, we denote the Euclidean distance from \( x \) to \( A \) by \( d(x, A) \).  For two sets \( A \) and \( B \), we define \( \overset{\rightarrow}{d}_H(A,B) \) and \( d_H(A,B) \) as the directed (one-sided) Hausdorff distance from \( A \) to \( B \) and the Hausdorff distance between \( A \) and \( B \), respectively. These are given by $\overset{\rightarrow}{d}_H(A,B) = \sup_{x \in A} d(x, B)$, $d_H(A,B) = \max\{\overset{\rightarrow}{d}_H(A,B), \overset{\rightarrow}{d}_H(B,A)\}$. Given a matrix \( A \in \mathbb{R}^{m \times n} \) and a subset of row indices \( I \subset [m] \), where for $m\in \mathbb{N}$ we have $[m]:=\{1,\cdots,m\}$, we denote by \( A_I \) the submatrix obtained by retaining only the rows of \( A \) indexed by \( I \) and deleting all other rows. Given a vector \( b \in \mathbb{R}^m \) and a subset \( I \subset [m] \), we define \( b_I \) analogously. We use \( \Rightarrow \), \( \xrightarrow{p} \), and \( \xRightarrow{p} \) to denote convergence in distribution, convergence in probability, and convergence in distribution in probability, respectively. We use $d_{\mathrm{Pr}}(\cdot,\cdot)$ to denote the Prohorov distance.

\section{Motivating Examples}
\label{SecExemp}

This section presents two examples in which the linear program \eqref{eqnLP} arises: one in a point-identified setting and the other in a partially identified setting. Additional motivating examples can be found in \cite{JHTR}, \cite{BG}, \cite{AV}, \cite{chen2026}, and \cite{CSS}.
\begin{example}[Welfare in the presence of peer effects]
\label{ExBh}

\citet{bhattacharya} studies how individuals should be assigned to peer groups
to maximize aggregate outcomes in the presence of heterogeneous peer effects.
In his leading example, a college assigns incoming freshmen to
double-occupancy dorm rooms to maximize mean first-year academic performance.
The value of the optimal assignment provides a useful first-best welfare
benchmark and can be used to assess whether the gains from reallocating
students justify its implementation costs.

Suppose students belong to one of three observable types
$i\in\{w,b,o\}$, and let $x=(x_{ww},x_{bb},x_{oo},x_{wb},x_{wo},x_{bo})\tr$, where $x_{ij}=x_{ji}$ is the fraction of rooms occupied by students of types
$i$ and $j$. Let $\rho_{ij}$ denote the expected average outcome of the two
students in an $(i,j)$ room, and let $\pi=(\pi_w,\pi_b,1-\pi_w-\pi_b)\tr$ denote the shares of the three types in the incoming class. The maximal
average outcome attainable through room assignment is the value of the LP
\begin{equation}
\label{eq:peerLP}
v=\max_x \langle \rho, x \rangle
\qquad\text{s.t.}\qquad
Ax=2\pi,\quad x\geq0,
\end{equation}
where
\[
A=
\begin{pmatrix}
2&0&0&1&1&0\\
0&2&0&1&0&1\\
0&0&2&0&1&1
\end{pmatrix}.
\]
The constraints ensure that the assignment respects the shares of each
student type; the factor of two arises because each room contains two
students.

Writing the equality constraints as pairs of inequalities, \eqref{eq:peerLP}
takes the form \eqref{eqnLP},
\[
v=\max_x \langle \rho, x \rangle
\qquad\text{s.t.}\qquad
Mx\leq\beta,
\]
where $M=(A^\top, -A^\top, -I_6)^\top$ is known, whereas $\rho$ must be estimated. The vector $\beta=(2\pi^\top,-2\pi^\top,0_6)^\top$ is
known when the type composition of the incoming class is observed and must
otherwise be estimated through $\pi$. For example, $\rho$ can be estimated
from a pilot experiment in which students are randomly assigned to rooms.

To illustrate why inference on $v$ is useful, let $v_{\mathrm{SQ}}$ denote
welfare under the status-quo assignment mechanism. Then $v-v_{\mathrm{SQ}}$ is the largest welfare gain attainable through reassignment. If implementing
a new assignment mechanism entails a per-student cost $c$, a planner may wish
to determine whether
\[
v-v_{\mathrm{SQ}}>c.
\]
Our inference method applies whether $v_{\mathrm{SQ}}$ is known or estimated.
In either case, $v-v_{\mathrm{SQ}}$ can be expressed as the value of a new LP
by augmenting the choice vector and absorbing $v_{\mathrm{SQ}}$ into the
objective function, using the same construction as in Example~\ref{ExKKLS}.
Thus, inference on whether the gains from an optimal reassignment exceed its
implementation costs is an instance of inference on the value of
\eqref{eqnLP}.
\end{example}

\begin{example}[Counterfactual welfare in dynamic discrete choice models]\label{ExKKLS}
\citet{KSS} and \citet{KKLS} study counterfactual policy effects in single-agent dynamic discrete choice models. We illustrate how their framework gives rise to \eqref{eqnLP} using their market-entry example. See Section \ref{SecDetEx_KKLS} of the Appendix for additional details.

In each period $t$, a firm chooses whether to operate in a market, $a_t\in\mathcal{A}=\{0,1\}$, where $0$ denotes staying out and $1$ denotes operating. The state is $z_t=(k_t,w_t)$, where $k_t=a_{t-1}$ records the firm's previous action and $w_t\in\{w_l,w_h\}$ is an exogenous profitability shock. Let $\pi_a\in\mathbb{R}^4$ denote the flow payoffs from action $a$ over the four possible states, and let $\pi=(\pi_0\tr,\pi_1\tr)\tr \in\mathbb{R}^8$.

Following \citet{KKLS}, these payoffs can be parameterized as
\begin{equation}\label{eqnKKLS}
\pi_0=
\begin{pmatrix}
oo\\
s
\end{pmatrix},
\qquad
\pi_1=
\begin{pmatrix}
vp-fc-ec\\
vp-fc
\end{pmatrix},
\end{equation}
where $oo$, $s$, $vp$, $fc$, and $ec$ are $2\times1$ vectors indexed by $w\in\{w_l,w_h\}$, representing, respectively, the outside option, scrap value, variable profit, fixed cost, and entry cost.

Flow payoffs are generally not point identified in dynamic discrete choice models (see \citealp{Rust,MT}). Point identification is usually obtained by normalizations, such as zero scrap values or zero fixed costs, and \citet{KSS} show that counterfactual predictions can be sensitive to them. \citet{KKLS} instead consider weaker restrictions that lead to partial identification. For example, they allow fixed costs and scrap values to vary with the profitability shock while imposing economically motivated sign and shape restrictions on the payoff components. Together with the model's linear restrictions on the payoffs, they imply that the identified set for $\pi$ is a polytope, $\Pi^I=\{\pi:R\pi\le r\}$, where $R$ and $r$ depend on estimable objects.

Consider the long-run average change in firm value induced by a counterfactual policy, $\theta_V = \tilde f\tr\tilde V - f\tr V$, where $f$ and $\tilde f$ are the baseline and counterfactual steady-state distributions of the state, and $V$ and $\tilde V$ are the corresponding vectors of firm values. For counterfactuals that leave the payoff and the transitions of the reference action unchanged, such as a subsidy to action $1$, $\theta_V$ can be written as $\theta_V=\gamma(\pi)\tr\pi+c_0(\pi)$, where $\gamma$ and $c_0$ depend on $\pi$ only through the counterfactual choice probabilities. Therefore, in general, $\theta_V$ is a nonlinear function of $\pi$. However, for some important counterfactuals, such as the lump-sum fixed cost subsidies and, when variable profits are known, the subsidies proportional to revenues considered in \citet{DRT}, the counterfactual choice probabilities are point identified \citep[see, e.g.,][]{KSS}. We show (see Proposition~\ref{prop:linear_thetaV} in Appendix~\ref{SecDetEx_KKLS}) that in these cases $\gamma(\pi)=\gamma$ and $c_0(\pi)=c_0$ are point identified, so that $\theta_V$ is affine in the partially identified payoff vector $\pi$.

Consequently, the identified bounds on $\theta_V$ are values of linear programs. The upper bound can be written in the form (see Remark~\ref{rem:affine_to_LP})
\[
v_0=\max_x \langle \rho, x \rangle
\qquad\text{subject to}\qquad
Mx\le\beta,
\]
with an analogous expression for the lower bound. Thus, inference on $\theta_V$ reduces directly to inference on the value of \eqref{eqnLP}. In this application, $M$, $\rho$, and $\beta$ are all unknown and must be estimated.

Testing the null hypothesis that the lower bound on $\theta_V$ is non-positive provides a robust conclusion about the effect of the policy: rejection implies that the counterfactual policy raises long-run average firm value for every payoff vector consistent with the data and the maintained model restrictions.
\end{example}

\section{Alternative Characterization of the Inference Problems}
\label{SecAC}

In this section, we use linear programming duality to obtain alternative characterizations of $v_0$ in \eqref{eqnLP}, as well as of the values above and below $v_0$. The following proposition gives the main result of the section, showing that these sets coincide with the values for which certain systems of linear inequalities are feasible. The result follows directly from strong duality for linear programming, so the proof is omitted.

\begin{proposition}
\label{PropChar}
Suppose that the value $v_0$ of the linear program \eqref{eqnLP} is finite. Then the following statements hold:
\begin{enumerate}
    \item \label{partUB} $v \in (-\infty, v_0]$ if and only if
        \begin{equation}
    \label{eqnUB}
    \left\{
    x \in \mathbb{R}^d
    \;\middle|\;
    Mx \leq \beta
    \ \text{and} \
    \langle \rho, x \rangle \geq v
    \right\}
    \neq \emptyset.
    \end{equation}

    \item \label{partLB} $v \in [v_0, +\infty)$ if and only if
       \begin{equation}
    \label{eqnLB}
    \left\{
    y \in \mathbb{R}^p
    \;\middle|\;
    y \geq 0,\;
    M^\top y = \rho,
    \ \text{and} \
    \langle \beta, y \rangle \leq v
    \right\}
    \neq \emptyset.
    \end{equation}

    \item \label{partTS} $v = v_0$ if and only if
       \begin{equation}
    \label{eqnTS}
    \left\{
    (x,y) \in \mathbb{R}^{d+p}
    \;\middle|\;
    y \geq 0,\;
    Mx \leq \beta,\;
    M^\top y = \rho,\;
    \langle \rho, x \rangle \geq v,\;
    \langle \beta, y \rangle \leq v
    \right\}
    \neq \emptyset.
    \end{equation}
\end{enumerate}
\end{proposition}

We use Proposition \ref{PropChar} to develop one- and two-sided inference procedures for $v_0$. We show that such inference can be carried out by inverting tests of hypotheses of the form \eqref{eqnHyp}.

Consider, for example, the construction of upper one-sided confidence intervals. Part~\ref{partUB} of Proposition~\ref{PropChar} shows that $v \leq v_0$ if and only if the system
\[
Mx \leq \beta
\quad \text{and} \quad
\langle \rho, x \rangle \geq v
\]
admits a solution. Equivalently, $v \leq v_0$ if and only if the hypothesis $H_0$ in \eqref{eqnHyp} holds with
\begin{equation}
\label{eqnIP1}
A=
\begin{bmatrix}
-M \\
\rho^\top
\end{bmatrix},
\qquad
b(v)=
\begin{bmatrix}
-\beta \\
v
\end{bmatrix},
\qquad
\text{and } \eta = x.
\end{equation}
Similarly, for lower one-sided confidence intervals, Part~\ref{partLB} of Proposition~\ref{PropChar} implies that $v \geq v_0$ if and only if the hypothesis $H_0$ in \eqref{eqnHyp} holds with
\begin{equation}
\label{eqnIP2}
A=
\begin{bmatrix}
M^\top \\
-M^\top \\
-\beta^\top \\
I_{p\times p}
\end{bmatrix},
\qquad
b(v)=
\begin{bmatrix}
\rho \\
-\rho \\
-v \\
0_{p\times 1}
\end{bmatrix},
\qquad
\text{and } \eta = y.
\end{equation}

Finally, for two-sided confidence intervals, Part~\ref{partTS} of Proposition~\ref{PropChar} implies that $v = v_0$ if and only if the hypothesis $H_0$ in \eqref{eqnHyp} holds with
\begin{equation}
\label{eqnIP3}
A=
\begin{bmatrix}
-M & 0_{p\times p} \\
\rho^\top & 0_{1\times p} \\
0_{d\times d} & M^\top \\
0_{d\times d} & -M^\top \\
0_{1\times d} & -\beta^\top \\
0_{p\times d} & I_{p\times p}
\end{bmatrix},
\qquad
b(v)=
\begin{bmatrix}
-\beta \\
v \\
\rho \\
-\rho \\
-v \\
0_{p\times 1}
\end{bmatrix},
\qquad
\text{and } \eta =
\begin{bmatrix}
x \\
y
\end{bmatrix}.
\end{equation}

In the next section, we describe a modification of the procedure in \cite{GM25} that we use to invert the tests in~\eqref{eqnHyp} and construct one- and two-sided confidence intervals for $v_0$. Alternatively, such confidence intervals can be obtained by inverting tests based on the procedures of \cite{CSS} and \cite{BPSSMT}. Our modification exploits the special LP structure of the systems in~\eqref{eqnHyp}; relative to a direct application of \cite{GM25}, it reduces computation and restores the power properties established below. The simulations also show that analogous box augmentations can substantially improve the power of \cite{CSS} and \cite{BPSSMT}.


\section{Construction of confidence intervals}
\label{SecInference}

The inference procedures for upper one-sided, lower one-sided, and two-sided confidence intervals share a common set of preliminary objects, which we introduce first.

Depending on the inferential goal, let $A$ and $b(v)$ denote the matrices in \eqref{eqnIP1}--\eqref{eqnIP3}, and let $\hat c=\operatorname{vec}\bigl([\hat A\ \hat b(v)]\bigr)$ be a $\sqrt N$-consistent estimator of $c=\operatorname{vec}\bigl([A\ b(v)]\bigr)$. Let $\Sigma$ denote the asymptotic variance matrix of $\hat{c}$ and define $D=\operatorname{diag}(\Sigma)^{1/2}$. Assumption~\ref{ass1} below formalizes the availability of such an estimator. Finally, let $\hat{\Sigma}$ and $\hat{D}$ denote consistent estimators of $\Sigma$ and $D$, respectively.

Let $\sigma_{A_{ij}}$ and $\sigma_{b_i}$ denote the asymptotic standard deviations of the estimators of the $(i,j)$-th entry of $A$ and the $i$-th entry of $b$, respectively, and let $\hat{\sigma}_{A_{ij}}$ and $\hat{\sigma}_{b_i}$ denote the corresponding entries of $\hat{D}$. Define the diagonal matrices $\Omega$ and $\hat{\Omega}$ by 
\begin{equation}\label{eq:Omega}
\Omega_{ii}
=
\sigma_{b_i}
\vee
\max_{j\in[d_A]}\sigma_{A_{ij}},
\qquad
\hat{\Omega}_{ii}
=
\hat{\sigma}_{b_i}
\vee
\max_{j\in[d_A]}\hat{\sigma}_{A_{ij}},
\end{equation}
for $i=1,\ldots,p_A$, where $p_A$ and $d_A$ denote the numbers of rows and columns of $A$, respectively.

Consider the auxiliary system 
\[
\mathcal{O}
:=
\{(x,y)\mid Mx\leq \beta,\ y\geq 0,\ M^\top y=\rho,\ \beta^\top y=\rho^\top x\},
\]
which can equivalently be written as
\begin{equation}
\label{eqnAux}
\mathcal{O}
=
\{z=(x,y)\mid \pi\leq \Pi z\},
\end{equation}
where
\[
\Pi=
\begin{pmatrix}
-M & 0 \\
0 & -I \\
0 & M^\top \\
0 & -M^\top \\
\rho^\top & -\beta^\top \\
-\rho^\top & \beta^\top
\end{pmatrix},
\qquad
\pi=
\begin{pmatrix}
-\beta \\
0 \\
\rho \\
-\rho \\
0 \\
0
\end{pmatrix}.
\]
Note that by the strong duality theorem of LP, $(x,y)\in {\cal O}$ if and only if $x$ is optimal for \eqref{eqnLP} and $y$ is optimal for its dual. Let the diagonal matrix $S$ be defined in relation to $\operatorname{vec}([\Pi\ \pi])$ in the same way that $\Omega$ is defined in relation to $\operatorname{vec}([A\ b(v)])$. Let $\hat{\Pi}$, $\hat{\pi}$, and $\hat{S}$ denote the corresponding sample counterparts. Note that the matrices $\Sigma$, $D$, $\Omega$, $\Pi$, $\pi$, and $S$, as well as their sample counterparts, do not depend on $v$.

Next, define
\begin{equation}
\label{eqnEAM7}
\hat{f}_N
=
\max\Bigl\{
\hat{\pi}^\top \lambda
\ \Big|\
\lambda\geq 0,\ 
\hat{\Pi}^\top\lambda=0,\ 
\mathbbm{1}^\top\hat{S}\lambda\leq 1
\Bigr\}.
\end{equation}

The nuisance parameter $\hat{z}=(\hat{x},\hat{y})$ is then obtained as the solution to the quadratic program (QP)
\begin{equation}
\label{eqnEAM6}
\hat{z}
:=
\arg\min
\Bigl\{
\|z\|^2
\ \Big|\
\hat{\pi}-\hat{\Pi}z
\leq
\Bigl(\hat{f}_N+\frac{\kappa_{1N}}{\sqrt{N}}\Bigr)\hat{S}\mathbbm{1}
\Bigr\}.
\end{equation}
The QP in~\eqref{eqnEAM6} does not depend on $v$ and therefore needs to be solved only once; its solution can subsequently be used for all values of $v$ on the grid. The tuning parameter $\kappa_{1N}$ is assumed to satisfy
\[
\kappa_{1N}=o(N^{1/4})
\qquad\text{and}\qquad
\kappa_{1N}\to\infty.
\]
The tuning parameters $\kappa_{2N}$ and $\kappa_{3N}$, defined below, are assumed to satisfy
\[
\kappa_N=o(N^{1/2})
\qquad\text{and}\qquad
\kappa_N\to\infty.
\]
In our simulations, we set $\kappa_{1N}=\log\log(N)$ and $\kappa_{2N}=\kappa_{3N}=\kappa_N=\sqrt{\log(N)}$.

\paragraph*{Upper one-sided intervals.}
For upper one-sided confidence intervals, let $A$ and $b(v)$ be defined as in equation~\eqref{eqnIP1}. We modify the system $b(v)\leq A\eta$ by imposing the additional \emph{box constraints} \footnote{All conclusions remain valid if the box-constraint constant $c=5$ is replaced by any fixed $c>1$. Table~\ref{tab:KKLS_vp} in Section~\ref{SecMC} examines sensitivity to this choice: larger values of $c$ improve coverage but reduce power, while smaller values may improve power at the cost of poorer finite-sample coverage.}
\begin{equation}
\label{eqnmod}
-\!\left(5\|\hat{x}\|_{\infty}+\frac{\kappa_{1N}}{N^{1/4}}\right)\mathbbm{1}
\leq
\eta
\leq
\!\left(5\|\hat{x}\|_{\infty}+\frac{\kappa_{1N}}{N^{1/4}}\right)\mathbbm{1}.
\end{equation}
Let $A_{\mathrm{aug}}$ and $b_{\mathrm{aug}}(v)$ denote the matrix and vector obtained by augmenting the original system with these bounds. Specifically,\footnote{Here, we think of $b_{\mathrm{aug}}$ as a function of the inputs $a\geq0$ and $v\in\mathbb{R}$.}
\begin{equation}
\label{eqnmod_mat}
A_{\mathrm{aug}}=
\begin{bmatrix}
-M\\
\rho^\top\\
I\\
-I
\end{bmatrix},
\qquad
b_{\mathrm{aug}}(\hat{a}_N,v)=
\begin{bmatrix}
-\beta\\
v\\
-\hat a_N\mathbbm{1}\\
-\hat a_N\mathbbm{1}
\end{bmatrix},
\end{equation}
where
\begin{equation}\label{eq:box}
\hat a_N
=
5
\|\hat{x}\|_{\infty}
+
\frac{\kappa_{1N}}{N^{1/4}}
,
\end{equation}
and $\hat{x}$ denotes the first component of $\hat{z}$, the solution to the quadratic program~\eqref{eqnEAM6}.

Next, let $\Omega_{\mathrm{aug}}$ be the standardizing matrix associated with the augmented system. It is obtained from $\Omega$ by setting to zero the diagonal entries corresponding to the additional rows introduced by \eqref{eqnmod}. Equivalently, we augment the original system $b(v)\leq A\eta$ with the bounds in \eqref{eqnmod}, while treating these added constraints as deterministic rather than estimated. In particular, when constructing $\hat{b}_{\mathrm{aug}}$ and $\hat{\Omega}_{\mathrm{aug}}$, we treat $\hat{a}_N$ as a deterministic quantity, just as we do $v$.

For each value of $v$ on a prespecified grid in $\mathbb{R}$, define the test statistic \footnote{In the simulations in Section~\ref{SecMC}, we describe a modified estimator $\hat{\Omega}_{\mathrm{aug}}$ that still satisfies the assumptions in Section~\ref{SecSP} but performs better in finite samples. See Section \ref{SecDetMC_AV_BM} \label{footnote:omega}}
\begin{equation}
\label{eqnEAM2}
\hat{T}_N(v)
:=
\max\Bigl\{
\hat{b}_{\mathrm{aug}}(\hat{a}_N,v)^\top \lambda
\ \Big|\
\lambda\geq 0,\ 
\hat{A}_{\mathrm{aug}}^\top\lambda=0,\ 
\mathbbm{1}^\top\hat{\Omega}_{\mathrm{aug}}\lambda\leq 1
\Bigr\}.
\end{equation}

For $\hat{e}_N:=\max\{\hat{a}_N-3\kappa_{1N}/N^{1/4},0\}$, let $\hat{\Delta}(v)$ denote the set 
\begin{equation}
\label{eqnEAM5}
\hat{\Delta}(v):=\left\{
\lambda\geq 0
\ \middle|\
-\frac{\kappa_{2N}}{\sqrt{N}}\mathbbm{1}
\leq
\hat{A}_{\mathrm{aug}}^\top\lambda
\leq
\frac{\kappa_{2N}}{\sqrt{N}}\mathbbm{1},
\;
\mathbbm{1}^\top\hat{\Omega}_{\mathrm{aug}}\lambda\leq 1,
\;
\hat{b}_{\mathrm{aug}}\!\left(\hat{e}_N,v\right)^\top\lambda
\geq
\hat{T}_N(v)-\frac{\kappa_{3N}}{\sqrt{N}}
\right\}.
\end{equation}

Let $\hat{H}_N(\cdot;v):=\mathcal{L}(v^* \mid \mathbb{W})$ denote the conditional distribution of $v^*$ given the data $\mathbb{W}$, where $v^*$ is the value of the LP
\begin{equation}
\label{eqnEAM4}
v^*
=
\max_{\lambda\in\hat{\Delta}(v)}
\left\langle
\zeta_{b_{\mathrm{aug}}}^*-\zeta_{A_{\mathrm{aug}}}^*\hat{x},
\lambda
\right\rangle.
\end{equation}
where $(\zeta_{b_{\mathrm{aug}}}^*,\zeta_{A_{\mathrm{aug}}}^*)$ denotes a bootstrap estimator of the distribution of \footnote{Here again, $\hat{a}_N$ is treated as deterministic. Consequently, the resulting asymptotic distribution depends on neither $\hat{a}_N$ nor $v$.}
\[
\left(
\sqrt{N}\bigl(\hat{b}_{\mathrm{aug}}(\hat{a}_N,v)-b_{\mathrm{aug}}(\hat{a}_N,v)\bigr),
\;
\sqrt{N}\bigl(\hat{A}_{\mathrm{aug}}-A_{\mathrm{aug}}\bigr)
\right).
\]
Such an estimator can be constructed directly from $\xi_N^*$ in Assumption~\ref{ass3}.

As above, when constructing $(\zeta_{b_{\mathrm{aug}}}^*,\zeta_{A_{\mathrm{aug}}}^*)$, we treat the constraints introduced in~\eqref{eqnmod} as deterministic. And since all deterministic constraints have
zero sampling variance, their corresponding components in
$(\zeta_{b_{\mathrm{aug}}}^*,\zeta_{A_{\mathrm{aug}}}^*)$ are identically zero. Thus, although the
multipliers associated with the deterministic constraints enter
$\hat{\Delta}(v)$, they do not contribute directly to the bootstrap
objective in \eqref{eqnEAM4}; all sampling variation in the bootstrap
process arises from the stochastic constraints. 

Given a significance level $\alpha\in(0,1/2)$, we reject the null hypothesis in favor of the alternative if and only if
\[
\sqrt{N}\hat{T}_N(v)
>
\hat{c}_N(1-\alpha;v),
\]
where $\hat{c}_N(1-\alpha;v)$ denotes the $(1-\alpha)$-quantile of $\hat{H}_N(\cdot;v)$.

We define the upper one-sided confidence interval as
\begin{equation}
\label{eqnUBCI}
CI_{1-\alpha}^U := (-\infty, \hat{v}_U],
\end{equation}
where $\hat{v}_U$ is the \textit{largest} value of $v$ such that $\sqrt{N}\hat{T}_N(v) \leq \hat{c}_N(1-\alpha;v)$.
\paragraph*{Lower one-sided intervals.}
The construction of lower one-sided confidence intervals is analogous to the upper one-sided case. The only difference is that $A$ and $b(v)$ are now given by equation~\eqref{eqnIP2}, and are augmented with the box-constraint
\begin{equation}
\label{eqnmod1}
-\hat{a}_N\mathbbm{1}
\leq \eta \leq
\hat{a}_N\mathbbm{1},
\end{equation}
where $\hat{a}_N=5\|\hat{y}\|_{\infty} + \frac{\kappa_{1N}}{N^{1/4}}$, and $\hat{y}$ denotes the second component of the solution $\hat{z}$ to the QP~\eqref{eqnEAM6}.

The test statistic, the feasible set $\hat{\Delta}$, and the rejection rule remain unchanged (with $A$ and $b$ now given by equation \eqref{eqnIP2}). The only modification appears in the construction of the simulated distribution: the quantity $v^*$ is now defined by
\begin{equation}
\label{eqnIPUI1}
v^*
=
\max_{\lambda\in\hat{\Delta}}
\left\langle
\zeta_{b_{\mathrm{aug}}}^*-\zeta_{A_{\mathrm{aug}}}^*\hat{y},
\lambda
\right\rangle.
\end{equation}
where, as above, $\hat{y}$ denotes the second component of $\hat{z}$.

We define the lower one-sided confidence interval as
\begin{equation}
\label{eqnLBCI}
CI_{1-\alpha}^L := [\hat{v}_L, \infty),
\end{equation}
where $\hat{v}_L$ is the \textit{smallest} value of $v$ such that $\sqrt{N}\hat{T}_N(v) \leq \hat{c}_N(1-\alpha;v)$.

\paragraph*{Two-sided intervals.}
For two-sided confidence intervals, let $A$ and $b(v)$ be defined as in equation~\eqref{eqnIP3}. We augment the system $b(v)\leq A\eta$ with the additional constraints
\begin{equation}
\label{eqnmod2}
-\hat{a}_N\mathbbm{1}
\leq \eta \leq
\hat{a}_N\mathbbm{1},
\end{equation}
where $\hat{a}_N=5\|\hat{z}\|_{\infty} + \frac{\kappa_{1N}}{N^{1/4}}$ and $\hat{z}$ is the solution to the QP \eqref{eqnEAM6}. The test statistic, feasible set $\hat{\Delta}$, and rejection rule remain unchanged.

The simulated distribution is constructed using
\begin{equation}
\label{eqnIPUI2}
v^*
=
\max_{\lambda\in\hat{\Delta}}
\left\langle
\zeta_{b_{\mathrm{aug}}}^*-\zeta_{A_{\mathrm{aug}}}^*\hat{z},
\lambda
\right\rangle.
\end{equation}
where, as above, $\hat{z}$ is the solution to equation~\eqref{eqnEAM6}.

We define the two-sided $(1-\alpha)$ confidence interval as
\begin{equation}
\label{eqnTSCI}
CI_{1-\alpha} := [\hat{v}_L,\hat{v}_U],
\end{equation}
where $\hat{v}_L$ and $\hat{v}_U$ are respectively the smallest and largest values of $v$ such that $\sqrt{N}\hat{T}_N(v) \leq \hat{c}_N(1-\alpha;v)$.

\begin{remark}
\label{remGM}
The inference procedures proposed above are computationally cheaper than a direct application of the method in \cite{GM25}. In particular, applying the procedure in \cite{GM25} directly to the hypotheses in equation~\eqref{eqnHyp} would require solving a quadratic program that depends on $v$ at every value of $v$ on the grid. By contrast, the procedures developed here require solving only a single quadratic program, which is independent of $v$ and can therefore be used for all grid values, as well as for upper one-sided, lower one-sided, and two-sided confidence intervals.

This modification yields a substantial computational simplification. As shown below, however, it does not come at the expense of statistical performance. Specifically, the resulting procedures remain asymptotically valid, and possess nontrivial power against local alternatives, and are consistent against fixed alternatives. 
\end{remark}
 
\begin{remark}
The method of \cite{CSS} also involves the estimation of a nuisance parameter analogous to that in equation~\eqref{eqnEAM6}. We believe that the arguments developed in this paper also apply to the approach of \cite{CSS}, and that their procedure can be modified so that the nuisance parameter needs to be computed only at a single value on the grid.\footnote{More precisely, the common value of the nuisance parameter can be obtained by solving equation (19) in \cite{CSS}, where the matrices $B$, $C$, and the vectors $\mu$ and $d$ that appear there are constructed from the system $e \leq Ez$, with $E$ and $e$ defined as above.}
\end{remark}


\section{Statistical properties}
\label{SecSP}

In this section, we establish the asymptotic validity of the inference procedures discussed in Section~\ref{SecInference}. We first introduce the assumptions required for pointwise asymptotic validity of our procedure. Conditions ensuring uniform validity are discussed further below.

Let $Q$ denote the true data-generating process, and define
\[
\mu(Q) := (\rho(Q)\tr, \beta(Q)\tr, \operatorname{vec}(M(Q))\tr)\tr.
\]
Here, $\rho(Q)$, $\beta(Q)$, and $M(Q)$ denote the values of the parameters $\rho$, $\beta$, and $M$, respectively, under the data-generating process $Q$. The following three assumptions require, respectively, that the LP value be finite, that a $\sqrt{N}$-consistent and asymptotically normal estimator of $\mu(Q)$ be available, and that its asymptotic variance be consistently estimable. The latter two conditions are satisfied, for instance, when $\rho$, $\beta$, and $M$ are expectations and the entries of $\hat{\rho}$, $\hat{\beta}$, and $\hat{M}$ are the corresponding sample means.\footnote{Asymptotic normality is not required for the validity of our procedure. What is necessary is that $\hat{\mu}$ has a nondegenerate asymptotic distribution that can be consistently estimated or simulated.}

\begin{assumption}
\label{ass0}
The LP with inputs $\mu(Q)$ has finite value.
\end{assumption}

\begin{assumption}
\label{ass1}
 There exists a consistent estimator $\hat{\mu}$ of $\mu(Q)$ and a nonzero symmetric matrix $V(Q)$, representing the asymptotic variance of $\hat{\mu}$, such that
\[
\sqrt{N}\bigl(\hat{\mu} - \mu(Q)\bigr)\ \Rightarrow \ \mathcal{N}\bigl(0, V(Q)\bigr).
\]
\end{assumption}

\begin{assumption}
\label{ass2}
There exists a consistent estimator $\hat{V}$ of $V(Q)$.
\end{assumption}

The next assumption requires the availability of a consistent bootstrap-like estimate of the asymptotic distributions $\mathcal{N}(0,V(Q))$.

\begin{assumption}
\label{ass3}
There exists a random variable $\zeta_N^*$, whose distribution is entirely determined by the realized sample $\mathbb{W}$, such that for all $\epsilon > 0$,
\[
\lim_{N \to \infty} \ Q\left( d_{\mathrm{Pr}}\left( \mathcal{L}(\zeta_N^* \mid \mathbb{W}), \mathcal{N}(0, V(Q)) \right) > \epsilon \right) = 0.
\]
\end{assumption}
Finally, we introduce an assumption that guarantees that the asymptotic distribution used to compute our critical values is nondegenerate. Let $\eta^*=x^*$, $\eta^*=y^*$, or $\eta^*=z^*$ according to whether $(A,b(v_0))$ is defined by \eqref{eqnIP1}, \eqref{eqnIP2}, or \eqref{eqnIP3}, respectively, where $z^*:=(x^*,y^*)$ is the minimum-norm element of the set $\mathcal{O}$ defined in \eqref{eqnAux}.\footnote{Since $(x,y)\in\mathcal{O}$ for every primal optimal solution $x$ and dual optimal solution $y$, it follows that $x^*$ and $y^*$ are the minimum-norm primal and dual optimal solutions, respectively.} Define
\[
\Delta_0(v_0):=
\arg\max
\Bigl\{
b_{\mathrm{aug}}(5\|\eta^*\|_{\infty},v_0)^\top \lambda
\ \Big|\
\lambda\geq 0,\ 
A_{\mathrm{aug}}^\top\lambda=0,\ 
\mathbbm{1}^\top \Omega_{\mathrm{aug}}\lambda\leq 1
\Bigr\},
\]
where $(A_{\mathrm{aug}},b_{\mathrm{aug}}(v))$ are constructed from $(A,b)$ in either \eqref{eqnIP1}, \eqref{eqnIP2}, or \eqref{eqnIP3}, as described in Section \ref{SecInference}. In Lemma \ref{lemDelta}, we show that $\Delta_0(v_0)$ always contains a nonzero element.

\begin{assumption}
\label{ass4}
Let $(A_{\mathrm{aug}},b_{\mathrm{aug}})$ be constructed from $(A,b)$ as in \eqref{eqnIP1}, \eqref{eqnIP2}, or \eqref{eqnIP3}, and let $(\zeta_{b_{\mathrm{aug}}},\zeta_{A_{\mathrm{aug}}})$ have the asymptotic distribution of \footnote{Here again, although the box constraints are random, we treat them as deterministic. Accordingly, the entries of $(\zeta_{b_{\mathrm{aug}}},\zeta_{A_{\mathrm{aug}}})$ corresponding to the box constraints are taken to be zero. Reasoning along similar lines also shows that the distribution of $(\zeta_{b_{\mathrm{aug}}},\zeta_{A_{\mathrm{aug}}})$ does not depend on $v_0$.}
\[
\left(
\sqrt{N}\bigl(\hat{b}_{\mathrm{aug}}(v_0)-b_{\mathrm{aug}}(v_0)\bigr),
\sqrt{N}\bigl(\hat{A}_{\mathrm{aug}}-A_{\mathrm{aug}}\bigr)
\right).
\]
Then
\[
\max_{\lambda\in\Delta_0(v_0)}
\operatorname{Var}
\!\left(
\left\langle
\zeta_{b_{\mathrm{aug}}}-\zeta_{A_{\mathrm{aug}}}\eta^*,
\lambda
\right\rangle
\right)>0.
\]
\end{assumption}
This assumption is satisfied, for example, if\footnote{Here, $(\zeta_b,\zeta_A)$ denotes the component of $(\zeta_{b_{\mathrm{aug}}},\zeta_{A_{\mathrm{aug}}})$ corresponding to the non-box constraints.} $\zeta_b-\zeta_A\eta^*$ is normally distributed with a positive definite covariance matrix. However, Assumption~\ref{ass4} is substantially weaker, as it only requires the covariance matrix to be nondegenerate in at least one direction. If Assumption~\ref{ass4} fails, our procedure continues to have asymptotically valid coverage provided that the rejection rule is modified to reject whenever
\[
\sqrt{N}\hat{T}_N(v)\geq \hat{c}_N(1-\alpha;v)+\tau,
\]
for any fixed $\tau>0$.

We now state one of the main results of this section, which establishes the asymptotic validity of the procedures proposed in Section~\ref{SecInference}. The proof is deferred to Section~\ref{Secthm1} of the Appendix.

\begin{theorem}
\label{thm1}

Suppose Assumptions~\ref{ass0}--\ref{ass4} hold. Let
\[
\hat C \in \left\{ CI^U_{1-\alpha},\, CI^L_{1-\alpha},\, CI_{1-\alpha} \right\}
\]
denote any of the confidence intervals defined in
Section~\ref{SecInference}.

Then, for every $\alpha \in (0,1/2)$,
\begin{equation}
\label{eqnthm1}
\underline{\lim}_{N\to\infty}
Q\!\left(v_0 \in \hat C\right)
\ge 1-\alpha.
\end{equation}

Moreover, the confidence intervals are asymptotically localized around
$v_0$ in the following sense:
\begin{align}
\label{eqnthm2}
\lim_{t\to\infty}
\underline{\lim}_{N\to\infty}
Q\!\left(
CI^U_{1-\alpha}
\cap
\bigl[v_0+t/\sqrt N,\infty\bigr]
=
\varnothing
\right)
&=1,
\\[0.5em]
\label{eqnthm3}
\lim_{t\to\infty}
\underline{\lim}_{N\to\infty}
Q\!\left(
CI^L_{1-\alpha}
\cap
\bigl(-\infty,v_0-t/\sqrt N\bigr]
=
\varnothing
\right)
&=1,
\\[0.5em]
\label{eqnthm4}
\lim_{t\to\infty}
\underline{\lim}_{N\to\infty}
Q\!\left(
CI_{1-\alpha}
\cap
\bigl[v_0-t/\sqrt N,\,
      v_0+t/\sqrt N\bigr]^c
=
\varnothing
\right)
&=1.
\end{align}

\end{theorem}

\begin{remark}
Equations~\eqref{eqnthm2}--\eqref{eqnthm4} show that both the one-sided and
two-sided testing procedures exhibit nontrivial power against local
alternatives and are consistent against fixed alternatives. Consequently,
the confidence intervals $CI^U_{1-\alpha}$, $CI^L_{1-\alpha}$, and
$CI_{1-\alpha}$ have length of order $N^{-1/2}$, matching the rate achieved
by confidence intervals based on Gaussian asymptotic approximations, such as
those proposed in \cite{BG} (in the ``regular'' case)\footnote{In fact, the rate obtained in \cite{BG} is slower than $N^{-1/2}$ in the ``non-regular'' case; see the comment following Theorem~1 therein.}, \cite{AV} and \cite{chen2026}.

We note that \eqref{eqnthm1} also holds if the confidence intervals
are obtained by inverting tests of the hypotheses in \eqref{eqnHyp} using
the methods of \citet{BPSSMT}, \citet{CSS}, or \citet{GM25}. However,
confidence intervals constructed by inverting these tests do not necessarily
satisfy \eqref{eqnthm2}--\eqref{eqnthm4}. Relative to the procedure in \cite{GM25}, the additional box constraints in~\eqref{eqnmod} are the key source of power in the modified method. In the proof of Theorem~\ref{thm1}, the box constraints allow us to bound the
right-hand side of \eqref{thmAVeqn36} using \eqref{thmAVeqnBC}, and thereby
show that the test statistic becomes large with high probability for
sufficiently large local deviations from the null. Without the additional box constraints, a naive implementation of the procedure in \cite{GM25} may be inconsistent (power does not converge to one) against certain fixed alternatives, leading to confidence intervals that are wider than necessary, despite maintaining asymptotically valid coverage (see Figure~\ref{fig:intro_nobox}).
\end{remark}

In Section~\ref{SecAss} of the Appendix, we describe classes of DGPs over
which the conclusions of Theorem~\ref{thm1} hold uniformly. Following
\citet{GM25}, the required conditions consist essentially of uniform
versions of Assumptions~\ref{ass0}--\ref{ass4}, together with a uniform
bound on the relevant Hoffman constants $\Lambda(A)$.

The need for this additional restriction can be understood by examining
the role of the estimated box constraint in the proof of
Theorem~\ref{thm1}. To extend the pointwise argument uniformly over a
class of DGPs, one needs a uniform counterpart to \eqref{pthm1eqn2} in Lemma~\ref{lemNP}; in
particular, the estimated box constraint must uniformly intersect the nuisance-parameter set $\{\eta\mid b(v)\leq A\eta\}$ with probability approaching one. Uniform
versions of Assumptions~\ref{ass1}--\ref{ass3} alone cannot guarantee
such a result.

To see why, suppose that the conclusion of Lemma~\ref{lemNP} held
uniformly over classes of DGPs satisfying
Assumptions~\ref{assU0}--\ref{assU3} in the Appendix. One could then use sample splitting
to estimate the box-constraint parameter $\hat a_N$ on one subsample and
perform inference on the other, treating the resulting box constraint as
fixed. Applying a procedure such as that of \cite{BPSSMT} on the second
subsample would then yield a nontrivial uniformly valid inference
procedure under Assumptions~\ref{assU0}--\ref{assU3} alone, contradicting
the corresponding impossibility result. Thus, the conclusion of
Lemma~\ref{lemNP} cannot hold uniformly under
Assumptions~\ref{assU0}--\ref{assU3} alone. Assumption~\ref{assU4}
provides one additional restriction under which the required uniform
control of the estimated box constraint can be obtained.

Note that the assumptions that we require for uniformity are weaker than those used to establish uniformity in \cite{BG}. Also, we do not require bounds on the magnitude of dual solutions, as imposed by Assumption~B3 of \cite{AV},\footnote{This assumption is needed in \cite{AV} to establish pointwise validity of the inference procedure, and \cite{AV} does not discuss conditions for uniformly valid inference.} which can be hard to justify, as such bounds depend in a nontrivial way on the potentially unknown inputs $\rho$, $\beta$ and $M$. Choosing the parameter $\overline{v}$ in Assumption B3 of \cite{AV} too
large is analogous to imposing a loose box constraint and therefore leads
to a less powerful procedure. Conversely, choosing $\overline{v}$ too
small is analogous to imposing an invalid box constraint and may result
in a procedure that fails to satisfy the desired coverage requirement. On the other hand, our procedure is computationally more demanding than either of theirs. In particular, it requires test inversion over a grid of candidate values of the LP, and, because the distribution of the test statistic is nonstandard, the corresponding critical values must be computed by simulation.

It is worth emphasizing that the validity of our results does not rely on
constraint qualifications such as LICQ or MFCQ; see \cite{BG} and \cite{KMS22}. Nor do we require the feasible
region to be compact. Our only requirement is that the identified value
$v_0$ be finite.\footnote{\citet{JHTR}, \citet{BG}, and \citet{AV} require
either that the feasible region of the linear program
$\{x \mid Mx \leq \beta\}$ be compact, or that there exist a constant
$C>0$ such that, without affecting the value of interest, the feasible
region can be restricted to the bounded set
$\{x \mid Mx \leq \beta,\ \|x\|_{\infty}\leq C\}$.}

As we show in the next two results, when these constraint qualifications
hold, the conclusions of Theorem \ref{thm1} can be strengthened. In
particular, Theorem \ref{thm2} shows that, under LICQ, the inequalities in
\eqref{eqnthm1} can be strengthened to equalities. Consequently, the
resulting confidence intervals are asymptotically exact and are therefore
not unnecessarily conservative (these results are similar to the second part of Theorem 4.19 in \cite{GM25}). Furthermore, Theorem \ref{thm3} shows
that, under MFCQ, the procedure of \cite{GM25}, which corresponds to
dropping the box constraints in our procedure,\footnote{More precisely,
setting $\hat{a}_N=\hat{e}_N=\infty$ yields the procedure of \cite{GM25},
except that the nuisance parameter $\hat{\eta}$, which is used in \eqref{eqnEAM4} to compute critical values, is estimated once using
\eqref{eqnEAM6}, rather than separately for each value of $v$ on the
grid (see Remark~\ref{remGM}).} has nontrivial power against local alternatives; that is,
\eqref{eqnthm2}--\eqref{eqnthm4} continue to hold. The proofs of Theorems~\ref{thm2} and~\ref{thm3} are deferred to
Sections~\ref{Secthm2} and~\ref{Secthm3} of the Appendix, respectively.

\begin{theorem}
\label{thm2}
Suppose Assumptions~\ref{ass0}--\ref{ass4} hold. In addition:\footnote{The nonzero requirement pertains to the baseline box construction in \eqref{eq:box}; Remark~\ref{rem:floor} gives a simple modification that also covers zero optimizers.}

\begin{itemize}
\item If the LP~\eqref{eqnLP} has a unique nonzero solution (in particular, if the associated dual LP satisfies the LICQ condition as defined in \cite{BG}), then inequality~\eqref{eqnthm1} holds with equality for the upper confidence intervals $CI^U_{1-\alpha}$.

\item If the dual of LP~\eqref{eqnLP} has a unique nonzero solution (in particular, if the primal LP satisfies the LICQ condition as defined in \cite{BG}), then inequality~\eqref{eqnthm1} holds with equality for the lower confidence intervals $CI^L_{1-\alpha}$.

\item If both the primal and dual LPs associated with~\eqref{eqnLP} have unique nonzero solutions (i.e., the \textit{regular} case in \cite{BG}), then inequality~\eqref{eqnthm1} holds with equality for all one-sided and two-sided confidence intervals.
\end{itemize}
\end{theorem}

\begin{remark}
To be more precise, the condition imposed in the first part of Theorem
\ref{thm2} is weaker than requiring the dual LP associated
with \eqref{eqnLP} to satisfy LICQ. Indeed, as defined in \cite{BG}, the
dual LP satisfies LICQ if the set of optimal solutions to \eqref{eqnLP}
is a singleton \textit{for every} constraint vector $\beta$ (and for fixed $M$ and $\rho$) for which the value of the primal/dual LP is finite. By contrast,
to strengthen \eqref{eqnthm1} to an equality for the upper confidence
interval on $v_0$, we require only that the optimal solution be unique
\textit{for the given} vector $\beta$. Analogous remarks apply
to the second and third parts of Theorem \ref{thm2}.
\end{remark}

\begin{remark}\label{rem:floor}
If the unique optimizer relevant for Theorem~\ref{thm2} is zero, then $\hat e_N$ can equal zero and the proposed procedure may be conservative. This can be avoided by replacing the corresponding box size with
\begin{equation}\label{eq:box_floor}
\hat a_N
=
5\max\{\|\hat\eta\|_\infty,1\}
+
\frac{\kappa_{1N}}{N^{1/4}},
\end{equation}
where $\hat\eta=\hat x$, $\hat y$, or $\hat z$ for the upper, lower, or two-sided interval, respectively. A straightforward modification of the proof shows that the conclusion of Theorem~\ref{thm2} then holds whenever the relevant optimizer is unique, including when it is zero; the remaining results can be modified analogously. Section~\ref{SecDetMC_AV_BM} illustrates the effect of this modification; see Figures~\ref{fig:AV_floor_upper} and~\ref{fig:AV_floor_lower}.
\end{remark}

In the following theorem, we use the linear-programming specialization of MFCQ as defined in
\citet{MFCQ} (see also Lemma \ref{lemMFCQ}). The connection between boundedness of the set of optimal solutions and MFCQ is established in Lemma \ref{lemMFCQ} in Appendix \ref{secAux}

\begin{theorem}
\label{thm3}
Suppose that Assumptions~\ref{ass0}--\ref{ass4} hold. Then the following
statements hold when the box constraints are dropped, i.e.,
$\hat{a}_N=\hat{e}_N=\infty$.

\begin{itemize}

\item If the set of optimal solutions to the primal LP~\eqref{eqnLP} is
bounded, or equivalently, if the feasible region of the
dual LP satisfies MFCQ, then \eqref{eqnthm2} continues to hold.

\item If the set of optimal solutions to the dual LP associated with
\eqref{eqnLP} is bounded, or equivalently, if the feasible region
of the primal LP satisfies MFCQ, then \eqref{eqnthm3} continues to hold.

\item If the sets of optimal solutions to both the primal and dual LPs
associated with \eqref{eqnLP} are bounded, or
equivalently, if the feasible regions of both the primal and dual LPs
satisfy MFCQ, then \eqref{eqnthm2}--\eqref{eqnthm4} continue to hold.

\end{itemize}
\end{theorem}

\begin{remark}
The representation of $b(v)\leq A\eta$ is not unique. Equivalent representations leave the underlying LP value unchanged but can generate different dual feasible and optimal solution sets. Thus, dual uniqueness and boundedness need not be representation invariant, which matters for the lower and two-sided intervals in Theorems~\ref{thm2} and~\ref{thm3}. When possible, it may therefore be useful to avoid redundant constraints and to retain equalities as equalities in the dual derivation leading to \eqref{eqnIP2} and \eqref{eqnIP3}, rather than replacing each equality by two inequalities; the latter introduces two nonnegative multipliers in place of one unrestricted multiplier and can create redundant dual directions.
\end{remark}

\begin{remark}
Theorem~\ref{thm3} has several implications for existing procedures.
First, if the feasible region of the primal LP~\eqref{eqnLP} is bounded,
then a direct application of the procedure in \cite{GM25} for constructing
one-sided upper confidence intervals has nontrivial power against local
alternatives; in particular, \eqref{eqnthm2} continues to hold without
the box constraints. Although we do not establish this result formally,
we conjecture that a similar conclusion may hold for the procedure in
\cite{CSS}. The same conclusion need not hold for the ``direct'' procedure proposed in \cite{BPSSMT}.
Their characterization of the null hypothesis checks for a violation of
the implication of Farkas' lemma in only one direction. Consequently,
their direct procedure may fail to reject with probability approaching one under
some fixed alternatives, even when the feasible region of
\eqref{eqnLP} is bounded.

Analogously, if the feasible region of the dual LP is bounded, the box
constraints can be dropped when constructing one-sided lower confidence
intervals, while \eqref{eqnthm3} continues to hold. Finally, if the
feasible region of the primal LP~\eqref{eqnLP} is bounded and satisfies
MFCQ, then the primal and dual optimal solution sets are both bounded.
Consequently, a direct application of the procedure in \cite{GM25} can
be used to construct both one- and two-sided confidence intervals with
length of order $N^{-1/2}$.
\end{remark}
\begin{remark}
Theorem~\ref{thm3} explains the asymmetry in the power curves in Figure~\ref{fig:intro_nobox}. The lack of power for $v<v_0$ reflects the unboundedness of the set of optimal dual solutions, whereas the strong power for $v>v_0$ is associated with the boundedness of the set of optimal primal solutions. Additional illustrations are provided in Figures~\ref{fig:lower_nobox}--\ref{fig:upper_box}.
\end{remark}

\subsection{Uniform Inference under Deterministic Box Constraints}
\label{SecDetBox}

In this section, we propose an inference procedure that is uniformly valid
over large classes of DGPs of the type commonly considered in the moment
inequality literature, namely, classes satisfying
Assumptions~\ref{assU1}--\ref{assU3} in the Appendix. The key additional structure is that
the systems in \eqref{eqnHyp} are known to contain deterministic
restrictions that restrict the nuisance parameter $\eta$ to a bounded set.
We show that, under this structure, a modified procedure satisfies the coverage conclusions of
Theorem~\ref{thm1} uniformly without imposing
Assumption~\ref{assU4}. The power conclusions of Theorem~\ref{thm1}
continue to hold pointwise for each DGP in the class.

This setting arises naturally in applications. For instance, in
Example~\ref{ExBh}, the elements of the feasible region correspond to
probability vectors and are therefore naturally restricted to lie in a
bounded set. In such settings, the procedure proposed below yields upper
confidence intervals that are uniformly valid over classes of DGPs
satisfying Assumptions~\ref{assU0}--\ref{assU3}.

As noted in the introduction, the impossibility results of \cite{BPSSMT}
and \cite{GM25} imply that nontrivial uniform inference on LP values
generally requires stronger conditions than
Assumptions~\ref{assU0}--\ref{assU3}. In particular, \citet{GM25} show that
testing $H_0:\exists\,\eta$ such that $b\leq A\eta$ is equivalent to
testing whether the value of an associated LP is zero. Hence, a
nontrivial uniformly valid procedure for inference on LP values would
induce a nontrivial uniformly valid test of such feasibility hypotheses.

These impossibility results further imply that, under
Assumptions~\ref{assU1}--\ref{assU3}, any uniformly valid test must have
trivial power against alternatives for which $0\leq A\eta$ admits a
nonzero solution. Equivalently, whenever $\{ \eta:b\leq A\eta\}$ is
nonempty, it must be bounded. Known deterministic restrictions that bound
$\eta$ eliminate precisely these problematic recession directions.

To describe the additional structure formally, let $A_Q$ and
$b_Q(v)$ be defined as in either \eqref{eqnIP1}, \eqref{eqnIP2}, or
\eqref{eqnIP3}, depending on the confidence interval under consideration.
We assume that, for $Q\in\mathcal{Q}$, the systems
\begin{equation}
\label{secUeqn1}
b_Q(v)\leq A_Q\eta
\end{equation}
contain a common subsystem of deterministic constraints given by
\begin{equation}
\label{eqnDBC}
b_D\leq A_D\eta,
\end{equation}
where $A_D$ and $b_D$ do not depend on $Q$, and these constraints define a bounded set $\mathcal{H}_D
:=
\left\{
\eta:
b_D\leq A_D\eta
\right\}$. The stochastic constraints are allowed to vary arbitrarily across
$Q\in\mathcal Q$.

We modify both the test statistic and the construction of the critical
value to exploit the known deterministic restrictions. In particular, we define the test statistic as
\begin{equation}
\label{secUeqn2}
\hat T_N(v)
=
\max_{\lambda\geq0}
\left\{
\hat b(v)^\top\lambda:
\hat A^\top\lambda=0,\ 
\mathbbm{1}^\top\hat\Omega\lambda\leq1
\right\},
\end{equation}
where $A$, $b(v)$, and $\Omega$ are the unaugmented objects discussed in Section \ref{SecInference}.

For the construction of the critical value, for each bootstrap draw define
\begin{equation}
\label{secUeqn3}
v^*
:=
\sup_{\eta\in\mathcal H_D}
\sup_{\lambda\in\hat{\Delta}(v)}
\left\langle
\zeta_b^*-\zeta_A^*\eta,
\lambda
\right\rangle,
\end{equation}
where
\[
\hat{\Delta}(v)
:=
\left\{
\lambda\geq0:
\begin{array}{l}
-\dfrac{\kappa_{1N}}{\sqrt N}\mathbbm{1}
\leq
\hat A^\top\lambda
\leq
\dfrac{\kappa_{1N}}{\sqrt N}\mathbbm{1},\\[2mm]
\mathbbm{1}^\top\hat\Omega\lambda\leq1,\\[1mm]
\hat b(v)^\top\lambda
\geq
\hat T_N(v)-\dfrac{\kappa_{2N}}{\sqrt N}
\end{array}
\right\}.
\]
We let $\hat c_N^D(1-\alpha;v)$ denote the conditional
$(1-\alpha)$-quantile of $v^*$ given the data, and reject the null
hypothesis at $v$ whenever
\[
\sqrt N\hat T_N(v)
>
\hat c_N^D(1-\alpha;v).
\]
Here, the tuning parameters $\kappa_{jN}$, for $j\in \{1,2\}$, are such that $\kappa_{jN}\to \infty$ and $\kappa_{jN}=o(\sqrt{N})$ (we set $\kappa_N=\sqrt{\log(N)}$ in simulations).
Thus, the critical value are now 
constructed by maximizing the bootstrap process over the compact set
$\mathcal H_D$ of nuisance parameter values satisfying the known
deterministic restrictions, rather than at an estimated value of $\eta$ as in Section \ref{SecInference}. The critical value is therefore obtained from
a least-favorable bootstrap distribution over all nuisance parameter values compatible with the deterministic restrictions.

The following theorem establishes that the modified procedure is uniformly
valid over classes of DGPs satisfying
Assumptions~\ref{assU0}--\ref{assU3}. Importantly, uniformity pertains to
the coverage statements. The power conclusions are pointwise, as in
Theorem~\ref{thm1}. The proof is deferred to
Section~\ref{Secthm4} of the Appendix.

\begin{theorem}
\label{thm4}
Let $\mathcal Q$ denote a class of DGPs, and consider the test statistics and
critical values in \eqref{secUeqn2} and \eqref{secUeqn3}. Suppose that
Assumptions~\ref{assU0}--\ref{assU3}, and \ref{assU6} hold. Then:

\begin{itemize}

\item If $A_Q$ and $b_Q(v)$ are defined as in \eqref{eqnIP1}, and the
corresponding system \eqref{secUeqn1} contains deterministic constraints that
bound $\eta$, then
\[
\liminf_{N\to\infty}
\inf_{Q\in\mathcal Q}
Q\left(
v_0(Q)\in CI^U_{1-\alpha}
\right)
\geq 1-\alpha,
\]
and the power conclusion in \eqref{eqnthm2} holds for every $Q\in\mathcal Q$.

\item If $A_Q$ and $b_Q(v)$ are defined as in \eqref{eqnIP2}, and the
corresponding system \eqref{secUeqn1} contains deterministic constraints that
bound $\eta$, then
\[
\liminf_{N\to\infty}
\inf_{Q\in\mathcal Q}
Q\left(
v_0(Q)\in CI^L_{1-\alpha}
\right)
\geq 1-\alpha,
\]
and the power conclusion in \eqref{eqnthm3} holds for every $Q\in\mathcal Q$.

\item If the systems \eqref{secUeqn1} corresponding to both
\eqref{eqnIP1} and \eqref{eqnIP2} contain deterministic constraints that
bound their respective nuisance parameters, then
\[
\liminf_{N\to\infty}
\inf_{Q\in\mathcal Q}
Q\left(
v_0(Q)\in \widehat C_{1-\alpha}
\right)
\geq 1-\alpha
\]
for
\[
\widehat C_{1-\alpha}
\in
\left\{
CI^U_{1-\alpha},
CI^L_{1-\alpha},
CI_{1-\alpha}
\right\},
\]
and the corresponding power conclusions in
\eqref{eqnthm2}--\eqref{eqnthm4} hold for every $Q\in\mathcal Q$.

\end{itemize}
\end{theorem}

\begin{remark}
In many applications, the identified set for the parameter of interest
$\theta$ is an interval $[v_L,v_U]$, where
\[
v_L
=
\min\left\{
\langle \rho, x \rangle:Mx\leq\beta
\right\},
\qquad
v_U
=
\max\left\{
\langle \rho, x \rangle:Mx\leq\beta
\right\}.
\]
Suppose that for the class of DGPs under consideration, the
feasible regions contain a common known set of deterministic constraints that bound $x$. This is, for instance, the case in Example~\ref{ExBh} above. Theorem~\ref{thm4} then
implies that uniformly valid inference on the identified set
$[v_L,v_U]$ is possible under
Assumptions~\ref{assU1}--\ref{assU3}. Indeed, a confidence region for the identified set can be constructed
using a lower confidence bound for $v_L$ and an upper confidence bound
for $v_U$. Since inference on both bounds involves directions that depend only on the primal variables, the deterministic bounds on $x$ are sufficient for the first part of Theorem~\ref{thm4} to apply. Consequently, the confidence region is uniformly valid and has excess length that is (pointwise) of order $N^{-1/2}$.

By contrast, inference in the opposite direction---for example, a lower
confidence bound for $v_U$---requires known deterministic bounds on the
corresponding dual variables for Theorem~\ref{thm4} to apply. Such bounds
may be more difficult to justify in applications. Thus, an important
implication of Theorem~\ref{thm4} is that uniformly valid confidence
regions for identified sets of the form $[v_L,v_U]$ can be constructed
under standard uniformity conditions whenever the feasible region contains
known deterministic constraints that bound the primal variable $x$.
\end{remark}

\begin{remark}
\citet{AV} obtains pointwise-valid inference under Assumptions A0 and A3 therein, which impose known deterministic bounds on the primal and dual variables. Under these restrictions, Theorem~\ref{thm4} yields nontrivial uniformly valid one- and two-sided inference under Assumptions~\ref{assU1}--\ref{assU3}, without conditions such as Assumption U2 of \cite{AV}, which is similar to Assumption~\ref{assU4}.

Computing \eqref{secUeqn3} may be demanding in high dimensions because it requires a nonconvex quadratic program, which is NP-hard in general. Moreover, because \eqref{secUeqn3} maximizes over all nuisance parameter values satisfying the deterministic restrictions, the resulting procedure can be conservative. We therefore view Theorem~\ref{thm4} primarily as establishing the existence of nontrivial uniformly valid inference procedures with reasonable pointwise power properties. Developing procedures that retain uniform validity while being less conservative and more computationally tractable is an interesting direction for future research.

We also note that, under the conditions of Theorem~\ref{thm4}, the screening procedure of \cite{BPSSMT} is uniformly valid and is \emph{consistent against fixed} alternatives, providing a computationally tractable alternative to the procedure developed in this section. It is less clear, however, whether their procedure has power against $N^{-1/2}$-local alternatives, as ours does. Investigating this question is left for future research.
\end{remark}

\section{Simulations}
\label{SecMC}
This section evaluates our procedure in two families of designs. The first follows \cite{AV}, while the second is based on Example~\ref{ExKKLS}. All methods are evaluated using $S=1,000$ Monte Carlo replications. We set $\kappa_{1N}=\log\log(N)$ when computing $\hat z$ in \eqref{eqnEAM6} and the box size, and $\kappa_{2N}=\kappa_{3N}=\sqrt{\log(N)}$ when constructing $\hat\Delta(v)$ in \eqref{eqnEAM5}. We use the box size in \eqref{eq:box_floor} and the alternative normalization described in footnote~\ref{footnote:omega} and Appendix~\ref{SecDetMC_AV_BM}. We use $B=1,000$ bootstrap replications for both the proposed method and the procedure of \cite{GM25}.

\subsection{Design from \cite{AV}}\label{SecMC_AV}
This design reproduces Example~(15) of
\cite{AV}:
\[\quad
M = \begin{pmatrix} (1+b) & -1 \\ -(1+\zeta) & 1 \\ -1 & 0 \\ 1 & 0 \end{pmatrix}, \qquad
\beta = \begin{pmatrix} -\nu \\ -\zeta \\ 1+\nu \\ 1 \end{pmatrix}, \qquad \rho = \begin{pmatrix} -1 \\ 0 \end{pmatrix}.\]

The data consist of $N$ i.i.d.\ draws $(U_i^b,U_i^\zeta, U_i^\nu)$ from $U[-1/2,1/2]$. The unknown parameters $b$, $\zeta$ and $\nu$ are estimated as $\hat b_N = b_0 + N^{-1}\sum_{i=1}^N U_i^b$, $\hat \zeta_N = \zeta_0 + N^{-1}\sum_{i=1}^N U_i^\zeta$, $\hat \nu_N = \nu_0 + N^{-1}\sum_{i=1}^N U_i^\nu$. We consider three subdesigns: $b_0 = 0$, $b_0 = -0.05$, and $b_0=-0.02$, while $\zeta_0=\nu_0=0$.  For $b_0=0$ the solution to the primal is unique, while the set of dual solutions is unbounded, and so Theorem~\ref{thm2} does not apply to the lower CI, i.e., coverage may be conservative. For negative $b_0$ the primal and dual LPs have unique solutions, with the primal solution $x^\ast=(0,0)\tr$. Hence, Theorem~\ref{thm2} does not apply to the upper CI so the resulting coverage may be conservative. As noted in Remark~\ref{rem:floor}, using the modified box constraint in \eqref{eq:box_floor} restores asymptotically exact coverage for the upper CI\@. Figure~\ref{fig:AV_floor} in Appendix~\ref{SecDetMC_AV_BM} illustrates the effect of this modification on coverage in this design.

We first compare our box-augmented method (BM) with \cite{AV}, and then with procedures for testing the feasibility of systems of inequalities, namely, \cite{GM25}, \cite{BPSSMT}, and \cite{CSS}.

\subsubsection{Comparison with \cite{AV}}\label{SecMC_AV_AV}

\begin{figure}[htbp]
    \centering
    \begin{subfigure}[t]{0.495\textwidth}
        \centering
        \includegraphics[width=\textwidth]{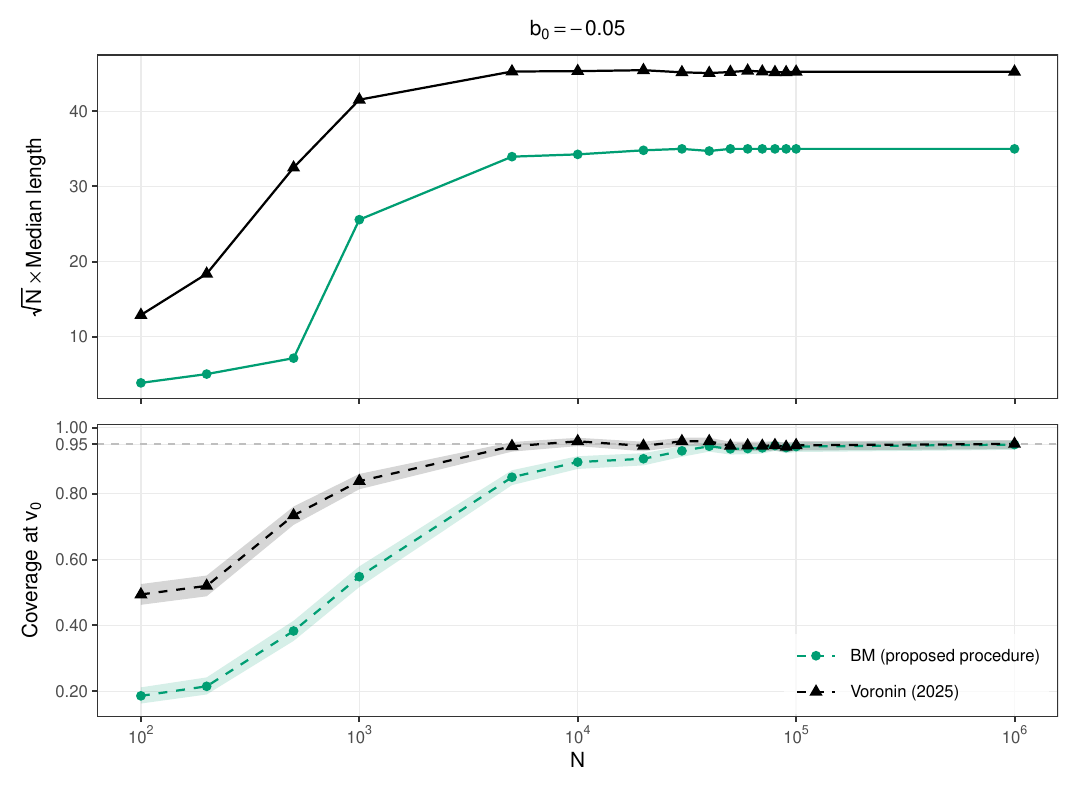}
        \caption{Coverage at $v_0$ for $b_0=-0.05$}
        \label{fig:AV_thm2_b0m005}
    \end{subfigure}
    \hfill
    \begin{subfigure}[t]{0.495\textwidth}
        \centering
        \includegraphics[width=\textwidth]{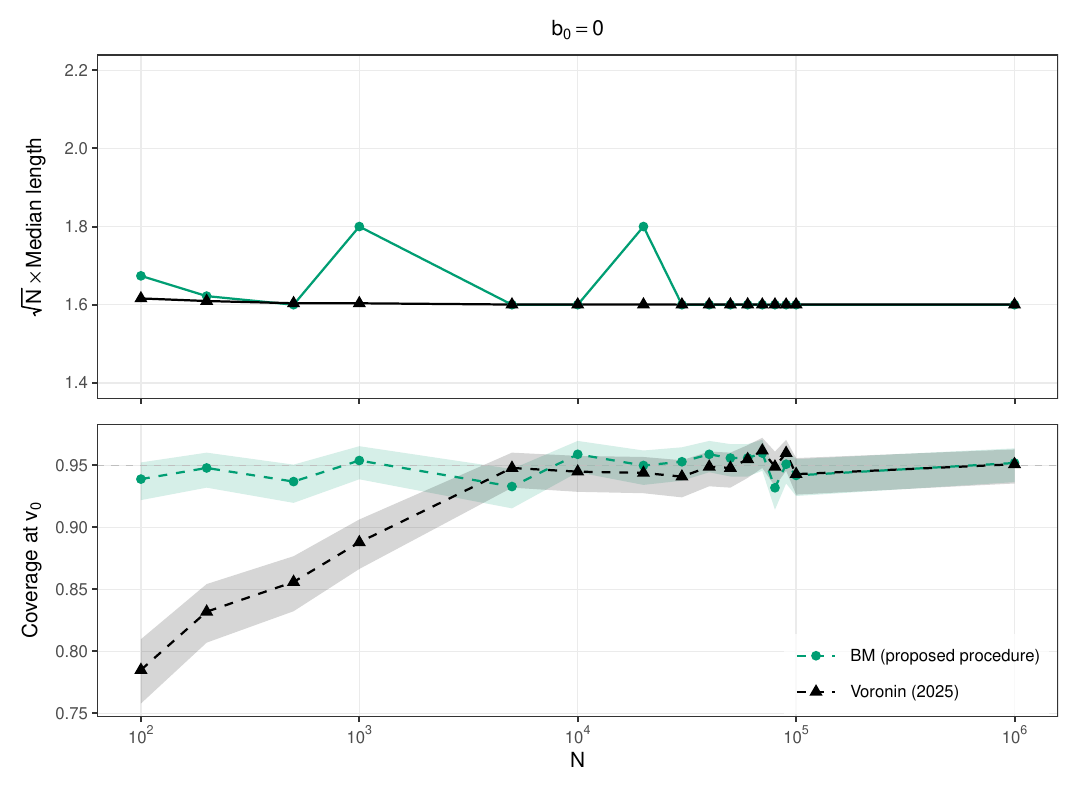}
        \caption{Coverage at $v_0$ for $b_0=0$.}
        \label{fig:AV_thm2_b00}
    \end{subfigure}
    \caption{\small Illustration of Theorems~\ref{thm1} and~\ref{thm2} for the two-sided CI under the simulations in Section~\ref{SecMC_AV}. The proposed procedure is implemented with $1,000$ bootstrap replications. Results are based on $1,000$ simulations. Shaded bands are pointwise $95\%$ \citet{Wilson} intervals reflecting Monte Carlo error.}
    \label{fig:AV_thm2}
\end{figure}

Figure~\ref{fig:AV_thm2} illustrates Theorems~\ref{thm1} and~\ref{thm2} for the two-sided CI. The bottom panels show that the proposed procedure attains asymptotic coverage of $0.95$ in both designs, as predicted by Theorem~\ref{thm1}. The top panels show that the median CI length, scaled by $\sqrt N$, stabilizes as $N$ increases, consistent with \eqref{eqnthm4}. For comparison, the panels also report coverage and median CI length for the procedure of \citet{AV}, which is based on a normal approximation.

The procedure of \citet{AV} requires two tuning parameters related to the dual solution: a penalty $w$, which diverges with the sample size, and a deterministic bound $\bar v$ satisfying Assumption~B3. The latter must be a valid upper bound on the norm of the relevant minimal-norm dual solution and, unlike $w$, remains fixed with $N$. Since such a bound is generally unknown, \citet{AV} proposes choosing $\bar v$ using an $\alpha$-quantile of the Tao--Vu distribution and recommends $\alpha=0.1$, which yields $\bar v=140.5$ in this design.

The choice of $\alpha$ is a tuning choice rather than a bound implied by the model. To illustrate the consequences of this choice, we also consider the $20\%$ quantile, which gives $\bar v=34.45$. This alternative choice remains valid for the designs $b_0=0$ and $b_0=-0.05$, and therefore performs well for the designs considered above. However, for $b_0=-0.02$, it violates Assumption~B3. Figure~\ref{fig:AV_thm2_b0m002} shows that the resulting confidence intervals substantially undercover. As shown in Appendix~\ref{SecDetMC_AV_AV}, this is not a finite-sample phenomenon: the asymptotic coverage of the nominal $95\%$ CI is $0.823$.

The same underlying issue can arise with the recommended $10\%$ quantile. In this design, the resulting value $\bar v=140.5$ violates Assumption~B3 for sufficiently small $|b_0|$. Demonstrating the resulting undercoverage directly, however, requires impractically large samples because the penalty $w$ must first become large enough for Assumption~A1 to hold. For example, at $b_0=-0.01$ and with $w\propto\ln\ln N$, this requires sample sizes of order $10^{36}$ or larger. We therefore use the $20\%$ quantile to illustrate the same sensitivity at feasible sample sizes.

The comparison highlights the role of Assumption~B3. The procedure of \citet{AV} requires the researcher to specify a valid deterministic bound on the dual solution, and an invalid choice can lead to asymptotic undercoverage. Our procedure does not require such prior knowledge: instead, the box constraint estimates the localization needed for inference, and the coverage results follow under the assumptions of Theorems~\ref{thm1}--\ref{thm2}.

\begin{figure}[htbp]
    \centering
    \includegraphics[width=\textwidth]{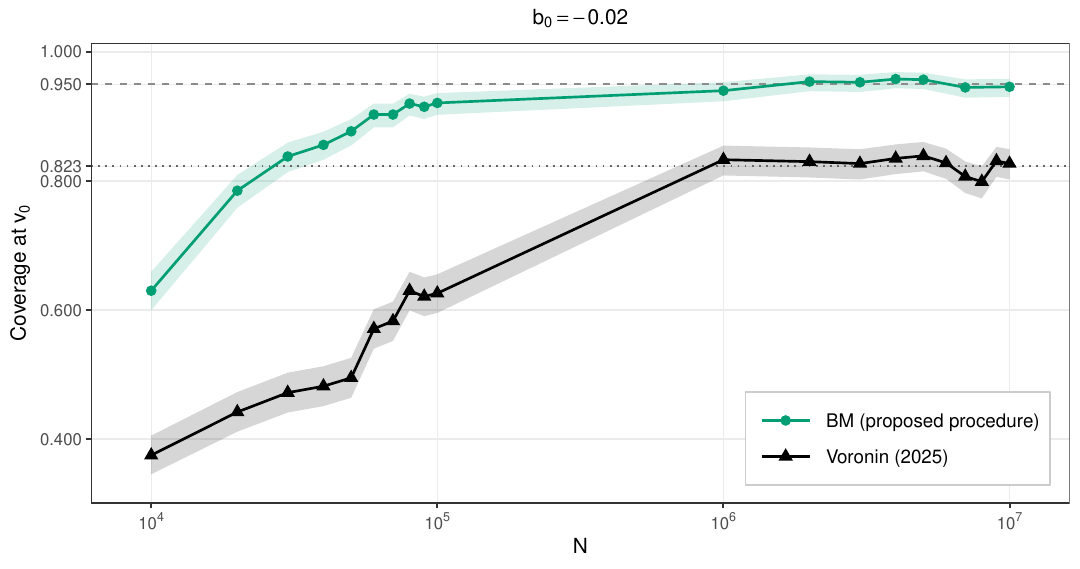}
    \caption{Coverage at $v_0$ for $b_0=-0.02$. $\bar{v}=34.45$, which corresponds to the $20\%$ quantile of the Tao--Vu distribution. The dotted line indicates the asymptotic coverage for \cite{AV} (see Remark~\ref{remAV} in Appendix~\ref{SecDetMC_AV_AV}). Shaded bands are pointwise $95\%$ \cite{Wilson} intervals for the simulated coverage, reflecting Monte Carlo error.}
    \label{fig:AV_thm2_b0m002}
\end{figure}
\subsubsection{Comparison with methods for testing system feasibility}
\label{SecComp}
We compare our procedure with the RGCC test of \citet{CSS}, the procedure of \citet{GM25}, and the screening test of \citet{BPSSMT}. For RGCC, we construct the two-sided CI as the intersection of the upper and lower one-sided CIs, each with level $1-\alpha/2$, because the direct two-sided system in \eqref{eqnIP3} does not satisfy the conditions of \citet{CSS}; see Remark~\ref{rem:css_ts_degen}. Details on the mappings into the frameworks of \citet{CSS} and \citet{BPSSMT} are provided in Appendices~\ref{SecDetMC_AV_CSS} and~\ref{SecDetMC_AV_BPSSMT}.

\begin{figure}[htbp]
    \centering
    \begin{subfigure}[t]{0.495\textwidth}
        \centering
        \includegraphics[width=\textwidth]{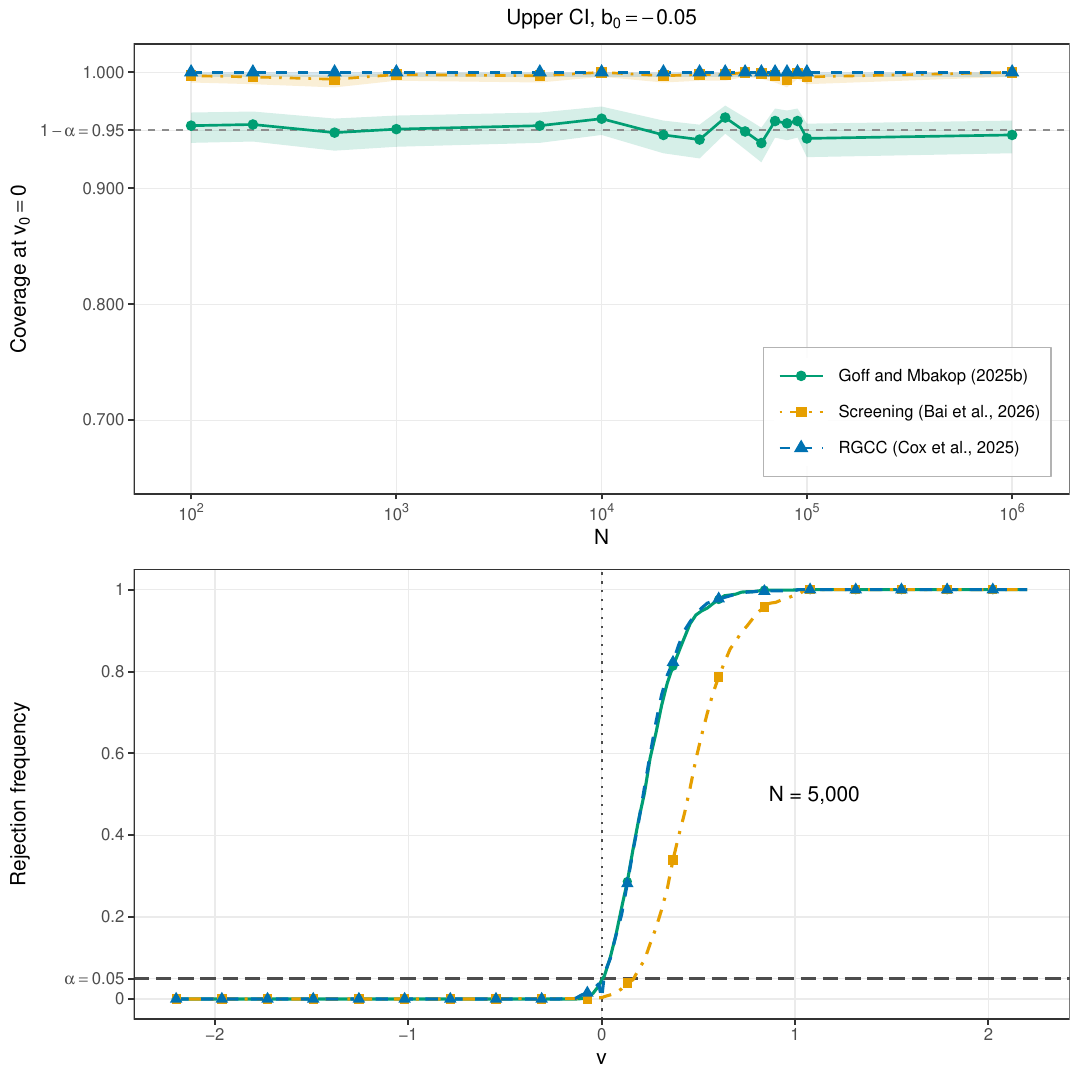}
        \caption{Upper CI}
        \label{fig:upper_b0m005}
    \end{subfigure}
    \hfill
    \begin{subfigure}[t]{0.495\textwidth}
        \centering
        \includegraphics[width=\textwidth]{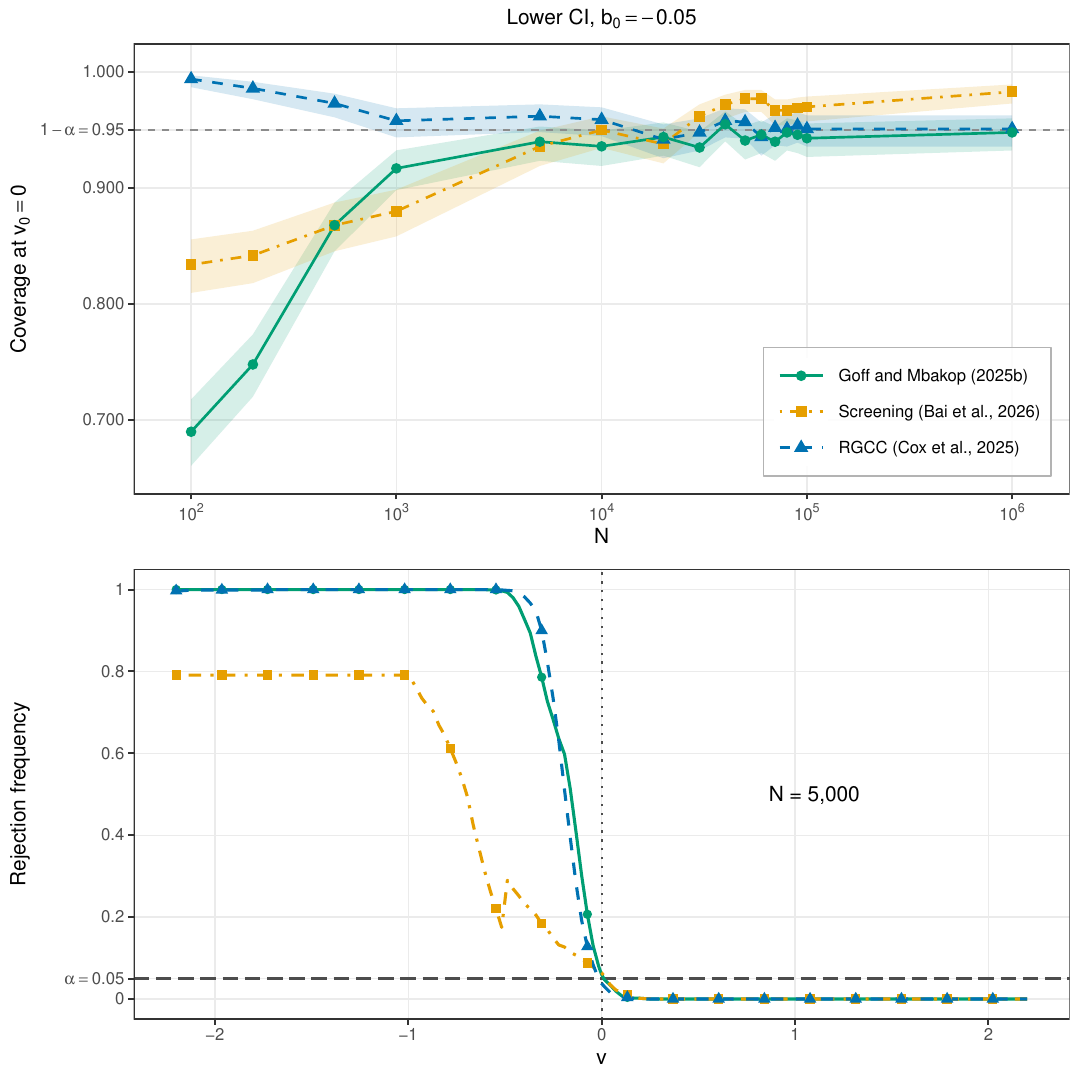}
        \caption{Lower CI}
        \label{fig:lower_b0m005}
    \end{subfigure}
    \caption{\small Coverage (top panels) and power at $N=5,000$ (bottom panels) for the one-sided systems in \eqref{eqnIP1}--\eqref{eqnIP2} under the design in Section~\ref{SecMC_AV} with $b_0=-0.05$. \citet{GM25} is implemented with $1,000$ bootstrap replications. Results are based on $1,000$ simulations with significance level $\alpha=0.05$. Shaded bands are pointwise $95\%$ \citet{Wilson} intervals reflecting Monte Carlo error.}
    \label{fig:AV_thm3_b0m005}
\end{figure}

The design with $b_0=-0.05$ satisfies the boundedness conditions relevant for Theorem~\ref{thm3}: the sets of optimal solutions to both the primal and dual LPs are bounded. Hence, the conclusions of the theorem continue to hold without the box augmentation. Figure~\ref{fig:AV_thm3_b0m005} illustrates this for the one-sided procedures, which allow a direct comparison with \citet{CSS}. All methods exhibit good coverage and power.

\begin{figure}[htbp]
    \centering
    \begin{subfigure}[t]{0.495\textwidth}
        \centering
        \includegraphics[width=\textwidth]{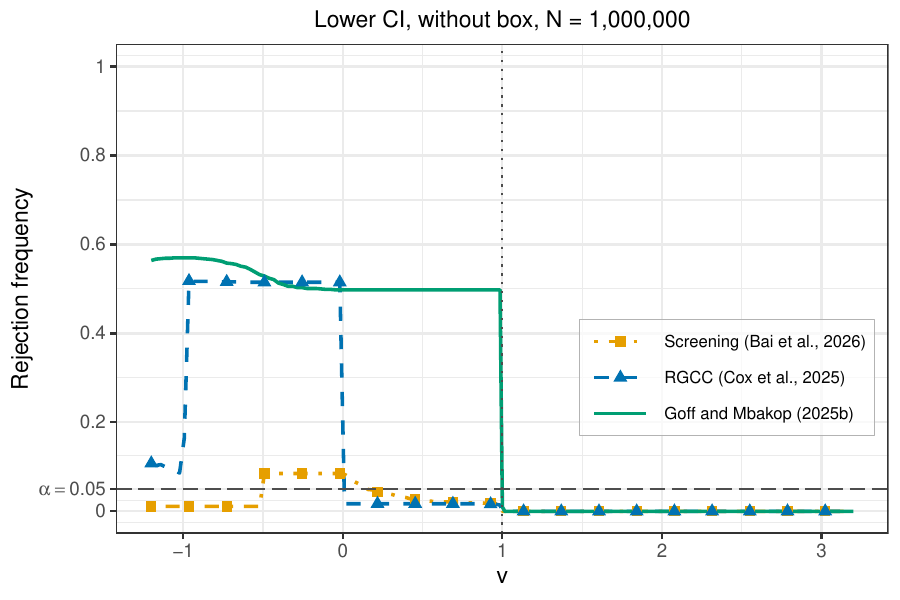}
        \caption{Lower one-sided tests, without the box.}
        \label{fig:lower_nobox}
    \end{subfigure}
    \hfill
    \begin{subfigure}[t]{0.495\textwidth}
        \centering
        \includegraphics[width=\textwidth]{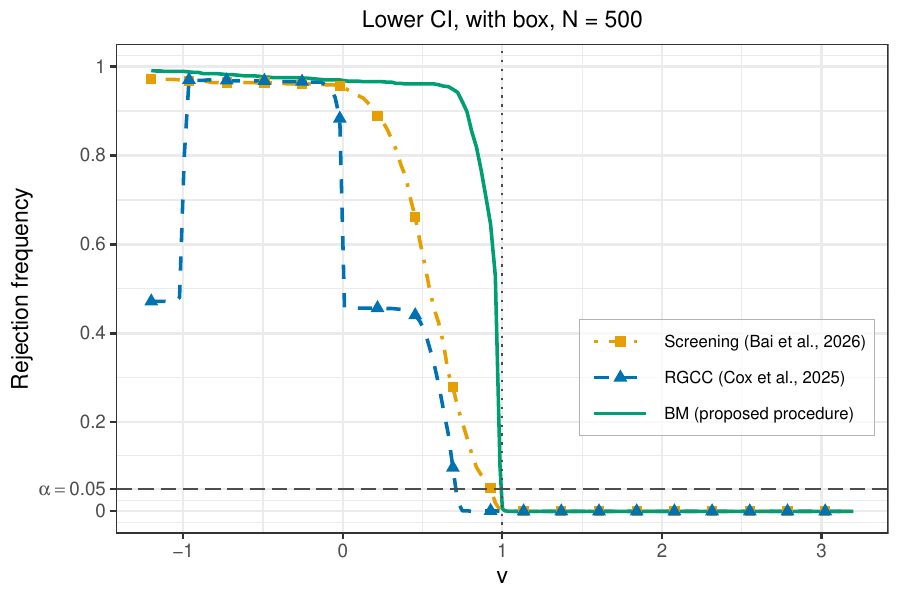}
        \caption{Lower one-sided tests, with the box.}
        \label{fig:lower_box}
    \end{subfigure}

    \medskip
    \begin{subfigure}[t]{0.495\textwidth}
        \centering
        \includegraphics[width=\textwidth]{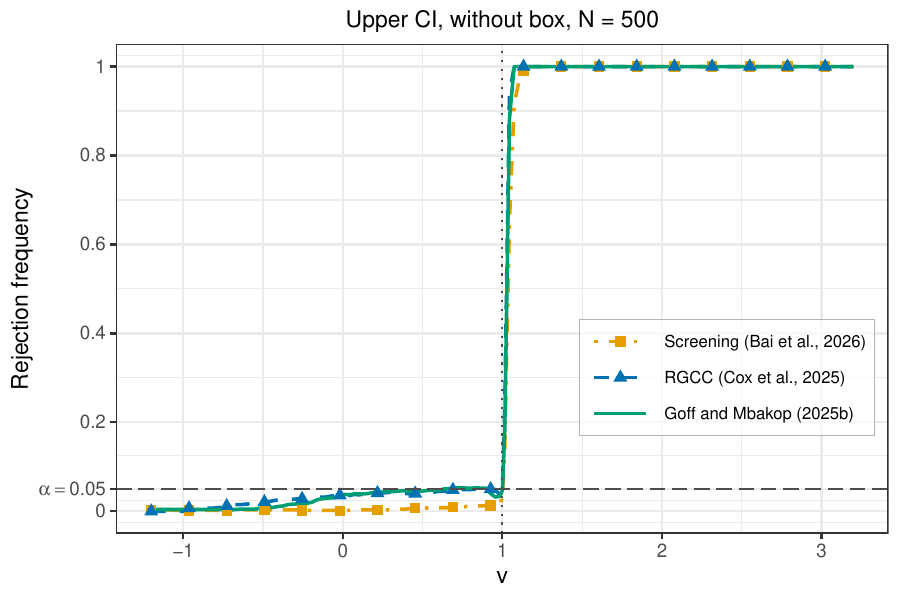}
        \caption{Upper one-sided tests, without the box.}
        \label{fig:upper_nobox}
    \end{subfigure}
    \hfill
    \begin{subfigure}[t]{0.495\textwidth}
        \centering
        \includegraphics[width=\textwidth]{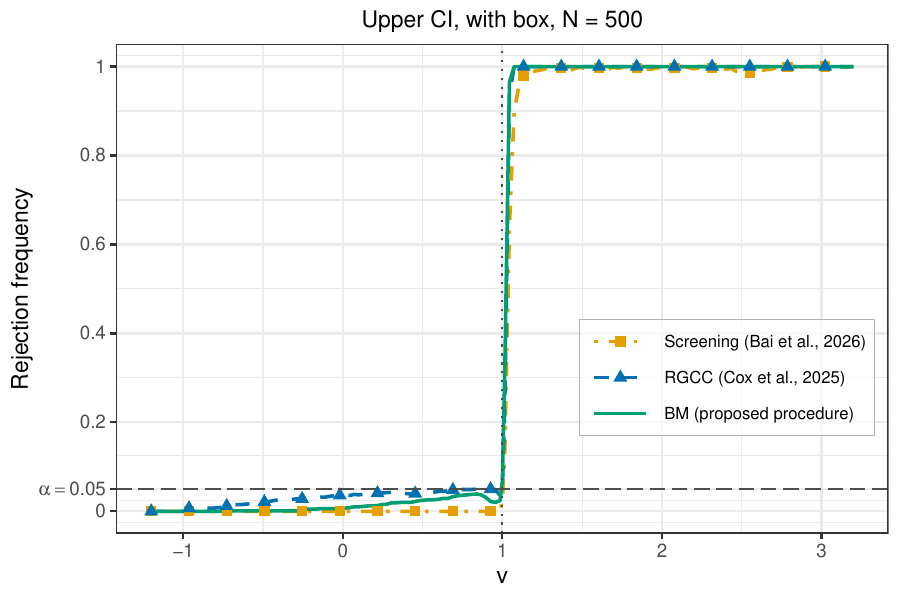}
        \caption{Upper one-sided tests, with the box.}
        \label{fig:upper_box}
    \end{subfigure}
    \caption{\small Power curves for the one-sided systems in \eqref{eqnIP1}--\eqref{eqnIP2} under the design in Section~\ref{SecMC_AV} with $b_0=0$, based on $1,000$ simulations. \citet{GM25} and BM are implemented with $1,000$ bootstrap replications. The significance level is $\alpha=0.05$.}
    \label{fig:onesided_b0}
\end{figure}

The case $b_0=0$ is different. Here the set of optimal dual solutions is unbounded, so Theorem~\ref{thm3} no longer guarantees the desired properties of the lower or two-sided procedures when the box constraints are omitted. Indeed, directly applying the procedures of \citet{BPSSMT}, \citet{CSS}, and \citet{GM25} to the lower system in \eqref{eqnIP2} can result in poor power, consistent with the impossibility results in \citet{GM25} and \citet{BPSSMT}; see Figure~\ref{fig:lower_nobox}. The box restrictions in \eqref{eqnmod} control the relevant unbounded directions and substantially improve power; see Figure~\ref{fig:lower_box}. A similar phenomenon arises for two-sided confidence intervals, as shown in Figures~\ref{fig:intro_nobox} and~\ref{fig:intro_box}. By contrast, the set of optimal primal solutions remains bounded, so Theorem~\ref{thm3} continues to apply to the upper procedure without box augmentation; accordingly, the box has little effect in Figures~\ref{fig:upper_nobox}--\ref{fig:upper_box}.

For \citet{BPSSMT}, the loss of power in the lower and two-sided systems is associated with the unboundedness on the dual side, while the box augmentation restores some of this power. For \citet{CSS}, additional regularity conditions fail in this design, and the relevant covariance matrix can be singular in some samples. These implementation details are discussed in Appendices~\ref{SecDetMC_AV_BPSSMT} and~\ref{SecDetMC_AV_CSS}. The main message of the experiment is that the procedures perform well without box augmentation when the boundedness conditions in Theorem~\ref{thm3} hold, but may have poor power when they fail; the box constraints are designed precisely to control the unbounded directions responsible for this deterioration.

\subsection{Design from Example \ref{ExKKLS}}\label{SecMC_KKLS}
We follow the simulation design from \cite{KKLS}, Appendix D. We simulate a single-agent firm entry/exit model with a finite state space, i.i.d.\ type-I extreme value payoff shocks, and a \textit{known} Markov transition process (hence, in this simulation $M$ is known). Data consist of $N$ markets observed over $T$ periods, with $(N,T)=(100,5)$ and $(N,T)=(1000,15)$. We study a counterfactual policy that gives rise to a linear objective, namely, a $20\%$ proportional subsidy to observed $vp$ (similar to the one in the empirical application of \citealp{KKLS}). We conduct inference on $v_0$, the minimal value of $\theta_V$ subject to Restriction 1, Restrictions 1--2, and Restrictions 1--3 (defined in Section~\ref{SecDetEx_KKLS}). Other details on the implementation can be found in Appendix~\ref{SecDetMC_KKLS}. The simulation results are presented in Table~\ref{tab:KKLS_vp}. Even at $(N,T)=(100,5)$, the method controls size, both with and without the box restrictions. In this design, the primal and dual solutions are unique, so the proposed method is expected to perform well with or without the box. The method of \cite{AV} slightly undercovers at $(N,T)=(100,5)$ but controls size at $(N,T)=(1000,15)$. 
\begin{table}[!ht]
\centering
\caption{Monte Carlo results: entry/exit example from \cite{KKLS}}
\label{tab:KKLS_vp}
\setlength{\tabcolsep}{3pt}
\small
\begin{adjustbox}{max width=\textwidth}
\begin{tabular}{l*{9}{c}}
\toprule
\multicolumn{10}{c}{\textbf{Revenue subsidy}} \\
\midrule
 & \multicolumn{3}{c}{Restriction 1} & \multicolumn{3}{c}{Restrictions 1--2} & \multicolumn{3}{c}{Restrictions 1--3} \\
\cmidrule(lr){2-4}\cmidrule(lr){5-7}\cmidrule(lr){8-10}
True identified set & \multicolumn{3}{c}{$[5.416,5.829]$} & \multicolumn{3}{c}{$[5.497,5.829]$} & \multicolumn{3}{c}{$[5.553,5.829]$} \\
Method & \multicolumn{2}{c}{BM} & Voronin & \multicolumn{2}{c}{BM} & Voronin & \multicolumn{2}{c}{BM} & Voronin \\
Box constraints  & Yes & No & -- & Yes & No & -- & Yes & No & -- \\
\midrule
\textit{$N=100$, $T=5$} & & & & & & & & & \\
Median lower bound & $5.109$ & $5.087$ & $5.212$ & $5.205$ & $5.201$ & $5.275$ & $5.250$ & $5.250$ & $5.383$ \\
Coverage & $1.000$ & $1.000$ & $.955$ & $1.000$ & $1.000$ & $.905$ & $1.000$ & $1.000$ & $.920$ \\
Median excess & $0.307$ & $0.330$ & $0.204$ & $0.292$ & $0.296$ & $0.222$ & $0.304$ & $0.304$ & $0.170$ \\
\midrule
\textit{$N=1000$, $T=15$} & & & & & & & & & \\
Median lower bound & $5.375$ & $5.375$ & $5.391$ & $5.459$ & $5.459$ & $5.469$ & $5.511$ & $5.511$ & $5.530$ \\
Coverage & $1.000$ & $1.000$ & $.953$ & $1.000$ & $1.000$ & $.951$ & $1.000$ & $1.000$ & $.951$ \\
Median excess & $0.042$ & $0.042$ & $0.025$ & $0.038$ & $0.038$ & $0.028$ & $0.043$ & $0.043$ & $0.023$ \\
\bottomrule
\end{tabular}
\end{adjustbox}

\begin{minipage}{0.95\linewidth}
\vspace{4pt}\footnotesize \textit{Note:} Coverage is the empirical coverage probability of the one-sided lower CI at $v_0=min\theta_V$ for nominal level $95\%$. Median lower bound, $\bar{L}$, is the median of the lower CI bounds over $1{,}000$ replications. Median excess length is the median of $v_0-L$. Columns BM report the proposed procedure with the box (set as in \eqref{eq:box_floor}), and without. AV reports the results of the method from \cite{AV}.
\end{minipage}
\end{table}


\section{Conclusion}

This paper develops inference procedures for the value of a linear program with estimated objective and constraints. Our starting point is a reduction of one- and two-sided inference on the LP value to testing the feasibility of systems of linear inequalities. While existing feasibility tests can therefore be used to construct asymptotically valid confidence intervals, a direct application may have poor power and produce excessively wide intervals. We relate this problem to unboundedness of the corresponding nuisance-parameter sets and propose estimated box constraints that localize these sets without asymptotically excluding the relevant solutions. Applied to the procedure of \cite{GM25}, this modification yields confidence intervals with asymptotically valid coverage and length of order $N^{-1/2}$.

We also characterize how standard constraint qualifications strengthen these conclusions, including conditions for asymptotically exact inference and conditions under which the box constraints can be omitted. In addition, we provide sufficient conditions for uniformly valid inference over classes of DGPs in the usual sense.

Several directions for future research remain. One is to develop alternatives to the uniformly valid procedure proposed here that are less conservative and more computationally tractable. It would also be useful to develop formal modifications of the procedures in \cite{CSS} and \cite{BPSSMT} that incorporate estimated box constraints and to establish their statistical properties. Other directions include improving the data-driven choice of tuning parameters and extending the analysis to high-dimensional settings in which the dimensions of the LP inputs are allowed to grow with the sample size.

\begin{singlespace}
\bibliographystyle{apalike}
\bibliography{biblio.bib}
\end{singlespace}

\begin{appendix}
\section{Proofs of Theorems \ref{thm1}, \ref{thm2}, \ref{thm3}, and \ref{thm4}}
\label{ThmP}

\subsection{Proof of Theorem \ref{thm1}}
\label{Secthm1}
\begin{proof}
We divide the proof into five steps. Steps~1--4 establish \eqref{eqnthm1} for the case $\hat{C}=CI^U_{1-\alpha}$. The proofs for the lower and two-sided confidence intervals are analogous and are therefore omitted. Step~5 establishes \eqref{eqnthm2}. The proofs of \eqref{eqnthm3} and \eqref{eqnthm4} are similar and are likewise omitted.

Since our procedure is a modification of that proposed in \cite{GM25}, Steps~1--4 follow the general strategy of the proofs of Theorems~4.9 and~4.19 in \cite{GM25}. However, several substantial modifications are required to accommodate the inclusion of the box constraints.

Let $z^*=(x^*,y^*)$ denote the minimum-norm element of $\mathcal{O}$ defined in Lemma~\ref{lemNP}. By LP strong duality, $x^*$ solves the upper system in \eqref{eqnIP1} for every $v\leq v_0$, while $y^*$ solves the lower system in \eqref{eqnIP2} for every $v\geq v_0$. At $v=v_0$, $(x^*,y^*)$ solves the two-sided system in \eqref{eqnIP3}.

Note that, given the definition of $CI^U_{1-\alpha}$, to prove \eqref{eqnthm1} (for upper confidence intervals) it suffices to show that
\begin{equation}
\label{pthmAVeqn1}
\limsup_{N\to \infty}
Q\!\left(
\sqrt{N}\hat{T}_N(v_0)>\hat{C}_N(1-\alpha;v_0)
\right)
\leq \alpha.
\end{equation}


\underline{Step 1.}
The objective of this step is to establish inequality \eqref{thmAVeqn7}, which will allow us to control the upper tails of the distribution of the test statistic.

Let $\phi(\mu,a,v)$ denote the value of the linear program associated with the inputs $\mu=(\rho,\beta,M)$, $a$, and $v$, defined by
\[
\phi(\mu,a,v)
:=
\max\Bigl\{
b_{\mathrm{aug}}(a,v)^\top\lambda
\;\Big|\;
\lambda \ge 0,\;
A_{\mathrm{aug}}^\top \lambda = 0,\;
\mathbbm{1}^\top \Omega_{\mathrm{aug}} \lambda \le 1
\Bigr\}.
\]
Here, $a$ denotes a generic value of the constant appearing in the box constraints and is not restricted to the particular choice $\hat{a}_N$ used in the definition of $b_{\mathrm{aug}}$ in \eqref{eqnmod_mat}.
A straightforward computation shows that \footnote{The variable $\lambda$ here is distinct from the one appearing in the preceding equation and has a different dimension.}
\begin{equation}
\label{thmAVeqn4}
\phi(\mu,a,v)=\max\bigl\{ b(v)^\top \lambda -a\|\omega\|_1 \;\Big|\; \lambda \ge 0,\; A^\top \lambda=\omega, \; \mathbbm{1}^\top\Omega \lambda\leq 1\bigr\},
\end{equation}
and by the strong duality theorem for LP, we have
\[
\phi(\mu,a,v)
=
\min\Bigl\{
t \;\Big|\;
t\Omega \mathbbm{1}+A\eta \ge b(v),\;
\|\eta\|_{\infty}\le a,\;
t\ge 0
\Bigr\},
\]
where $A$ and $b(v)$ are as in \eqref{eqnIP1}. In particular, we have
\begin{equation}
\label{pthmAVeqn2}
\phi(\mu,a,v)=0
\quad \text{whenever} \quad
v\leq v_0
\ \text{and}\
\|x^*\|_{\infty}\leq a.
\end{equation}

Let $\xi=(\xi_A,\xi_b,\xi_{\Omega})$ denote a generic perturbation of the inputs $(A,b,\Omega)$. Here, $\xi_A$, $\xi_b$, and $\xi_{\Omega}$ have the same dimensions as $A$, $b$, and $\Omega$, respectively, and $\xi_{\Omega}$, like $\Omega$, is restricted to be a diagonal matrix. Given $\xi$, let the resulting perturbed inputs $(A_N(\xi),b_N(\xi,v),\Omega_N(\xi))$ be defined by
\begin{equation}
\label{thmAVeqn5}
(A_N(\xi),b_N(\xi,v),\Omega_N(\xi))=\Bigl(A+\frac{\xi_A}{\sqrt{N}}, b(v)+\frac{\xi_b}{\sqrt{N}}, \Omega +\frac{\xi_{\Omega}}{\sqrt{N}}\Bigr).
\end{equation}
We assume throughout that the perturbation matrices $\xi_{\Omega}$, like the matrices $\Omega$, have their $i$th diagonal entry equal to zero whenever the $i$th row corresponds to a deterministic constraint.\footnote{Recall that the $i$th diagonal entry of $\Omega$ is equal to zero whenever the $i$th row of $A$ and the $i$th component of $b$ are deterministic, that is, known and not subject to estimation.} Likewise, any entries of $\xi_A$ and $\xi_b$ corresponding to deterministic components of $A$ and $b(v)$ are assumed to be zero. Under this convention, the perturbed inputs in \eqref{thmAVeqn5} can be viewed as generic realizations of estimators of $(A,b,\Omega)$.

Let $\phi_N(\mu(\xi),a,v)$ denote the value of the LP in \eqref{thmAVeqn4} obtained by replacing the inputs with their perturbed counterparts defined in \eqref{thmAVeqn5}. Let $L_N(\lambda,\omega,\eta,t;\xi,a,v)$ denote the Lagrangian associated with this perturbed LP. A straightforward calculation shows that
\[
L_N(\lambda,\omega,\eta,t;\xi,a,v)
=
b_N(\xi,v)^\top\lambda
-
a\|\omega\|_1
+
\eta^\top\!\bigl(\omega-A_N(\xi)^\top\lambda\bigr)
+
t\bigl(1-\mathbbm{1}^\top\Omega_N(\xi)\lambda\bigr).
\]

Similarly, let $L(\lambda,\omega,\eta,t;a,v)$ denote the Lagrangian associated with the unperturbed LP in \eqref{thmAVeqn4}. It is given by
\[
L(\lambda,\omega,\eta,t;a,v)
=
b(v)^\top\lambda
-
a\|\omega\|_1
+
\eta^\top\!\bigl(\omega-A^\top\lambda\bigr)
+
t\bigl(1-\mathbbm{1}^\top\Omega\lambda\bigr).
\]
Let $\Delta_{0,N}(\xi,a,v)$ and $S_{0,N}(\xi,a,v)$ denote, respectively, the sets of optimal solutions to the perturbed LP in \eqref{thmAVeqn4} (obtained by replacing $\mu$ with $\mu_N(\xi)$) and its dual. Similarly, let $\Delta_0(a,v)$ and $S_0(a,v)$ denote the corresponding sets of optimal solutions for the unperturbed primal and dual LPs.

For
\[
(\lambda_{\xi},\omega_{\xi},\eta_{\xi},t_{\xi})
\in
\Delta_{0,N}(\xi,a,v)\times S_{0,N}(\xi,a,v)
\]
and
\[
(\lambda_0,\omega_0,\eta_0,t_0)
\in
\Delta_0(a,v)\times S_0(a,v),
\]
and whenever $\phi(\mu,a,v)$ and $\phi_N(\mu(\xi),a,v)$ are finite, the saddle-point property of the Lagrangian implies that 
\begin{equation}
\label{thmAVeqn6}
\begin{aligned}
\sqrt{N}\Bigl(\phi_N(\mu(\xi),a,v)-\phi(\mu,a,v)\Bigr)
&=
\sqrt{N}\Bigl(
L_N(\lambda_{\xi},\omega_{\xi},\eta_{\xi},t_{\xi};\xi,a,v)
-
L(\lambda_0,\omega_0,\eta_0,t_0;a,v)
\Bigr)
\\
&\leq
\sqrt{N}\Bigl(
L_N(\lambda_{\xi},\omega_{\xi},\eta_0,t_0;\xi,a,v)
-
L(\lambda_{\xi},\omega_{\xi},\eta_0,t_0;a,v)
\Bigr)
\\
&=
\xi_b^\top\lambda_{\xi}
-
\eta_0^\top\xi_A^\top\lambda_{\xi}
-
t_0\,\mathbbm{1}^\top\xi_{\Omega}\lambda_{\xi}.
\end{aligned}
\end{equation}

Equation \eqref{pthmAVeqn2} implies that for $v\leq v_0$ and $a\geq \|x^*\|_{\infty}$,  $\phi(\mu,a,v)=0$. Moreover, $(\eta_0, t_0) \in S_{0}(a,v_0)$ implies $t_0 = 0$, and we may choose $\eta_0=x^*$. Therefore, for $v\leq v_0$ and $a\geq \|x^*\|_{\infty}$, the inequality above implies that\footnote{ Note that Farkas' lemma implies that $\phi(\mu,a,v)$ and $\phi_N(\mu(\xi),a,v)$, \textit{for all} perturbations $\xi$ such that $\Omega_N(\xi)$ and $\Omega$ have the same nonzero diagonal entries, are finite whenever $a\geq \|x^*\|_{\infty}$ and $v\leq v_0$ (see Lemma \ref{lemAux}).}
\begin{equation}
\label{thmAVeqn7}
\sqrt{N}\,\phi_N(\mu_N(\xi),a,v)
\leq
\xi_b^\top\lambda_{\xi}
-
{x^*}^\top\xi_A^\top\lambda_{\xi}
=
\langle \xi_b-\xi_Ax^*,\lambda_{\xi}\rangle
\end{equation}
for all $(\lambda_{\xi},\omega_{\xi})\in \Delta_{0,N}(\xi,a,v)$.

Assumptions~\ref{ass1}--\ref{ass3} and Lemma~\ref{lemNP} imply the existence of sequences
\[
\alpha_N \downarrow 0,\qquad
M_N \uparrow \infty,\qquad
\kappa_N \uparrow \infty,
\]
such that
\[
M_N=o(\kappa_N),\qquad
M_N\alpha_N=o(1),\qquad
\kappa_N=o(N^{1/4}),\qquad
M_N\kappa_N/N^{1/4}=o(1),
\]
and the event
\[ E_N := \left\{ \mathbb{W} \;\middle|\; \begin{array}{l} \|\hat{\Omega} - \Omega\| \leq \alpha_N,\quad d_{\mathrm{Pr}}\bigl(\mathcal{L}(\zeta^* \mid \mathbb{W}), \mathcal{N}(0, V)\bigr) \leq \alpha_N,\\[0.5em] \sqrt{N} \|\hat{\mu} - \mu\| \leq M_N,\quad \|\hat{z} - z^*\|_{\infty} \leq\kappa_N^{3/4}/N^{1/4}\end{array} \right\}, \]
satisfies
\[
\lim_{N\to\infty}\delta_{0N}=0,
\qquad\text{where}\qquad
\delta_{0N}:=Q(E_N^c),
\]
and additionally 
\[
d_{\mathrm{Pr}}
\Bigl(
\sqrt{N}(\hat{\mu}-\mu),
\mathcal{N}(0,V)
\Bigr)
\leq \alpha_N.
\]

Let
\[
\mathcal{X}_N
:=
\left\{
\xi
\;\middle|\;
\|\xi_b\|\leq M_N,\;
\|\xi_A\|\leq M_N,\;
\|\xi_{\Omega}\|\leq \sqrt{N}\alpha_N
\right\}.
\]

Throughout the remainder of the proof, we use $C$ to denote a generic positive constant whose value may vary from one line to the next.


\underline{Step 2.}
The objective of this step is to show that there exists a sequence $\gamma_N=o(1)$ such that, on the event $E_N$, for all sufficiently large $N$, and for every $v\leq v_0$ and $a(\xi)$ such that $|a(\xi)-5\|x^*\|_{\infty}|\leq \kappa_N/N^{1/4}$, $a(\xi)>0$, we have\footnote{Here, we allow the box-constraint parameter to depend freely on $\xi$. Moreover, with a slight abuse of notation, we write $\lambda\in\Delta_{0,N}(\xi)$ to mean that there exists $\omega$ such that $(\lambda,\omega)\in\Delta_{0,N}(\xi)$. This convention will be used throughout the remainder of the proof.}
\begin{equation}
\label{thmAVeqn8}
\sup_{\xi\in\mathcal{X}_N}
\sup_{\lambda\in\Delta_{0,N}(\xi,a(\xi),v)}
\left\{
\langle \xi_b-\xi_Ax^*,\lambda\rangle
-
\sup_{\lambda\in\hat{\Delta}}
\langle \xi_b-\xi_A\hat{x},\lambda\rangle
\right\}
\leq
\gamma_N.
\end{equation}
Here $\hat{x}$ is as defined in ~\eqref{eqnEAM6} and a straightforward calculation shows that, in accordance with \eqref{thmAVeqn8}, $\hat{\Delta}$ in \eqref{eqnEAM5} can equivalently be written as\footnote{\label{footnoteAV1}Formally, the set $\hat{\Delta}$ defined here is the projection of the set $\hat{\Delta}$ in \eqref{eqnEAM5} onto the coordinates corresponding to the non-box constraints; by abuse of notation, we continue to denote this projection by $\hat{\Delta}$. This distinction is immaterial for the computation of $v^*$ because we treat the box constraints as deterministic, implying that the entries of $\zeta_{b_{\mathrm{aug}}}^*-\zeta_{A_{\mathrm{aug}}}^*\hat{x}$ corresponding to these constraints are zero.}
\[
\hat{\Delta}=\left\{
\lambda\geq 0
\;\middle|\;
-\frac{\kappa_N}{\sqrt{N}}\mathbbm{1}
\leq
\hat{A}^\top\lambda-\omega
\leq
\frac{\kappa_N}{\sqrt{N}}\mathbbm{1},
\;
\mathbbm{1}^\top\hat{\Omega}\lambda\leq 1,
\;
\hat{b}(v)^\top\lambda-\hat{e}_N\|\omega\|_1
\geq
\hat{T}_N(v)-\frac{\kappa_{N}}{\sqrt{N}}
\right\}.
\]

Let
\[
\underline{\sigma}
:=
\min\{\Omega_{ii}\mid \Omega_{ii}>0\}
\]
denote the smallest positive diagonal entry of $\Omega$. Let $U\subseteq[p_A]$ denote the indices of the rows of $A$ that correspond to nondeterministic constraints, i.e., $U=\{i\in [p_A]\mid \Omega_{ii}>0\}$. Next, let $\mathcal{P}_U:\mathbb{R}^{p_A}\longrightarrow\mathbb{R}^{p_A}$
denote the projection operator that retains only the components corresponding to the nondeterministic constraints. More precisely, for any $\lambda\in\mathbb{R}^{p_A}$,
\[
[\mathcal{P}_U(\lambda)]_i
=
\begin{cases}
0,
& \text{if the $i$th row of $A$ and the $i$th component of $b$ are deterministic},\\[0.3em]
\lambda_i,
& \text{otherwise.}
\end{cases}
\]

On the event $E_N$, for all sufficiently large $N$ such that $\alpha_N \leq \underline{\sigma}/2$, we have $[\hat{\Omega}]_{ii} > \underline{\sigma}/2$ for all $i \in U$. The same reasoning shows that for all such $N$ and for all $\xi \in \mathcal{X}_N$, we also have $[\Omega_N(\xi)]_{ii} \geq \underline{\sigma}/2$ for all $i \in U$. As a consequence, since $\|\Omega_N(\xi) \lambda\|_1 \leq 1$ for all $\lambda \in \Delta_{0,N}(\xi,a,v)$, it follows that
\begin{equation}
\label{thmAVeqn9}
\sup_{\xi \in \mathcal{X}_N} \sup_{\lambda \in \Delta_{0,N}(\xi,a,v)} \|{\cal P}_U(\lambda)\|_1 \leq \frac{2}{\underline{\sigma}}.
\end{equation}
In the derivations that follow, we assume that N is sufficiently large that $\alpha_N\leq \underline{\sigma}/2$.

 Since $\|\hat{x}-x^*\|\leq \kappa_N/N^{1/4}$ on the event $E_N$, inequality \eqref{thmAVeqn9} implies that
\begin{equation}
\label{thmAVeqn10}
\begin{aligned}
\sup_{\xi\in\mathcal{X}_N}
\sup_{\lambda\in\Delta_{0,N}(\xi,a(\xi),v)}
\Bigl\{
\langle \xi_b-\xi_A x^*,\lambda\rangle
-
\langle \xi_b-\xi_A\hat{x},\lambda\rangle
\Bigr\}
&=
\sup_{\xi\in\mathcal{X}_N}
\sup_{\lambda\in\Delta_{0,N}(\xi,a(\xi),v)}
\Bigl\{
\langle \xi_A(\hat{x}-x^*),\lambda\rangle
\Bigr\}
\\
&=
\sup_{\xi\in\mathcal{X}_N}
\sup_{\lambda\in\Delta_{0,N}(\xi,a(\xi),v)}
\Bigl\{
\langle \xi_A(\hat{x}-x^*),\mathcal{P}^u(\lambda)\rangle
\Bigr\}
\\
&\leq
\frac{2M_N}{\underline{\sigma}}
\frac{\kappa_N}{N^{1/4}}.
\end{aligned}
\end{equation}
 Note that, since $M_N$ is chosen such that $M_N (\kappa_N/N^{1/4}) = o(1)$, the right-hand side of inequality~\eqref{thmAVeqn10} is asymptotically negligible.

We now show that for all sufficiently large $N$, the following holds on the event $E_N$ and for all $\xi \in \mathcal{X}_N$:\footnote{Here, by abuse of notation, we use $\mathcal{P}_U(\Delta_{0,N}(\xi,a,v))$ to denote the set $\{\mathcal{P}_U(\lambda) \mid (\lambda, \omega) \in \Delta_{0,N}(\xi,a,v)\}$.}
\begin{equation}
\label{thmAVeqn11}
\overset{\rightarrow}{d}_H\bigl(\mathcal{P}_U(\Delta_{0,N}(\xi,a(\xi),v)), \mathcal{P}_U(\hat{\Delta})\bigr) \leq C \alpha_N.
\end{equation}
Assuming that inequality~\eqref{thmAVeqn11} holds, a straightforward computation shows that on the event $E_N$ we have
\begin{equation}
\label{thmAVeqn12}
\sup_{\xi \in \mathcal{X}_N} \sup_{\lambda \in \Delta_{0,N}(\xi,a(\xi),v)} 
\left\{ 
\langle \xi_b - \xi_A \hat{x}, \lambda \rangle 
- \sup_{\lambda \in \hat{\Delta}} \langle \xi_b - \xi_A \hat{x}, \lambda \rangle 
\right\}
\leq C \alpha_N M_N.
\end{equation}

Combining inequalities~\eqref{thmAVeqn10} and~\eqref{thmAVeqn12} yields inequality~\eqref{thmAVeqn8} with
\[
\gamma_N = C M_N \frac{\kappa_N}{N^{1/4}} + C \alpha_N M_N = o(1).
\]
It therefore remains to establish inequality~\eqref{thmAVeqn11} in order to complete Step~2.

Using inequality~\eqref{thmAVeqn9} together with the formulation of $\phi$ in~\eqref{thmAVeqn4}, we obtain that, on the event $E_N$, for all sufficiently large $N$, and for every $(\lambda,\omega)\in\Delta_{0,N}(\xi,a(\xi),v)$,
\begin{equation}
\label{thmAVeqn13}
\|\hat{A}^\top\lambda-\omega\|
=
\bigl\|(\hat{A}-A_N(\xi))^\top\mathcal{P}_U(\lambda)\bigr\|
\leq
\frac{CM_N}{\sqrt{N}},
\end{equation}
where the equality follows from the fact that
$A_N(\xi)^\top\lambda=\omega$ for every
$(\lambda,\omega)\in\Delta_{0,N}(\xi,a(\xi),v)$.
Hence, for all sufficiently large $N$,
\begin{equation}
\label{thmAVeqn14}
\Delta_{0,N}(\xi,a(\xi),v)
\subseteq
\left\{
(\lambda,\omega)
\;\middle|\;
-\frac{\kappa_N}{\sqrt{N}}\mathbbm{1}
\leq
\hat{A}^\top\lambda-\omega
\leq
\frac{\kappa_N}{\sqrt{N}}\mathbbm{1},
\;
\lambda\geq 0
\right\}.
\end{equation}

Moreover, on the event $E_N$, for all sufficiently large $N$, and for every $(\lambda,\omega)\in\Delta_{0,N}(\xi,a(\xi),v)$,
\[
\|\hat{\Omega}\lambda\|_1
\leq
\bigl\|(\hat{\Omega}-\Omega_N(\xi))\mathcal{P}_U(\lambda)\bigr\|_1
+
\|\Omega_N(\xi)\lambda\|_1
\leq
1+C\alpha_N.
\]

Motivated by the preceding inequality, define the rescaling operator
\[
\Gamma_N(\lambda,\omega)
=
\bigl(
\Gamma_{1N}(\lambda),
\Gamma_{2N}(\omega)
\bigr),
\]
where
\[
\Gamma_{1N}(\lambda)
:=
(1+C\alpha_N)^{-1}\lambda,
\qquad
\Gamma_{2N}(\omega)
:=
(1+C\alpha_N)^{-1}\omega.
\]
It then follows from \eqref{thmAVeqn14} and the preceding inequality that, on the event $E_N$, for all sufficiently large $N$, and for every $(\lambda,\omega)\in\Delta_{0,N}(\xi,a(\xi),v)$,
\begin{equation}
\label{thmAVeqn15}
\Gamma_N(\lambda,\omega)
\in
\left\{
(\lambda,\omega)
\;\middle|\;
-\frac{\kappa_N}{\sqrt{N}}\mathbbm{1}
\leq
\hat{A}^\top\lambda-\omega
\leq
\frac{\kappa_N}{\sqrt{N}}\mathbbm{1},
\;
\|\hat{\Omega}\lambda\|_1\leq 1,
\;
\lambda\geq 0
\right\},
\end{equation}
and
\begin{equation}
\label{thmAVeqn16}
\bigl\|
\mathcal{P}_U\bigl(
\lambda-\Gamma_{1N}(\lambda)
\bigr)
\bigr\|
\leq
C\alpha_N.
\end{equation}

On the event $E_N$, for every $(\lambda,\omega)\in\Delta_{0,N}(\xi,a(\xi),v)$, we have
\[
b_N(\xi,v)^\top\lambda-a(\xi)\|\omega\|_1\geq 0,
\]
since $(\lambda,\omega)=(0,0)$ is feasible for the perturbed problem. Furthermore, the condition $|a(\xi)-5\|x^*\|_{\infty}|\leq \kappa_N/N^{1/4}$ implies that $a(\xi)\geq 5\|x^*\|{\infty}-\kappa_N/N^{1/4}$. It follows readily that on the event $E_N$ and for all sufficiently large N, $a(\xi)\geq \max\{0, 5\|x^*\|{\infty}-\kappa_N/N^{1/4}\}> \max\{0,5\|\hat{x}\|_{\infty}-2\kappa_N/N^{1/4}\}=\hat{e}_N$. Hence, for all sufficiently large $N$,
\begin{equation}
\label{thmAVeqn17}
\begin{aligned}
\langle \hat{b}(v),\Gamma_{1N}(\lambda)\rangle
&-\hat{e}_N\|\Gamma_{2N}(\omega)\|_1
\geq \langle \hat{b}(v),\Gamma_{1N}(\lambda)\rangle
-a(\xi)\|\Gamma_{2N}(\omega)\|_1\\
&=\langle \hat{b}(v)-b_N(\xi,v),
\mathcal{P}_U\bigl(\Gamma_{1N}(\lambda)\bigr)\rangle
+
\langle b_N(\xi,v),\Gamma_{1N}(\lambda)\rangle
-a(\xi)\|\Gamma_{2N}(\omega)\|_1
\\
&\geq
\langle \hat{b}(v)-b_N(\xi,v),
\mathcal{P}_U\bigl(\Gamma_{1N}(\lambda)\bigr)\rangle
\\
&\geq
-\frac{CM_N}{\sqrt{N}}.
\end{aligned}
\end{equation}
where the second inequality follows from the positive homogeneity of the objective function and the fact that
$\bigl(\Gamma_{1N}(\lambda),\Gamma_{2N}(\omega)\bigr)$ is a positive scalar multiple of $(\lambda,\omega)$.

We now show that, on the event $E_N$, for all sufficiently large $N$ and for every $v\leq v_0$,
\begin{equation}
\label{thmAVeqn18}
\hat{T}_N(v)\leq \frac{CM_N}{\sqrt{N}}.
\end{equation}

Since the statistic $\hat{T}_N(v)$ is nondecreasing in $v$, it is sufficient to establish \eqref{thmAVeqn18} for the case $v=v_0$. By the strong duality theorem for LP,
\[
\hat{T}_N(v_0)
=
\min\Bigl\{
t
\;\Big|\;
t\hat{\Omega}\mathbbm{1}
+\hat{A}x
\geq
\hat{b}(v_0),\;
\|x\|_{\infty}\leq \hat{a}_N,\;
t\geq 0
\Bigr\}.
\]
On the event $E_N$, we have $\|x^*\|_{\infty}\leq \hat{a}_N$. We have (as $b(v_0)\leq Ax^*$)
\[
\begin{aligned}
(\hat{b}(v_0)-\hat{A}x^*)_+
&=
\bigl(
\hat{b}(v_0)-b(v_0)
+
b(v_0)-Ax^*
+
(A-\hat{A})x^*
\bigr)_+
\\
&\leq
\bigl(\hat{b}(v_0)-b(v_0)\bigr)_+
+
\bigl((A-\hat{A})x^*\bigr)_+
\\
&\leq
\frac{CM_N}{\sqrt{N}}\hat{\Omega}\mathbbm{1},
\end{aligned}
\]
where the last inequality follows from the fact that
$\hat{\Omega}_{ii}\geq \underline{\sigma}/2$ for all $i\in U$ and all sufficiently large $N$. These observations, together with the definition of $\hat{T}_N(v_0)$, imply inequality~\eqref{thmAVeqn18}.

Equations~\eqref{thmAVeqn17} and~\eqref{thmAVeqn18}, together with the fact that $M_N=o(\kappa_N)$, imply that, on the event $E_N$, for every $(\lambda,\omega)\in\Delta_{0,N}(\xi,a(\xi),v)$ and all sufficiently large $N$,
\begin{equation}
\label{thmAVeqn19}
\langle \hat{b}(v),\Gamma_{1N}(\lambda)\rangle
-\hat{a}_N\|\Gamma_{2N}(\omega)\|_1
\geq
\hat{T}_N(v)-\frac{\kappa_N}{\sqrt{N}}.
\end{equation}
Combining this inequality with~\eqref{thmAVeqn15} implies that
\[
\bigl\{
\Gamma_{1N}(\lambda)
\mid
(\lambda,\omega)\in\Delta_{0,N}(\xi,a(\xi),v)
\bigr\}
\subseteq
\hat{\Delta}.
\]
Together with~\eqref{thmAVeqn16}, this containment establishes inequality~\eqref{thmAVeqn11}, which completes the proof of inequality~\eqref{thmAVeqn8}.


\underline{Step 3} In this step, we use equation~\eqref{thmAVeqn8} together with a coupling argument to upper bound the test statistic $\sqrt{N}\hat{T}_N(v)$ by another statistic whose distribution can be consistently estimated.

Let
\[
\hat{\xi}^*=(\hat{\xi}^*_A,\hat{\xi}^*_b,\hat{\xi}^*_{\Omega},\hat{a}^*_N)
=
(\sqrt{N}(\hat{A}^*-A),\sqrt{N}(\hat{b}^*(v)-b(v)),\sqrt{N}(\hat{\Omega}^*-\Omega),\hat{a}^*_N)
\]
be an independent copy of the random element
\[
\hat{\xi}
=
(\sqrt{N}(\hat{A}-A),\sqrt{N}(\hat{b}(v)-b(v)),\sqrt{N}(\hat{\Omega}-\Omega),\hat{a}_N),
\]
computed from a sample $\mathbb{W}_N^*$, where $\mathbb{W}_N^*$ is independent of and identically distributed as $\mathbb{W}_N$. Define the event
\[
F_N^*
:=
\left\{
\mathbb{W}_N^*
\,\middle|\,
\|\hat{\xi}_A^*\|>M_N
\;\text{or}\;
\|\hat{\xi}_b^*\|>M_N
\;\text{or}\;
\|\hat{\xi}_{\Omega}^*\|>\sqrt{N}\alpha_N
\;\text{or}\;
\left|\hat{a}_N^*-5\|x\|_{\infty}\right|>\frac{\kappa_N}{N^{1/4}}
\right\}.
\]
By the choice of $M_N$ and $\alpha_N$, Assumptions~\ref{ass1} and~\ref{ass2}, and Lemma~\ref{lemNP}, the sequence
\[
\delta_{1N}:=Q(F_N^*)
\]
satisfies $\delta_{1N}\to 0$ as $N\to\infty$.

Equations~\eqref{thmAVeqn7} and~\eqref{thmAVeqn8} imply that, for all sufficiently large $N$, for $v\leq v_0$, and on the event $E_N$,
\footnote{Here we use the fact that $\hat{a}_N^*\geq \|x^*\|_{\infty}$ on the complement of $F_N^*$, which implies that $\phi(\mu,\hat{a}_N^*,v))=0$ for $v\leq v_0$.}
\begin{equation}
\label{thmAVeqn20}
\sqrt{N}\,\phi_N(\mu_N(\hat{\xi}^*),\hat{a}_N^*,v)
\leq
\sup_{\lambda\in\hat{\Delta}}
\langle \hat{\xi}_b^*-\hat{\xi}_A^*\hat{x},\lambda\rangle
+\gamma_N
+\infty\cdot\mathbbm{1}_{\{F_N^*\}}.
\end{equation}
In all arguments that follow, we assume that $v\leq v_0$. The left-hand side of \eqref{thmAVeqn20} has the same distribution as $\sqrt{N}\hat{T}_N(v)$, and note that the right-hand side depends on $v$ only through the set $\hat{\Delta}$. We next show that the distribution of the right-hand side can be consistently estimated by a bootstrap procedure.

Let \( J_N(\cdot; v) \) denote the CDF of \( \sqrt{N} \hat{T}_N(v) \). Define
\[
\hat{J}_N(\cdot; v) := \mathcal{L} \left( \sup_{\lambda\in\hat{\Delta}}
\langle \hat{\xi}_b^*-\hat{\xi}_A^*\hat{x},\lambda\rangle
+\gamma_N
+\infty\cdot\mathbbm{1}_{\{F_N^*\}} \,\middle|\, \mathbb{W}_N \right)
\]
as the conditional (on \( \mathbb{W}_N \)) CDF of the right-hand side of inequality \eqref{thmAVeqn20}, viewed as a random element taking values in \( \mathbb{R} \cup \{\infty\} \), and define
\[
\hat{G}_N(\cdot; v) := \mathcal{L} \left( \sup_{\lambda\in\hat{\Delta}}
\langle \hat{\xi}_b^*-\hat{\xi}_A^*\hat{x},\lambda\rangle
 \,\middle|\, \mathbb{W}_N \right)
\]
as the conditional CDF of the first term on the right-hand side of inequality \eqref{thmAVeqn20}.

Similarly, define
\[
\hat{G}_\infty(\cdot; v)
:=
\mathcal{L}
\left(
\sup_{\lambda \in \hat{\Delta}}
\langle \zeta_b-\zeta_Ax^*, \lambda \rangle
\,\middle|\,
\mathbb{W}_N
\right),
\]
where $(\zeta_A,\zeta_b)$ denotes the asymptotic distribution of $(\sqrt{N}(\hat{A}-A),\sqrt{N}(\hat{b}-b))$, which can be obtained from $\mathcal{N}(0,V(Q))$ in Assumption~\ref{ass1}. Let $\hat{H}_N(\cdot;v)$ denote the distribution of $v^*$ as defined in \eqref{eqnEAM4}. Specifically, \footnote{Here, as in Footnote~\ref{footnoteAV1}, with a slight abuse of notation, $\sup_{\lambda \in \hat{\Delta}}\langle \zeta_b-\zeta_Ax^*,\lambda\rangle$, where $\hat{\Delta}$ denotes the projected set described in \eqref{thmAVeqn8}, is equal to $v^*$ in \eqref{eqnEAM4}.}
\[
\hat{H}_N(\cdot; v)
:=
\mathcal{L}
\left(
\sup_{\lambda \in \hat{\Delta}}
\langle \zeta_b^*-\zeta_A^*\hat{x}, \lambda \rangle
\,\middle|\,
\mathbb{W}_N
\right),
\]
where $(\zeta^*_A,\zeta^*_b)$ is constructed from the bootstrap estimator $\zeta^*_N$ of $\mathcal{N}(0,V(Q))$ given in Assumption~\ref{ass3}.

Arguing as in the derivation of \eqref{thmAVeqn9}, it is straightforward to show that, on the event $E_N$, for all sufficiently large $N$ such that $\alpha_N\leq \underline{\sigma}/2$,
\begin{equation}
\label{thmAVeqn21}
\sup_{\lambda \in \hat{\Delta}} \|{\cal P}_U(\lambda)\|_1 \leq \frac{2}{\underline{\sigma}}.
\end{equation}
It thus follows that the function $g$ defined by
\begin{equation}
\label{thmAVeqn22}
g(w)=\sup_{\lambda\in\hat{\Delta}}\langle w,\mathcal{P}_U(\lambda)\rangle
\end{equation}
is uniformly (on $E_N$) Lipschitz continuous for all sufficiently large N.

By Strassen's coupling characterization of the Prohorov metric,
\[
\delta_{2N}
:=
\sup_{\mathbb{W}\in E_N}
d_{\mathrm{Pr}}\!\left(\hat{J}_N(\cdot;v),\hat{G}_N(\cdot;v)\right)
\leq
\gamma_N+\delta_{1N}
\longrightarrow 0.
\]

Moreover, by Assumptions~\ref{ass1} and~\ref{ass3}, Strassen's theorem, and the fact that $\|\hat{x}\|$ is uniformly bounded on $E_N$,
\[
\delta_{3N}
:=
\sup_{\mathbb{W}\in E_N}
d_{\mathrm{Pr}}
\!\left(
\hat{\xi}_b^*-\hat{\xi}_A^*\hat{x},
\zeta_b^*-\zeta_A^*\hat{x}
\right)
\longrightarrow 0.
\]

Similarly, by Assumption~\ref{ass3}, Strassen's theorem, and the consistency of $\hat{x}$ for $x^*$ on $E_N$,
\[
\delta_{4N}
:=
\sup_{\mathbb{W}\in E_N}
d_{\mathrm{Pr}}
\!\left(
\zeta_b^*-\zeta_A^*\hat{x},
\zeta_b-\zeta_Ax^*
\right)
\longrightarrow 0.
\]

Since the function $g$ defined in \eqref{thmAVeqn22} is uniformly Lipschitz continuous on $E_N$ for all sufficiently large $N$, it follows from Strassen's theorem and the fact that $\delta_{3N}\to 0$ that
\[
\delta_{5N}
:=
\sup_{\mathbb{W}\in E_N}
d_{\mathrm{Pr}}
\!\left(
\hat{G}_N(\cdot;v),\hat{H}_N(\cdot;v)
\right)
\longrightarrow 0.
\]
Similarly, Strassen's theorem and the fact that $\delta_{4N}\to 0$ imply that
\begin{equation}
\label{thmAVeqn23}
\delta_{6N}
:=
\sup_{\mathbb{W}\in E_N}
d_{\mathrm{Pr}}
\!\left(
\hat{H}_N(\cdot;v),\hat{G}_{\infty}(\cdot;v)
\right)
\longrightarrow 0.
\end{equation}
The triangle inequality for the Prohorov metric then implies
\begin{equation}
\label{thmAVeqn24}
\delta_{7N}
:=
\sup_{\mathbb{W}\in E_N}
d_{\mathrm{Pr}}
\!\left(
\hat{J}_N(\cdot;v),\hat{G}_{\infty}(\cdot;v)
\right)
\leq
\delta_{2N}+\delta_{5N}+\delta_{6N}
\longrightarrow 0.
\end{equation}

Finally, define
\[
\delta_N:=\max\{\delta_{6N},\delta_{7N}\}.
\]
Let $\alpha\in(0,1/2)$ and let $\hat{c}_N(1-\alpha;v)$ be the critical value defined in Section~\ref{SecInference}. For now, suppose that there exists a constant $C>0$ such that, for all sufficiently large $N$ and on the event $E_N$,
\begin{equation}
\label{thmAVeqn25}
\hat{G}_{\infty}(\hat{C}_N(1-\alpha;v_0)+\delta_N;v_0)
-
\hat{G}_{\infty}(\hat{C}_N(1-\alpha;v_0)-\delta_N;v_0)
\leq C\delta_N.
\end{equation}
We will establish \eqref{thmAVeqn25} in Step 4 below. 

For all sufficiently large $N$ and on the event $E_N$, Equations~\eqref{thmAVeqn23}, \eqref{thmAVeqn24}, and \eqref{thmAVeqn25} imply that
\begin{equation}
\label{thmAVeqn26}
\begin{aligned}
J_N(\hat{c}_N(1-\alpha;v_0);v_0)
&\geq
\hat{J}_N(\hat{c}_N(1-\alpha;v_0);v_0)\\
&\geq
\hat{G}_{\infty}(\hat{c}_N(1-\alpha;v_0)-\delta_N;v_0)-\delta_N\\
&\geq
\hat{G}_{\infty}(\hat{c}_N(1-\alpha;v_0)+\delta_N;v_0)-C\delta_N\\
&\geq
\hat{H}_{N}(\hat{c}_N(1-\alpha;v_0);v_0)-C\delta_N\\
&\geq
1-\alpha-C\delta_N.
\end{aligned}
\end{equation}
where the last inequality follows from the definition of $\hat{c}_N(1-\alpha;v_0)$, which implies that $\hat{H}_{N}(\hat{c}_N(1-\alpha;v_0);v_0)\geq 1-\alpha$. 

Let $q_N(\cdot;v_0)$ denote the quantile function associated with $J_N(\cdot;v_0)$. Inequality~\eqref{thmAVeqn26} implies that on the event $E_N$ and for all sufficiently large $N$,
\begin{equation}
\label{thmAVeqn27}
q_N(1-\alpha-C\delta_N;v_0)
\leq
\hat{c}_N(1-\alpha;v_0).
\end{equation}
Therefore,
\begin{equation}
\label{thmAVeqn28}
\begin{aligned}
Q\!\left(\sqrt{N}\hat{T}_N(v_0)>\hat{c}_N(1-\alpha;v_0)\right)
&\leq
Q\!\left(\{\sqrt{N}\hat{T}_N(v_0)>\hat{c}_N(1-\alpha;v_0)\}\cap E_N\right)
+\delta_{0N}\\
&\leq
Q\!\left(\sqrt{N}\hat{T}_N(v_0)>q_N(1-\alpha-C\delta_N;v_0)\right)
+\delta_{0N}\\
&\leq
\alpha+C\delta_N+\delta_{0N}.
\end{aligned}
\end{equation}
It follows that
\[
\limsup_{N\to\infty}
Q\!\left(\sqrt{N}\hat{T}_N(v_0)>\hat{c}_N(1-\alpha;v_0)\right)
\leq \alpha,
\]
which proves \eqref{pthmAVeqn1}.

\underline{Step 4} We now prove \eqref{thmAVeqn25}. Let $\Delta_0(v_0)$ be as defined in Assumption~\ref{ass4}. Equation~\eqref{pthmAVeqn2} implies that $\phi(\mu,5\|x^*\|_{\infty},v_0)=0$. Therefore,
\[
\Delta_0(v_0)
=
\left\{
\lambda\geq 0
\;\middle|\;
A_{\mathrm{aug}}^\top\lambda=0,\;
\mathbbm{1}^\top\Omega_{\mathrm{aug}}\lambda\leq 1,\;
\left\langle
b_{\mathrm{aug}}(5\|x^*\|_{\infty},v_0),\lambda
\right\rangle=0
\right\}.
\]
A straightforward computation shows that the projection of the augmented optimal face $\Delta_0(v_0)$ onto the coordinates corresponding to the non-box constraints is given by
\[
\Delta_0(v_0)
=
\left\{
\lambda\geq 0
\;\middle|\;
A^\top\lambda-\omega=0,\;
\mathbbm{1}^\top\Omega\lambda\leq 1,\;
\left\langle b(v_0),\lambda\right\rangle
-5\|x^*\|_{\infty}\|\omega\|_1=0
\right\},
\]
where, by an abuse of notation, we continue to denote this projected set by $\Delta_0(v_0)$. The latter set is precisely $\Delta_{0,N}(0,5\|x^*\|_{\infty},v_0)$. Therefore, by equation~\eqref{thmAVeqn11}, there exists a constant $C>0$ such that, for all sufficiently large $N$ and on the event $E_N$,
\[
\overset{\rightarrow}{d}_H
\left(
\mathcal{P}_U(\Delta_0(v_0)),
\mathcal{P}_U(\hat{\Delta})
\right)
\leq C\alpha_N.
\]
By Assumption~\ref{ass4}, there exist $c>0$ and $\lambda_0\in\Delta_0(v_0)$ such that
\[
\operatorname{Var}
\left(
\left\langle
\zeta_b-\zeta_Ax^*,
\lambda_0
\right\rangle
\right)\geq c.
\]
where $(\zeta_b,\zeta_A)$ are as defined in $\hat{G}_{\infty}(\cdot;v)$ and denote the projection of $(\zeta_{b_{\mathrm{aug}}},\zeta_{A_{\mathrm{aug}}})$ onto the components corresponding to the non-box constraints. The components corresponding to the box constraints are equal to zero.

Combining the Hausdorff distance bound and the variance bound above, we obtain that, for all sufficiently large $N$ and on the event $E_N$, there exists $\hat{\lambda}\in\hat{\Delta}$ such that
\[
\left\|
\mathcal{P}_U(\lambda_0-\hat{\lambda})
\right\|
\leq C\alpha_N,
\]
and
\begin{equation}
\label{thmAVeqn29}
\operatorname{Var}
\left(
\left\langle
\zeta_b-\zeta_Ax^*,
\hat{\lambda}
\right\rangle
\right)
=
\operatorname{Var}
\left(
\left\langle
\zeta_b-\zeta_Ax^*,
\mathcal{P}_U(\hat{\lambda})
\right\rangle
\right)
\geq c/2.
\end{equation}
Equation~\eqref{thmAVeqn25} follows from \eqref{thmAVeqn21}, \eqref{thmAVeqn29}, and the anti-concentration result in Proposition~B.1 of \cite{GM25}.

\underline{Step 5} In this step, we prove that
\begin{equation}
\label{thmAVeqn30}
\lim_{t\to\infty}\lim_{N\to\infty}
Q\!\left(
\sqrt{N}\hat{T}_N(v_0+h/\sqrt{N})
\leq
\hat{c}_N(1-\alpha;v_0+h/\sqrt{N})
\ \text{for some } h\geq t
\right)
=0,
\end{equation}
which yields equation \eqref{eqnthm2}.
We first note that, by equation~\eqref{thmAVeqn4} and the fact that the feasible multipliers $\lambda$ are nonnegative,
\begin{equation}
\label{thmAVeqn31}
\hat{T}_N(v_0+t/\sqrt{N})
\leq
\hat{T}_N(v_0+h/\sqrt{N})
\qquad
\text{for all } h\geq t.
\end{equation}

Next, observe that, for every $v\in\mathbb{R}$,\footnote{Recall that $\hat{\Delta}$ depends on $v$, although we keep the dependence implicit for notational simplicity.}
\[
\hat{\Delta}
\subseteq
\tilde{\Delta}
:=
\left\{
\lambda\geq0
\;\middle|\;
\mathbbm{1}^{\top}\hat{\Omega}\lambda\leq1
\right\}.
\]
Consequently,
\[
\sup_{\lambda\in\hat{\Delta}}
\langle \zeta_b^*-\zeta_A^*\hat{x}, \lambda \rangle
\leq
\sup_{\lambda\in\tilde{\Delta}}
\langle \zeta_b^*-\zeta_A^*\hat{x}, \lambda \rangle.
\]

Let $\tilde{c}_N(1-\alpha)$ denote the $(1-\alpha)$-quantile of
\[
\tilde{v}^*
:=
\sup_{\lambda\in\tilde{\Delta}}
\langle \zeta_b^*-\zeta_A^*\hat{x}, \lambda \rangle.
\]
Equation~\eqref{thmAVeqn21} implies that, for all sufficiently large $N$ and on the event $E_N$, the quantiles $\tilde{c}_N(1-\alpha)$ are well defined and finite, as $\langle \zeta_b^*-\zeta_A^*\hat{x},\lambda\rangle
=
\langle \zeta_b^*-\zeta_A^*\hat{x},\mathcal{P}_U(\lambda)\rangle$.

Since $\tilde{\Delta}$ does not depend on $v$, it follows that
\begin{equation}
\label{thmAVeqn32}
\hat{c}_N(1-\alpha;v)
\leq
\tilde{c}_N(1-\alpha)
\qquad
\text{for all } v\in\mathbb{R}.
\end{equation}

Combining \eqref{thmAVeqn31} and \eqref{thmAVeqn32} yields
\[
\begin{aligned}
\Bigl\{
&\sqrt{N}\hat{T}_N\!\left(v_0+\frac{h}{\sqrt{N}}\right)
\leq
\hat{c}_N\!\left(1-\alpha;v_0+\frac{h}{\sqrt{N}}\right)
\ \text{for some } h\geq t
\Bigr\}\\
&\subseteq
\Bigl\{
\sqrt{N}\hat{T}_N\!\left(v_0+\frac{t}{\sqrt{N}}\right)
\leq
\tilde{c}_N(1-\alpha)
\Bigr\}.
\end{aligned}
\]
Thus, to establish \eqref{thmAVeqn30}, it remains to show that
\[
\lim_{t\to\infty}\lim_{N\to\infty}
Q\!\left(
\sqrt{N}\hat{T}_N\!\left(v_0+\frac{t}{\sqrt{N}}\right)
\leq
\tilde{c}_N(1-\alpha)
\right)
=0.
\]
Since $Q(E_N)\to1$, it is sufficient to prove that
\begin{equation}
\label{thmAVeqn33}
\lim_{t\to\infty}\lim_{N\to\infty}
Q\!\left(
\left\{
\sqrt{N}\hat{T}_N\!\left(v_0+\frac{t}{\sqrt{N}}\right)
\leq
\tilde{c}_N(1-\alpha)
\right\}
\cap E_N
\right)
=0.
\end{equation}

Lemma~B.5 of \cite{GM25} (see also Lemma~4 of \cite{ARP}) implies that, for all sufficiently large $N$ and on the event $E_N$,
\[
d_H
\left(
\mathcal{P}_U(\tilde{\Delta}),
\mathcal{P}_U(\tilde{\Delta}_0)
\right)
\leq
C\alpha_N,
\]
where $\tilde{\Delta}_0:=\left\{
\lambda\geq0
\;\middle|\;
\mathbbm{1}^{\top}\Omega \lambda\leq1
\right\}$.
Consequently, on $E_N$, the random function
\[
\hat{\tilde{g}}(w)
:=
\sup_{\lambda\in\tilde{\Delta}}
\left\langle
w,
\mathcal{P}_U(\lambda)
\right\rangle
\]
converges uniformly on every bounded subset of its domain to
\[
\tilde{g}(w)
:=
\sup_{\lambda\in\tilde{\Delta}_0}
\left\langle
w,
\mathcal{P}_U(\lambda)
\right\rangle.
\]
Since
\[
(\zeta_b^*,\zeta_A^*,\hat{x})
\Rightarrow
(\zeta_b,\zeta_A,x^*),
\]
the continuous mapping theorem yields
\[
\sup_{\lambda\in\tilde{\Delta}}
\langle
\zeta_b^*-\zeta_A^*\hat{x},
\lambda
\rangle
\Rightarrow
\sup_{\lambda\in\tilde{\Delta}_0}
\langle
\zeta_b-\zeta_Ax^*,
\lambda
\rangle.
\]

By Assumption~\ref{ass4} and the anti-concentration result in Proposition~B.1 of \cite{GM25},\footnote{Note that we have the inclusion $\Delta_0(v_0)\subseteq \tilde{\Delta}_0$} the limiting distribution is nondegenerate and admits a bounded density in a neighborhood of each of its $(1-\alpha)$-quantiles, for every $\alpha\in(0,1/2)$. Therefore,
\begin{equation}
\label{thmAVeqn34}
\tilde{c}_N(1-\alpha)
\overset{p}{\longrightarrow}
\tilde{c}(1-\alpha),
\end{equation}
where $\tilde{c}(1-\alpha)$ denotes the $(1-\alpha)$-quantile of
\[
\sup_{\lambda\in\tilde{\Delta}_0}
\langle
\zeta_b-\zeta_Ax^*,
\lambda
\rangle.
\]

Let
\[
\hat{\xi}
:=
\left(
\sqrt{N}(\hat{A}-A),
\sqrt{N}(\hat{b}-b),
\sqrt{N}(\hat{\Omega}-\Omega)
\right)
\]
denote the perturbation associated with the realized data. In terms of $\hat{\xi}$, we can write
\[
\sqrt{N}\hat{T}_N(v_0+t/\sqrt{N})
=
R_N(t)+D_N(t),
\]
where
\[
\begin{aligned}
R_N(t)
&:=
\sqrt{N}\Bigl(
\phi_N(\mu(\hat{\xi}),\hat{a}_N,v_0+t/\sqrt{N})
-
\phi(\mu,\hat{a}_N,v_0+t/\sqrt{N})
\Bigr),\\
D_N(t)
&:=
\sqrt{N}\phi(\mu,\hat{a}_N,v_0+t/\sqrt{N}).
\end{aligned}
\]
We first derive a lower bound for the drift term $D_N(t)$. Specifically, we show that, for some $C>0$, all sufficiently large $N$, and on $E_N$,
\begin{equation}
\label{thmAVeqn35}
D_N(t)\geq Ct,\qquad t\geq0.
\end{equation}
Indeed, since $\|x^*\|_{\infty}\leq \hat{a}_N$ on the event $E_N$, by \eqref{lemLBeqn1} of Lemma~\ref{lemLB}, there exists a constant $C>0$ such that
\[
\phi(\mu,\hat{a}_N,v_0+h)\geq Ch
\qquad\forall h\geq0.
\]
Setting $h=t/\sqrt{N}$ and using the definition of $D_N(t)$ then yields
\eqref{thmAVeqn35}.

We now derive a lower bound for $R_N(t)$. For notational convenience, let $v_t:=v_0+\frac{t}{\sqrt N}$.
By Lemma~\ref{lemAux} for all sufficiently large $N$, on the event
\[
\left\{
\sqrt{N}\hat{T}_N\!\left(v_0+\frac{t}{\sqrt{N}}\right)
\leq
\tilde{c}_N(1-\alpha)
\right\}
\cap E_N
\]
We must have $\phi(\mu,\hat{a}_N,v_t)<\infty$ and thus $R_N(t)<\infty$. Moreover, by the saddle-point property of the Lagrangian,

\begin{equation}
\label{thmAVeqn36}
\begin{aligned}
R_N(t)
&=
\sqrt N\Bigl(
\phi_N(\mu(\hat{\xi}),\hat{a}_N,v_t)
-
\phi(\mu,\hat{a}_N,v_t)
\Bigr)
\\
&=
\sqrt N\Bigl(
L_N(\lambda_{\hat{\xi}},\omega_{\hat{\xi}},
\eta_{\hat{\xi}},s_{\hat{\xi}};
\hat{\xi},\hat{a}_N,v_t)
-
L(\lambda_0,\omega_0,\eta_0,s_0;
\hat{a}_N,v_t)
\Bigr)
\\
&\geq
\sqrt N\Bigl(
L_N(\lambda_0,\omega_0,
\eta_{\hat{\xi}},s_{\hat{\xi}};
\hat{\xi},\hat{a}_N,v_t)
-
L(\lambda_0,\omega_0,
\eta_{\hat{\xi}},s_{\hat{\xi}};
\hat{a}_N,v_t)
\Bigr)
\\
&=
\sqrt{N}(\hat{b}(v_t)-b(v_t))^\top\lambda_0
-
\eta_{\hat{\xi}}^\top(\sqrt{N}(\hat{A}-A))^\top\lambda_0
-
\sqrt{N}s_{\hat{\xi}}\,
\mathbbm{1}^\top(\hat{\Omega}-\Omega)\lambda_0,
\end{aligned}
\end{equation}
where $(\lambda_0,\omega_0)\in\Delta_0(\hat{a}_N,v_t)$ and
$(\eta_{\hat{\xi}},s_{\hat{\xi}})
\in S_{0,N}(\hat{\xi},\hat{a}_N,v_t)$.
By the dual characterization of $\phi$ that follows \eqref{thmAVeqn4}, we have

\[
s_{\hat{\xi}}
=
\min\Bigl\{
s \;\Big|\;
s\hat{\Omega} \mathbbm{1}+\hat{A}\eta \ge \hat{b}(v_t),\;
\|\eta\|_{\infty}\le \hat{a}_N,\;
s\ge 0
\Bigr\}=\phi_N(\mu(\hat{\xi}),\hat{a}_N,v_t).
\]
We first derive an upper bound on $\sqrt{N}s_{\hat{\xi}}$. Arguing as in \eqref{thmAVeqn6}, we get
\[
\begin{aligned}
0\leq \sqrt{N}s_{\hat{\xi}}
&=
\sqrt N\Bigl(
\phi_N(\mu(\hat{\xi}),\hat{a}_N,v_t)
-
\phi(\mu,\hat{a}_N,v_t)
\Bigr)
+\sqrt{N}\phi(\mu,\hat{a}_N,v_t)
\\
&\leq
\sqrt{N}(\hat{b}(v_t)-b(v_t))^\top\lambda_{\hat{\xi}}
-
\eta_0^\top(\sqrt{N}(\hat{A}-A))^\top\lambda_{\hat{\xi}}
+\sqrt{N}\phi(\mu,\hat{a}_N,v_t)\\
&=
\sqrt{N}(\hat{b}(v_t)-b(v_t))^\top\mathcal{P}_U(\lambda_{\hat{\xi}})
-
\eta_0^\top(\sqrt{N}(\hat{A}-A))^\top\mathcal{P}_U(\lambda_{\hat{\xi}})
+\sqrt{N}\phi(\mu,\hat{a}_N,v_t)
\end{aligned}.
\]

Assumption \ref{ass1} and equations \eqref{thmAVeqn9} and \eqref{thmAVeqn11} imply that 
\[
\sqrt{N}(\hat{b}(v_t)-b(v_t))^\top\mathcal{P}_U(\lambda_{\hat{\xi}})
-
\eta_0^\top(\sqrt{N}(\hat{A}-A))^\top\mathcal{P}_U(\lambda_{\hat{\xi}})=O_p(1)
\]
and 
Equation \eqref{lemLBeqn4} from Lemma \ref{lemLB} implies that
\[
\sqrt{N}\phi(\mu,\hat{a}_N,v_t)\leq Ct
\]
for some $C>0$.
It then follows from the inequalities above and Assumption \ref{ass2} that the last term in \eqref{thmAVeqn36} satisfies 
\[
\sqrt{N}s_{\hat{\xi}}\,
\mathbbm{1}^\top(\hat{\Omega}-\Omega)\lambda_0=o_p(1)
\]
for any fixed $t>0$. On the other hand, the remaining term in \eqref{thmAVeqn36} 
\[
\chi_N:=\sqrt{N}(\hat{b}(v_t)-b(v_t))^\top\lambda_0
-
\eta_{\hat{\xi}}^\top
\bigl(\sqrt{N}(\hat{A}-A)\bigr)^\top\lambda_0
\]
does not depend on $v_t$ (and thus on $t$), and the box-constraint implies that
\begin{equation}
\label{thmAVeqnBC}
\begin{aligned}
\|\chi_N\|
&\leq
\|\sqrt{N}(\hat{b}(v_t)-b(v_t))\|
\|\lambda_0\|+
C\hat{a}_N
\|\sqrt{N}(\hat{A}-A)\|
\|\lambda_0\| \\
&\overset{d}\longrightarrow \|\zeta_b\|\|\lambda_0\|
+
2C\|x^*\|_{\infty}
\|\zeta_A\|\|\lambda_0\|
\end{aligned}
\end{equation}
where the last line follows from the continuous mapping theorem. Thus $\{\chi_N\}$ is uniformly bounded in probability.

In conclusion, for any fixed $t>0$
\[
\begin{aligned}
Q\!\left(
\left\{
\sqrt{N}\hat{T}_N(v_t)
\leq
\tilde{c}_N(1-\alpha)
\right\}
\cap E_N
\right)
&=
Q\!\left(
\left\{
R_N(t)+D_N(t)
\leq
\tilde{c}_N(1-\alpha)
\right\}
\cap E_N
\right)
\\
&\leq
Q\!\left(
\left\{
Ct-\chi_N-\tilde{c}(1-\alpha)+o_p(1)\leq 0
\right\}
\cap E_N
\right)
\end{aligned}
\]
where we have used \eqref{thmAVeqn34}. Now for any $\epsilon>0$, since $\chi_N$ does not depend on $t$, we can clearly pick $t>0$ sufficiently large
that
\[
\limsup_{N\to \infty}Q\!\left(
\left\{
Ct-\chi_N-\tilde{c}(1-\alpha)+o_p(1)\leq 0
\right\}
\cap E_N
\right)\leq \epsilon.
\]
This completes the proof of \eqref{thmAVeqn33}.
\end{proof}

\subsection{Proof of Theorem \ref{thm2}}
\label{Secthm2}
\begin{proof}
We first establish the claim under the assumption that the LP \eqref{eqnLP} has a unique nonzero solution (Case 1). The proof for the case in which the dual LP has a unique nonzero solution is analogous and is therefore omitted. We then consider the case in which both the primal and dual LPs have unique nonzero solutions (Case 2).

\underline{\textbf{Case 1:}} Suppose that the LP \eqref{eqnLP} has a unique solution $x^* \neq 0$. To establish the claim, it suffices to show that
\begin{equation}
\label{pthmregeqn1}
\liminf_{N\to\infty}
Q\left(
\sqrt{N}\hat{T}_N(v)>\hat{c}_N(1-\alpha;v)
\quad \forall v\geq v_0
\right)
\geq \alpha,
\end{equation}
where $\hat{T}_N$ and $\hat{c}_N$ denote the test statistic and critical value, respectively, used to construct the upper one-sided confidence interval in Section \ref{SecInference}. Indeed, \eqref{pthmregeqn1} implies that the probability that $v_0$ is excluded from the upper one-sided confidence interval is asymptotically at least $\alpha$. Combining this result with \eqref{eqnthm1}, which provides the corresponding upper bound on the probability of exclusion, yields
\[
\lim_{N\to\infty}
Q\left(v_0\in CI^U_{1-\alpha}\right)
=
1-\alpha.
\]

We establish \eqref{pthmregeqn1} by considering two exhaustive cases. First, suppose that there exists $h_0>0$ such that
\begin{equation}
\label{pthmregeqn2}
\phi\left(
\mu,2\lVert x^*\rVert_{\infty},v_0+h_0
\right)
<\infty,
\end{equation}
where $\phi$ is defined in \eqref{thmAVeqn4}. We then consider the complementary case in which
\begin{equation}
\label{pthmregeqn3}
\phi\left(
\mu,2\lVert x^*\rVert_{\infty},v_0+h
\right)
=
\infty
\qquad
\text{for all } h>0.
\end{equation}
We break the proof into several steps. Step 1 through 4 below prove \eqref{pthmregeqn1} when \eqref{pthmregeqn2} holds, and step 5 establishes \eqref{pthmregeqn1} when \eqref{pthmregeqn3} holds.

\underline{Step 1} Suppose that \eqref{pthmregeqn2} holds. In this step, we reduce the task of proving \eqref{pthmregeqn1} to showing that equation \eqref{pthmregeqn5} below holds. Fix $\epsilon>0$. It follows from \eqref{eqnthm2} that, for some $h_{\epsilon}>0$ sufficiently large,
\[
\liminf_{N\to\infty}
Q\left(
\sqrt{N}\hat{T}_N(v_0+h)
>
\hat{c}_N(1-\alpha;v_0+h)
\quad
\forall\, h\geq \frac{h_{\epsilon}}{\sqrt{N}}
\right)
\geq 1-\epsilon.
\]
We thus have
\[
\begin{aligned}
&Q\left(
\bigcap_{h\geq0}
\left\{
\sqrt{N}\hat{T}_N(v_0+h)
>
\hat{c}_N(1-\alpha;v_0+h)
\right\}
\right)
\\
&=
Q\left(
\bigcap_{h\geq h_{\epsilon}/\sqrt{N}}
\left\{
\sqrt{N}\hat{T}_N(v_0+h)
>
\hat{c}_N(1-\alpha;v_0+h)
\right\}
\right.
\\
&\qquad\left.
\cap
\bigcap_{0\leq h<h_{\epsilon}/\sqrt{N}}
\left\{
\sqrt{N}\hat{T}_N(v_0+h)
>
\hat{c}_N(1-\alpha;v_0+h)
\right\}
\right)
\\
&\geq
Q\left(
\bigcap_{0\leq h<h_{\epsilon}/\sqrt{N}}
\left\{
\sqrt{N}\hat{T}_N(v_0+h)
>
\hat{c}_N(1-\alpha;v_0+h)
\right\}
\right)
\\
&\quad-
Q\left(
\bigcup_{h\geq h_{\epsilon}/\sqrt{N}}
\left\{
\sqrt{N}\hat{T}_N(v_0+h)
\leq
\hat{c}_N(1-\alpha;v_0+h)
\right\}
\right).
\end{aligned}
\]
which yields 
\begin{equation}
\begin{aligned}
\label{pthmregeqn4}
&\liminf_{N\to\infty}
Q\left(
\bigcap_{h\geq 0}
\left\{
\sqrt{N}\hat{T}_N(v_0+h)
>
\hat{c}_N(1-\alpha;v_0+h)
\right\}
\right)
\\
&\qquad\geq
\liminf_{N\to\infty}
Q\left(
\bigcap_{0\leq h<h_{\epsilon}/\sqrt{N}}
\left\{
\sqrt{N}\hat{T}_N(v_0+h)
>
\hat{c}_N(1-\alpha;v_0+h)
\right\}
\right)
-\epsilon.
\end{aligned}
\end{equation}
We focus on the first term on the right-hand side of \eqref{pthmregeqn4}. First, note that since $\hat{T}_N(v)$ is nonnegative and nondecreasing in $v$, and given the definition of $\hat{c}_N(1-\alpha;v)$, we have
\[
\left\{
\sqrt{N}\hat{T}_N(v_0)
\geq
\tilde{c}_N\left(
1-\alpha;
v_0+\frac{h_{\epsilon}}{\sqrt{N}}
\right)
\right\}
\subseteq
\bigcap_{0\leq h<h_{\epsilon}/\sqrt{N}}
\left\{
\sqrt{N}\hat{T}_N(v_0+h)
>
\hat{c}_N(1-\alpha;v_0+h)
\right\},
\]
where $\tilde{c}_N(1-\alpha;v)$ denotes the $(1-\alpha)$-quantile of
\[
\sup_{\lambda\in\tilde{\Delta}(v)}
\left\langle
\zeta_b^*-\zeta_A^*\hat{x},
\lambda
\right\rangle,
\]
and
\[
\tilde{\Delta}(v):=\left\{\lambda\geq0:\exists\,w,\ 
\begin{array}{l}
-\dfrac{\kappa_N}{\sqrt N}\mathbbm{1}\leq\hat A^\top\lambda-w\leq\dfrac{\kappa_N}{\sqrt N}\mathbbm{1},\quad
\mathbbm{1}^\top\hat\Omega\lambda\leq1,\\[1mm]
\hat b(v)^\top\lambda-\hat e_N\lVert w\rVert_1\geq-\dfrac{\kappa_N}{\sqrt N}
\end{array}
\right\}.
\]
Note that we have used the fact that $\hat{\Delta}(v)\subseteq\tilde{\Delta}(v)$, which implies
\[
\sup_{\lambda\in\hat{\Delta}(v)}
\left\langle
\zeta_b^*-\zeta_A^*\hat{x},
\lambda
\right\rangle
\leq
\sup_{\lambda\in\tilde{\Delta}(v)}
\left\langle
\zeta_b^*-\zeta_A^*\hat{x},
\lambda
\right\rangle.
\]
Consequently, the $(1-\alpha)$-quantile of the random variable on the left-hand side, denoted by $\hat{c}_N(1-\alpha;v)$, is no greater than the corresponding quantile of the random variable on the right-hand side, denoted by $\tilde{c}_N(1-\alpha;v)$.

As $\epsilon>0$ is arbitrary, the result follows if we establish 
\begin{equation}
\label{pthmregeqn5}
\liminf_{N\to\infty}
Q\left(
\left\{
\sqrt{N}\hat{T}_N(v_0)
>
\tilde{c}_N\left(
1-\alpha;
v_0+\frac{h_{\epsilon}}{\sqrt{N}}
\right)
\right\}
\right)\geq \alpha.
\end{equation}

\underline{Step 2} Let $E_N$, $\alpha_N$, $\kappa_N$ and $M_N$ be defined as in the paragraph that follows equation \eqref{thmAVeqn7}. In the remainder of the proof, we let $C>0$ denote a generic constant whose value can vary from one line to the next. Set $(\hat{\zeta}_b,\hat{\zeta}_A):=(\sqrt{N}(\hat{b}(v_0)-b(v_0)), \sqrt{N}(\hat{A}-A))$. The goal of this step is to derive inequality \eqref{pthmregeqn10} which plays a key role in proving \eqref{pthmregeqn5}.

Arguing as in the proof of equation \eqref{thmAVeqn36}, on $E_N$ and for all sufficiently large N, since $\hat{a}_N\geq \|x^*\|_{\infty}$ and thus $\phi(\mu,\hat{a}_N,v_0)=0$, we have

\begin{equation}
\label{pthmregeqn6}
\begin{aligned}
\sqrt{N}\hat{T}_N(v_0)
&=
\sqrt N\Bigl(
\phi(\hat{\mu},\hat{a}_N,v_0)
-
\phi(\mu,\hat{a}_N,v_0)
\Bigr)
\\
&\geq
\langle \hat{\zeta}_b-\hat{\zeta}_A\hat{\eta}, \lambda_0\rangle
-
\sqrt{N}\hat{T}_N(v_0)\,
\mathbbm{1}^\top(\hat{\Omega}-\Omega)\lambda_0,
\end{aligned}
\end{equation}
where $\hat{\eta}\in S_{0,N}(\hat{\mu},\hat{a}_N,v_0) $ is any solution of the dual LP associated with $\hat{T}_N$ given by
\begin{equation}
\label{pthmregeqn7}
S_{0,N}(\hat{\mu},\hat{a}_N,v_0):=\bigl\{\eta \mid  \hat{T}_N(v_0)\hat{\Omega} \mathbbm{1}+\hat{A}\eta \ge \hat{b}(v_0), \quad 
\|\eta\|_{\infty}\le \hat{a}_N\bigr\},
\end{equation}
and $\lambda_0\in \Delta_0(\hat{a}_N,v_0)$ is any optimal solution of the primal LP associated with $\phi(\mu,\hat{a}_N,v_0)$ given by
\begin{equation}
\label{pthmregeqn8}
\Delta_0(\hat{a}_N,v_0):=\bigl\{\lambda\geq 0 \mid  \mathbbm{1}^\top \Omega\lambda\leq 1,\quad  b(v_0)^\top \lambda-\hat{a}_N\|A^\top \lambda\|_1\geq 0\bigr\}.
\end{equation}
On $E_N$, we have 
\[
|\mathbbm{1}^\top(\hat{\Omega}-\Omega)\lambda_0|\leq \|\mathbbm{1}\|\|\hat{\Omega}-\Omega\|\|\mathcal{P}_U(\lambda_0)\|\leq C\alpha_N,
\]
which yields when combined with inequalities \eqref{thmAVeqn9} and \eqref{thmAVeqn18}
\begin{equation}
\label{pthmregeqn9}
|\sqrt{N}\hat{T}_N(v_0)\,
\mathbbm{1}^\top(\hat{\Omega}-\Omega)\lambda_0|\leq \gamma_{1N}:=CM_N\alpha_N,
\end{equation}
for all sufficiently large $N$.
Note that $\gamma_{1N}=o(1)$ by our choice of $M_N$ and $\alpha_N$. Moreover, as $\hat{a}_N\leq 5\|x^*\|_{\infty}+1$ on $E_N$ (for all sufficiently large N) and $\hat{T}_N(v_0)=o(1)$,  if $\{\eta_N\}_{N\geq 1}$ is any sequence such that $\eta_N\in S_{0,N}(\hat{\mu},\hat{a}_N,v_0)$, then $\{\eta_N\}_{N\geq 1}$ must have a further subsequence converging to a limit $\eta_0$ which must satisfy
\[
b(v_0)\leq A\eta_0.
\]
By assumption, the system $b(v_0)\leq A\eta$ has the unique solution given by $x^*$. We thus conclude that 
\begin{equation}
\label{pthmregeqn15}
\overset{\rightarrow}{d}_H\bigl(S_{0,N}(\hat{\mu},\hat{a}_N,v_0),\{x^*\}\bigr)=o(1)
\end{equation}
 on the event $E_N$, and using \eqref{thmAVeqn9}
\[
|\langle \hat{\zeta}_A(\hat{\eta}-x^*),\lambda_0\rangle|\leq \| \hat{\eta}-x^*\|\|\hat{\zeta}_A\|\|\mathcal{P}_U(\lambda_0)\|\leq \gamma_{2N}:=C M_N\overset{\rightarrow}{d}_H\bigl(S_{0,N}(\hat{\mu},\hat{a}_N,v_0),\{x^*\}\bigr).
\]
Assuming (w.l.o.g.) that $M_N\uparrow \infty$ is chosen to further satisfy 
\[
M_N \overset{\rightarrow}{d}_H\bigl(S_{0,N}(\hat{\mu},\hat{a}_N,v_0),\{x^*\}\bigr)=o(1),
\]
combining the latter inequalities with \eqref{pthmregeqn6} and \eqref{pthmregeqn9} then yields 
\begin{equation}
\label{pthmregeqn10}
\sqrt{N}\hat{T}_N(v_0)
\geq
\sup_{\lambda_0\in \Delta_0(\hat{a}_N,v_0)}\langle \hat{\zeta}_b-\hat{\zeta}_Ax^*,\lambda_0\rangle
-\gamma_{3N}
\end{equation}
where $\gamma_{3N}=o(1)$.

\underline{Step 3.} The goal of this step is to establish inequality \eqref{pthmregeqn12}, which shows that points in $\tilde{\Delta}(v_N)$ are close to points in the feasible region of the program on the right-hand side of \eqref{pthmregeqn10}. 

Define $v_N:=v_0+h_{\epsilon}/\sqrt{N}$. We now show that on $E_N$ and for all sufficiently large N
\begin{equation}
\label{pthmregeqn12}
\overset{\rightarrow}{d}_H\bigl( \tilde{\Delta}(v_N),\Delta_{0}(\hat{a}_N,v_0) \bigr)\leq C\kappa_N/\sqrt{N}.
\end{equation}
First note that for any $\lambda\geq 0$, since $b(v_0)\leq Ax^*$, we have
\[
b(v_0)^\top \lambda -\hat{a}_N\|A^\top \lambda\|_1\leq \lambda^\top A x^*-\hat{a}_N\|A^\top \lambda\|_1\leq (\|x^*\|_{\infty}-\hat{a}_N)\|A^\top \lambda\|_1.
\]
Since on $E_N$ and for all sufficiently large $N$ we have $\hat{a}_N-\|x^*\|_{\infty}> \|x^*\|$, the latter inequality implies that 
\[
\|A^\top \lambda\|_1=0 \quad \forall \lambda \in \Delta_0(\hat{a}_N,v_0), 
\]
which yields an alternative description of $\Delta_0(\hat{a}_N,v_0)$ free of the random box-constraint parameter $\hat{a}_N$:
\begin{equation}
\label{pthmregeqn13}
\Delta_0(\hat{a}_N,v_0)=\Delta_0(v_0):=\bigl\{\lambda \geq 0 \mid  \mathbbm{1}^\top \Omega\lambda \leq 1,\quad A^\top \lambda=0, \quad b(v_0)^\top \lambda\geq 0\bigr\}.
\end{equation}
In particular, for sufficiently large N and on $E_N$, the set $\Delta_0(\hat{a}_N,v_0)$ is deterministic and equal to $\Delta_0(v_0)$.

Similarly, \eqref{pthmregeqn2} and \eqref{lemAuxeqn2} imply that on $E_N$ and for all sufficiently large N, we have $\hat{T}_N(v_N)<\infty$, and the dual characterization of $\hat{T}_N$ implies that there exists $\hat{\eta}$ such that
\begin{equation}
\label{pthmregeqn16}
\hat{b}(v_N)\leq \hat{A}\hat{\eta}+\hat{T}_N(v_N)\hat{\Omega}\mathbbm{1}\quad \text{and}\quad \|\hat{\eta}\|_{\infty}\leq \hat{a_N}.
\end{equation}
Suppose for now that on $E_N$ and for all sufficiently large N we have
\begin{equation}
\label{pthmregeqn14}
\hat{T}_N(v_N)\leq CM_N/\sqrt{N}.
\end{equation}
Then arguing as in the derivation of \eqref{pthmregeqn15}, we conclude that $\|\hat{\eta}-x^*\|_{\infty}<\|x^*\|_{\infty}$ for all sufficiently large $N$, and \eqref{pthmregeqn16} and \eqref{pthmregeqn14} give for $\lambda\geq 0$ such that $\mathbbm{1}^\top \hat{\Omega}\lambda\leq 1$
\begin{equation}
\label{pthmregeqn17}
\begin{aligned}
\hat b(v_N)^\top\lambda-\hat e_N\lVert \hat{A}^\top \lambda\rVert_1&\leq \lambda^\top \hat{A}\hat{\eta}-\hat e_N\lVert \hat{A}^\top \lambda\rVert_1+\hat{T}_N(v_N)\lambda^\top\hat{\Omega}\mathbbm{1}\\
&\leq CM_N/\sqrt{N}+(\|\hat{\eta}\|_{\infty}-\hat{e}_N)\|\hat{A}^\top\lambda\|_1.
\end{aligned}
\end{equation}
For $\lambda\in \tilde{\Delta}(v_N)$, since $\|\hat{A}^\top \lambda-w\|_{\infty}\leq \kappa_N/\sqrt{N}$, we have on $E_N$
\[
\hat b(v_N)^\top\lambda-\hat e_N\lVert w\rVert_1\leq \hat b(v_N)^\top\lambda-\hat e_N\lVert\hat{A}^\top \lambda\rVert_1+\hat e_N \lVert w-\hat{A}^\top \lambda\rVert_1\leq \hat b(v_N)^\top\lambda-\hat e_N\lVert\hat{A}^\top \lambda\rVert_1+C\kappa_N/\sqrt{N}.
\]
Thus for $\lambda\in \tilde{\Delta}(v_N)$ the inequality 
\[
\hat b(v_N)^\top\lambda-\hat e_N\lVert w \rVert_1\geq -\kappa_N/\sqrt{N}
\]
when combined with the preceding inequality implies that
\begin{equation}
\label{pthmregeqn18}
\hat b(v_N)^\top\lambda-\hat e_N\lVert \hat{A}^\top \lambda \rVert_1\geq -C\kappa_N/\sqrt{N}
\end{equation}
for all $\lambda \in \tilde{\Delta}(v_N)$. As \eqref{pthmregeqn17} and \eqref{pthmregeqn18} must hold for all sufficiently large N on $E_N$ and for all $\lambda \in \tilde{\Delta}(v_N)$, we conclude that 
\[
\lVert \hat{A}^\top \lambda \rVert_{\infty}\leq C\kappa_N/\sqrt{N}
\]
and

\begin{equation}
\label{pthmregeqn19}
\tilde{\Delta}(v_N)\subseteq \Delta''(v_N):=\left\{\lambda\geq0:\|\hat{A}^\top \lambda\|_{\infty}\leq\frac{C\kappa_N}{\sqrt N},\ 
\mathbbm{1}^\top\hat\Omega\lambda\leq1,\ 
\hat b(v_N)^\top\lambda \geq-\frac{C\kappa_N}{\sqrt N}\right\}.
\end{equation}

To see that \eqref{pthmregeqn14} holds, we proceed as in the argument that follows \eqref{thmAVeqn36}. Note that
\begin{equation*}
\begin{aligned}
\sqrt{N}\hat{T}_N(v_N)&=\sqrt N\Bigl(
\phi_N(\hat{\mu},\hat{a}_N,v_N)
-
\phi(\mu,\hat{a}_N,v_N)
\Bigr)
+\sqrt{N}\phi(\mu,\hat{a}_N,v_N)\\
&\leq \langle \hat{\zeta}_b-\hat{\zeta}_A x^*,\hat{\lambda} \rangle +\sqrt{N}\phi(\mu,\hat{a}_N,v_N)
\end{aligned}
\end{equation*}
where $\hat{\lambda}$ is any element of $\Delta_{0,N}(\hat{\mu},\hat{a}_N,v_N)$ defined by
\[
 \Delta_{0,N}(\hat{\mu},\hat{a}_N,v_N):=\arg\max\{\hat{b}(v_N)^\top \lambda-\hat{a}_N \|\hat{A}^\top \lambda\|_1 \mid \lambda \geq 0,\ \mathbbm{1}^\top \hat{\Omega}\lambda\leq1 \}.
\]
On $E_N$, using \eqref{thmAVeqn9}, we have
\[
|\langle \hat{\zeta}_b-\hat{\zeta}_A x^*,\hat{\lambda} \rangle|\leq CM_N\|\mathcal{P}_U(\hat{\lambda})\|\leq CM_N.
\]
Also, \eqref{lemLBeqn4} implies that  on $E_N$ for all sufficiently large N we have 
\[
\sqrt{N}\phi(\mu,\hat{a}_N,v_N)\leq C.
\]
Combining the latter two inequalities yields \eqref{pthmregeqn14}.

We claim that for all $\lambda\in \Delta''(v_N)$ ias defined in \eqref{pthmregeqn19}, we have
\begin{equation}
\label{pthmregeqn20}
\hat b(v_0)^\top\lambda \geq-\frac{C\kappa_N}{\sqrt N}.
\end{equation}
Indeed, first note that 
\begin{equation}
\label{pthmregeqn22}
\hat{b}(v_N)=\hat b(v_0)+h_{\epsilon}/\sqrt{N}e^*=(1-\frac{h_{\epsilon}/h_0}{\sqrt{N}})\hat{b}(v_0)+\frac{h_{\epsilon}/h_0}{\sqrt{N}} \hat{b}(v_0+h_0).
\end{equation}
where (given \eqref{eqnIP1}) $e^*$ denotes the vector of the same dimension as $b(v)$ with all entries equal to zero except a one in the last coordinate. We first show that on $E_N$ and for all sufficiently large N, we have
\begin{equation}
\label{pthmregeqn21}
\sup_{\lambda \in \Delta''(v_N)} \langle \hat b(v_0+h_0), \lambda \rangle \leq C.
\end{equation}
Using the strong duality theorem of LP, we have
\begin{equation*}
\begin{aligned}
&\max\bigl\{\langle \hat b(v_0+h_0), \lambda \rangle\mid \lambda\geq0,\ \|\hat{A}^\top \lambda\|_{\infty}\leq\frac{C\kappa_N}{\sqrt N},\ 
\mathbbm{1}^\top\hat\Omega\lambda\leq1\bigr\}\\
&=\min\bigl\{ t+\frac{C\kappa_N}{\sqrt N}\|\eta\|_1\mid t\geq0,\ \hat{b}(v_0+h_0)\leq \hat{A}\eta+t\hat{\Omega}\mathbbm{1}\bigr\}.
\end{aligned}
\end{equation*}
As \eqref{pthmregeqn2} holds, there exists $\eta_0$ such that
\[
\|\eta_0\|_{\infty}\leq 5\|x^*\|_{\infty} \quad \text{and}\quad b(v_0+h_0)\leq A\eta_0+t_0\Omega\mathbbm{1}
\]
where $t_0:=\phi(\mu,5\|x^*\|_{\infty},v_0+h_0)$. On the event $E_N$ and for all sufficiently large N, we have
\begin{equation*}
\begin{aligned}
\hat{b}(v_0+h_0)-\hat{A}\eta_0&=\hat{b}(v_0+h_0)-b(v_0+h_0)-(\hat{A}-A)\eta_0+b(v_0+h_0)-A\eta_0\\
&\leq \bigl(C\frac{M_N}{\sqrt{N}}+t_0\bigr)\hat{\Omega}\mathbbm{1}.
\end{aligned}
\end{equation*}
It then follows from strong duality that 
\[
\sup_{\lambda \in \Delta''(v_N)} \langle \hat b(v_0+h_0), \lambda \rangle\leq C\frac{M_N}{\sqrt{N}}+t_0+\frac{C\kappa_N}{\sqrt N}\|\eta_0\|_1
\]
which yields \eqref{pthmregeqn21}. Equations \eqref{pthmregeqn22} and \eqref{pthmregeqn21} then imply that on $E_N$ and for all sufficiently large N, we have for all $\lambda\in \Delta''(v_N)$
\[
-C\kappa_N/\sqrt{N}\leq \langle\hat{b}(v_N),\lambda \rangle\leq (1-\frac{h_{\epsilon}/h_0}{\sqrt{N}}) \langle\hat{b}(v_0),\lambda \rangle +\frac{C}{\sqrt{N}}
\]
which yields \eqref{pthmregeqn20} as desired.

On the event $E_N$ and for all sufficiently large N, using inequality \eqref{thmAVeqn9}, it can easily be shown that \eqref{pthmregeqn20} implies that 
\[
b(v_0)^\top\lambda \geq-\frac{C\kappa_N}{\sqrt N} \quad \forall \lambda \in \tilde{\Delta}(v_N).
\]
Also, as we have shown that 
\[
\|\hat{A}^\top \lambda\|_{\infty}\leq C\kappa_N/\sqrt{N}\quad \forall \lambda \in \tilde{\Delta}(v_N),
\]
it can also be easily shown that
\[
\|A^\top \lambda\|_{\infty}\leq C\kappa_N/\sqrt{N}\quad \forall \lambda \in \tilde{\Delta}(v_N).
\]
Finally, as 
\[
\mathbbm{1}^\top \hat{\Omega}\lambda\leq 1 \quad \forall  \lambda \in \tilde{\Delta}(v_N)
\]
we have 
\[
(\mathbbm{1}^\top \Omega\lambda-1)_+\leq \kappa_N/\sqrt{N}  \quad \forall  \lambda \in \tilde{\Delta}(v_N).
\]
Recalling the expression for $\Delta_0(\hat{a}_N,v_N)$ in \eqref{pthmregeqn13}, \eqref{pthmregeqn12} then follows from Hoffman's theorem (see \cite{AH}).

\underline{Step 4.} We now complete the proof of \eqref{pthmregeqn5}. We first show that, on the event $E_N$ and for all sufficiently large $N$,
\begin{equation}
\label{pthmregeqn23}
\overset{\rightarrow}{d}_H
\bigl(
\mathcal{P}_U(\Delta_0(v_0)),\mathcal{P}_U(\hat{\Delta}(v_N))
\bigr)
\leq
C\alpha_N,
\end{equation}
where $\hat{\Delta}(v)$ is as defined in \eqref{eqnEAM5}.
Combining \eqref{pthmregeqn23} with \eqref{pthmregeqn12} and the inclusion
$\hat{\Delta}(v_N)\subseteq\tilde{\Delta}(v_N)$ then yields
\begin{equation}
\label{pthmregeqn24}
d_H
\bigl(
\mathcal{P}_U(\Delta_0(v_0)),\mathcal{P}_U(\hat{\Delta}(v_N))
\bigr)
\leq
C\max\{\frac{\kappa_N}{\sqrt{N}},\alpha_N\}.
\end{equation}

To establish \eqref{pthmregeqn23}, fix $\lambda\in\Delta_0(v_0)$. On the event $E_N$ and for all sufficiently large $N$, we have
\[
\lVert\hat{A}^{\top}\lambda\rVert_{\infty}
=
\lVert(\hat{A}-A)^{\top}\lambda\rVert_{\infty}
\leq
C\frac{M_N}{\sqrt{N}},
\]
and
\[
\mathbbm{1}^{\top}\hat{\Omega}\lambda
=
\mathbbm{1}^{\top}(\hat{\Omega}-\Omega)\lambda
+
\mathbbm{1}^{\top}\Omega\lambda
\leq
1+C\alpha_N.
\]
Moreover,
\[
\begin{aligned}
\langle\hat{b}(v_N),\lambda\rangle
&=
\langle\hat{b}(v_N)-b(v_N),\lambda\rangle
+
\langle b(v_N),\lambda\rangle
\\
&\geq
-C\frac{M_N}{\sqrt{N}}
+
\langle b(v_0),\lambda\rangle
\\
&\geq
-C\frac{M_N}{\sqrt{N}},
\end{aligned}
\]
where the first inequality uses the fact that $b(v)^\top\lambda$ is nondecreasing in $v$ for every $\lambda\geq0$.

Now define
\[
\Gamma_N(\lambda)
:=
\frac{\lambda}{1+C\alpha_N}.
\]
It follows that, on $E_N$ and for all sufficiently large $N$,
\[
\lVert\hat{A}^{\top}\Gamma_N(\lambda)-0\rVert_{\infty}
\leq
\frac{\kappa_N}{\sqrt{N}},
\qquad
\mathbbm{1}^{\top}\hat{\Omega}\Gamma_N(\lambda)
\leq1,
\]
and
\[
\langle\hat{b}(v_N),\Gamma_N(\lambda)\rangle
-\hat{e}_N\lVert 0\rVert_1
\geq
-\frac{\kappa_N}{\sqrt{N}}.
\]
Hence, taking $w=0$ in the definition of $\hat{\Delta}(v_N)$, we obtain
\[
\Gamma_N(\lambda)\in\hat{\Delta}(v_N).
\]

Finally, since
\[
\lVert \mathcal{P}_U(\lambda-\Gamma_N(\lambda))\rVert
=
\frac{C\alpha_N}{1+C\alpha_N}\lVert\mathcal{P}_U(\lambda)\rVert
\leq C\alpha_N,
\]
uniformly over $\lambda\in\Delta_0(v_0)$, it follows that
\[
\overset{\rightarrow}{d}_H
\bigl(
\mathcal{P}_U(\Delta_0(v_0)),\mathcal{P}_U(\hat{\Delta}(v_N))
\bigr)
\leq
C\alpha_N.
\]
The definition of $\alpha_N$ then yields \eqref{pthmregeqn23}.

Define the random function $\hat{g}$ by
\[
\hat{g}(x)
:=
\sup_{\lambda\in\mathcal{P}_U(\hat{\Delta}(v_N))}
\langle x,\lambda\rangle,
\]
and define the function $g$ by
\[
g(x)
:=
\sup_{\lambda\in\mathcal{P}_U(\Delta_0(v_0))}
\langle x,\lambda\rangle.
\]
As in \eqref{thmAVeqn22}, on the event $E_N$ and for all sufficiently large $N$, the functions $\hat{g}$ and $g$ are uniformly Lipschitz continuous. Moreover, \eqref{pthmregeqn24} implies that $\hat{g}$ converges uniformly to $g$ on every fixed compact set. It then follows from Assumption \ref{ass3} and the continuous mapping theorem that
\begin{equation}
\label{pthmregeqn25}
\sup_{\mathbb{W}\in E_N}
d_{\mathrm{Pr}}\left(
\mathcal{L}\left(
\left.
\sup_{\lambda\in\hat{\Delta}(v_N)}
\left\langle
\zeta_b^*-\zeta_A^*\hat{x},\lambda
\right\rangle
\,\right|\, \mathbb{W}
\right),
\mathcal{L}\left(
\sup_{\lambda\in\Delta_0(v_0)}
\left\langle
\zeta_b-\zeta_A x^*,\lambda
\right\rangle
\right)
\right)
\longrightarrow 0.
\end{equation}

Assumption \ref{ass4} implies that the distribution of
\[
\sup_{\lambda\in\Delta_0(v_0)}
\left\langle
\zeta_b-\zeta_A x^*,\lambda
\right\rangle
\]
is continuous and strictly increasing at its $(1-\alpha)$-quantile for every $\alpha\in(0,1/2)$. Let $c(1-\alpha;v_0)$ denote this quantile. It then follows from \eqref{pthmregeqn25} that
\begin{equation}
\label{pthmregeqn26}
\sup_{\mathbb{W}\in E_N}
\left|
\hat{c}_N(1-\alpha;v_N)
-
c(1-\alpha;v_0)
\right|
\longrightarrow 0.
\end{equation}

By a similar argument, using Assumption \ref{ass1}, the random variable on the right-hand side of \eqref{pthmregeqn10} converges in distribution to
\[
\sup_{\lambda\in\Delta_0(v_0)}
\left\langle
\zeta_b-\zeta_A x^*,\lambda
\right\rangle.
\]
Equivalently,
\begin{equation}
\label{pthmregeqn27}
d_{\mathrm{Pr}}\left(
\mathcal{L}\left(
\sup_{\lambda\in\Delta_0(v_0)}
\left\langle
\hat{\zeta}_b-\hat{\zeta}_A x^*,\lambda
\right\rangle
\right),
\mathcal{L}\left(
\sup_{\lambda\in\Delta_0(v_0)}
\left\langle
\zeta_b-\zeta_A x^*,\lambda
\right\rangle
\right)
\right)
\longrightarrow 0.
\end{equation}

Fix $\delta>0$ sufficiently small. It follows from \eqref{pthmregeqn10}, \eqref{pthmregeqn26}, and \eqref{pthmregeqn27} that
\begin{equation*}
\begin{aligned}
&\liminf_{N\to\infty}
Q\left(
\sqrt{N}\hat{T}_N(v_0)
>
\hat{c}_N(1-\alpha;v_N)
\right)
\\
&\quad=
\liminf_{N\to\infty}
Q\left(
\left\{
\sqrt{N}\hat{T}_N(v_0)
>
\hat{c}_N(1-\alpha;v_N)
\right\}
\cap E_N
\right)
\\
&\quad\geq
\liminf_{N\to\infty}
Q\left(
\left\{
\sup_{\lambda\in\Delta_0(v_0)}
\left\langle
\hat{\zeta}_b-\hat{\zeta}_A x^*,\lambda
\right\rangle
>
\hat{c}_N(1-\alpha;v_N)+\gamma_{3N}
\right\}
\cap E_N
\right)
\\
&\quad\geq
\liminf_{N\to\infty}
Q\left(
\left\{
\sup_{\lambda\in\Delta_0(v_0)}
\left\langle
\hat{\zeta}_b-\hat{\zeta}_A x^*,\lambda
\right\rangle
>
c(1-\alpha;v_0)+\delta
\right\}
\cap E_N
\right)
\\
&\quad\geq
Q\left(
\sup_{\lambda\in\Delta_0(v_0)}
\left\langle
\zeta_b-\zeta_A x^*,\lambda
\right\rangle
>
c(1-\alpha;v_0)+\delta
\right)
\\
&\quad=
\alpha-O(\delta).
\end{aligned}
\end{equation*}
Since $\delta>0$ is arbitrary, letting $\delta\downarrow0$ yields
\[
\liminf_{N\to\infty}
Q\left(
\sqrt{N}\hat{T}_N(v_0)
>
\hat{c}_N(1-\alpha;v_N)
\right)
\geq
\alpha,
\]
as desired. This concludes the proof of \eqref{pthmregeqn5}.

\underline{Step 5.} Suppose now that \eqref{pthmregeqn3} holds. By
\eqref{lemInfEqn1} and \eqref{lemAuxeqn1},
\[
\hat{T}_N(v_0+h)=+\infty
\qquad
\text{for all } h>0.
\]
Consequently,
\[
\begin{aligned}
&Q\left(
\bigcap_{h\geq0}
\left\{
\sqrt{N}\hat{T}_N(v_0+h)
>
\hat{c}_N(1-\alpha;v_0+h)
\right\}
\right)
\\
&\qquad=
Q\left(
\sqrt{N}\hat{T}_N(v_0)
>
\hat{c}_N(1-\alpha;v_0)
\right).
\end{aligned}
\]

It therefore remains to show that
\[
\liminf_{N\to\infty}
Q\left(
\sqrt{N}\hat{T}_N(v_0)
>
\hat{c}_N(1-\alpha;v_0)
\right)
\geq\alpha.
\]

The argument is analogous to that used in Steps 2--4. In particular,
\eqref{pthmregeqn10} continues to hold. Moreover, since
$\hat{\Delta}(v_0)\subseteq\tilde{\Delta}(v_0)$,
\[
\sup_{\lambda\in\hat{\Delta}(v_0)}
\left\langle
\zeta_b^*-\zeta_A^*\hat{x},
\lambda
\right\rangle
\leq
\sup_{\lambda\in\tilde{\Delta}(v_0)}
\left\langle
\zeta_b^*-\zeta_A^*\hat{x},
\lambda
\right\rangle,
\]
where $\tilde{\Delta}(v_0)$ is defined as in the discussion following
\eqref{pthmregeqn4}.

Repeating the arguments of Steps 3 and 4 with $v_N$ replaced by $v_0$
yields, on the event $E_N$,
\[
d_H\left(
\mathcal{P}_U(\Delta_0(v_0)),
\mathcal{P}_U(\hat{\Delta}(v_0))
\right)
=o(1).
\]
Indeed, the argument is simpler in this case because no comparison
between $b(v_N)$ and $b(v_0)$ is required.

It follows, exactly as in the derivation of \eqref{pthmregeqn25} and
\eqref{pthmregeqn26}, that the conditional distribution of
\[
\sup_{\lambda\in\hat{\Delta}(v_0)}
\left\langle
\zeta_b^*-\zeta_A^*\hat{x},
\lambda
\right\rangle
\]
converges uniformly on $E_N$ to the distribution of
\[
\sup_{\lambda\in\Delta_0(v_0)}
\left\langle
\zeta_b-\zeta_Ax^*,
\lambda
\right\rangle,
\]
and hence
\[
\sup_{\mathbb{W}\in E_N}
\left|
\hat{c}_N(1-\alpha;v_0)
-
c(1-\alpha;v_0)
\right|
\longrightarrow0.
\]
Combining this result with \eqref{pthmregeqn10} and arguing as in the
final part of Step 4 gives
\[
\liminf_{N\to\infty}
Q\left(
\sqrt{N}\hat{T}_N(v_0)
>
\hat{c}_N(1-\alpha;v_0)
\right)
\geq\alpha.
\]
This establishes \eqref{pthmregeqn1} under \eqref{pthmregeqn3} and
completes the proof of Case 1.

\vspace{10mm}

\underline{\textbf{Case 2:}} We now consider the case in which the primal and dual LPs
have unique nonzero solutions, denoted by $x^*$ and $y^*$, respectively,
and set
\[
z^*:=(x^*,y^*).
\]
We aim to show that
\begin{equation}
\label{pthmregcase2eqn1}
\liminf_{N\to\infty}
Q\left(v_0\notin CI_{1-\alpha}\right)
\geq\alpha,
\end{equation}
where $CI_{1-\alpha}$ denotes the two-sided confidence interval for $v_0$.

By the definition of $CI_{1-\alpha}$, we have
\[
\left\{v_0\notin CI_{1-\alpha}\right\}
=
\mathcal R_N^+\cup\mathcal R_N^-,
\]
where
\[
\mathcal R_N^+
:=
\bigcap_{v\geq v_0}
\left\{
\sqrt N\hat T_N(v)>
\hat c_N(1-\alpha;v)
\right\},
\]
and
\[
\mathcal R_N^-
:=
\bigcap_{v\leq v_0}
\left\{
\sqrt N\hat T_N(v)>
\hat c_N(1-\alpha;v)
\right\}.
\]

As in Case 1, we establish the result by considering two exhaustive
cases. First, suppose that there exists $h_0>0$ such that
\begin{equation}
\label{pthmregeqn28}
\max\left\{
\phi\left(\mu,2\lVert z^*\rVert_\infty,v_0+h_0\right),
\phi\left(\mu,2\lVert z^*\rVert_\infty,v_0-h_0\right)
\right\}
<\infty,
\end{equation}
where $\phi$ is defined as in \eqref{thmAVeqn4}, with $A$ and $b(v)$
given by \eqref{eqnIP3}. We then consider the complementary case in which
\begin{equation}
\label{pthmregeqn29}
\max\left\{
\phi\left(\mu,2\lVert z^*\rVert_\infty,v_0+h\right),
\phi\left(\mu,2\lVert z^*\rVert_\infty,v_0-h\right)
\right\}
=\infty
\qquad
\forall\,h>0.
\end{equation}

\underline{Step 1.} Suppose that \eqref{pthmregeqn28} holds. Fix
$\epsilon>0$. By \eqref{eqnthm4}, there exists $h_\epsilon>0$ such that
\begin{equation}
\label{pthmregcase2eqn2}
\begin{aligned}
\liminf_{N\to\infty}
Q\left(v_0\notin CI_{1-\alpha}\right)
\geq&
\liminf_{N\to\infty}
Q\left(
\mathcal R_{N,\epsilon}^+
\cup
\mathcal R_{N,\epsilon}^-
\right)
-\epsilon,
\end{aligned}
\end{equation}
where
\[
\mathcal R_{N,\epsilon}^+
:=
\bigcap_{0\leq h<h_\epsilon/\sqrt N}
\left\{
\sqrt N\hat T_N(v_0+h)
>
\hat c_N(1-\alpha;v_0+h)
\right\},
\]
and
\[
\mathcal R_{N,\epsilon}^-
:=
\bigcap_{0\leq h<h_\epsilon/\sqrt N}
\left\{
\sqrt N\hat T_N(v_0-h)
>
\hat c_N(1-\alpha;v_0-h)
\right\}.
\]

Using the notation introduced in the paragraph containing
\eqref{plemLBeqn1}, write
\[
\hat T_N(v)
=
\max\left\{
\phi_U(\hat\mu,\hat a_N,v),
\phi_L(\hat\mu,\hat a_N,v)
\right\}.
\]
Since $\phi_U(\hat\mu,\hat a_N,v)$ is nondecreasing in $v$ and
$\phi_L(\hat\mu,\hat a_N,v)$ is nonincreasing in $v$, it follows that
\begin{equation}
\label{pthmregcase2eqn3}
\hat T_N(v_0+h)
\geq
\phi_U(\hat\mu,\hat a_N,v_0)
\qquad
\forall\,h\geq0,
\end{equation}
and
\begin{equation}
\label{pthmregcase2eqn4}
\hat T_N(v_0-h)
\geq
\phi_L(\hat\mu,\hat a_N,v_0)
\qquad
\forall\,h\geq0.
\end{equation}

We next derive a common asymptotic approximation for the two-sided
critical values over the shrinking neighborhood
\[
\left[
v_0-\frac{h_\epsilon}{\sqrt N},
v_0+\frac{h_\epsilon}{\sqrt N}
\right].
\]

\underline{Step 2.} Define the population multiplier sets
\[
\Delta_{0,U}(v_0)
:=
\left\{
\lambda_U\geq0:
A_U^\top\lambda_U=0,\ 
\mathbbm 1^\top\Omega_U\lambda_U\leq1,\ 
\langle b_U(v_0),\lambda_U\rangle\geq0
\right\},
\]
and
\[
\Delta_{0,L}(v_0)
:=
\left\{
\lambda_L\geq0:
A_L^\top\lambda_L=0,\ 
\mathbbm 1^\top\Omega_L\lambda_L\leq1,\ 
\langle b_L(v_0),\lambda_L\rangle\geq0
\right\}.
\]
Since $b_U(v_0)\leq A_Ux^*$ and $b_L(v_0)\leq A_Ly^*$, the last
inequalities in the preceding displays necessarily hold with equality.

Define
\[
\hat Z_{N,U}
:=
\sup_{\lambda_U\in\Delta_{0,U}(v_0)}
\left\langle
\hat\zeta_{b,U}-\hat\zeta_{A,U}x^*,
\lambda_U
\right\rangle,
\]
and
\[
\hat Z_{N,L}
:=
\sup_{\lambda_L\in\Delta_{0,L}(v_0)}
\left\langle
\hat\zeta_{b,L}-\hat\zeta_{A,L}y^*,
\lambda_L
\right\rangle.
\]
Arguing as in the derivation of \eqref{pthmregeqn10}, on the event $E_N$
and for all sufficiently large $N$, we have
\begin{equation}
\label{pthmregcase2eqn5}
\sqrt N\phi_U(\hat\mu,\hat a_N,v_0)
\geq
\hat Z_{N,U}
-o(1),
\end{equation}
and
\begin{equation}
\label{pthmregcase2eqn6}
\sqrt N\phi_L(\hat\mu,\hat a_N,v_0)
\geq
\hat Z_{N,L}
-o(1).
\end{equation}
Indeed, the argument is identical to that in Case 1, since $x^*$ and
$y^*$ are the unique solutions of the upper and lower systems,
respectively, at $v_0$.

Now define
\[
\Delta_0(v_0)
:=
\left\{
\lambda=(\lambda_L,\lambda_U)\geq0:
\begin{array}{l}
A^\top\lambda=0,\qquad
\mathbbm 1^\top\Omega\lambda\leq1,\\[1mm]
\langle b(v_0),\lambda\rangle\geq0
\end{array}
\right\}.
\]
As above, the last inequality necessarily holds with equality.

Using the block structure of the two-sided system and arguing as in the
derivation of \eqref{plemLBeqn1}, we obtain
\begin{equation}
\label{pthmregcase2eqn7}
\begin{aligned}
&\max\left\{
\sup_{\lambda_U\in\Delta_{0,U}(v_0)}
\left\langle
\hat\zeta_{b,U}-\hat\zeta_{A,U}x^*,
\lambda_U
\right\rangle,
\right.
\\[-1mm]
&\hspace{25mm}\left.
\sup_{\lambda_L\in\Delta_{0,L}(v_0)}
\left\langle
\hat\zeta_{b,L}-\hat\zeta_{A,L}y^*,
\lambda_L
\right\rangle
\right\}
\\
&\qquad=
\sup_{\lambda\in\Delta_0(v_0)}
\left\langle
\hat\zeta_b-\hat\zeta_Az^*,
\lambda
\right\rangle.
\end{aligned}
\end{equation}
The equality follows from positive homogeneity and the common
normalization $\mathbbm 1^\top\Omega\lambda\leq1$.

Combining \eqref{pthmregcase2eqn5}--\eqref{pthmregcase2eqn7} yields
\begin{equation}
\label{pthmregcase2eqn8}
\sqrt N\hat T_N(v_0)
\geq
\sup_{\lambda\in\Delta_0(v_0)}
\left\langle
\hat\zeta_b-\hat\zeta_Az^*,
\lambda
\right\rangle
-o(1).
\end{equation}

Let
\[
Z
:=
\sup_{\lambda\in\Delta_0(v_0)}
\left\langle
\zeta_b-\zeta_Az^*,
\lambda
\right\rangle,
\]
and let $c(1-\alpha;v_0)$ denote its $(1-\alpha)$-quantile.
Assumption \ref{ass1} and the continuous mapping theorem imply that
\begin{equation}
\label{pthmregcase2eqn9}
\sup_{\lambda\in\Delta_0(v_0)}
\left\langle
\hat\zeta_b-\hat\zeta_Az^*,
\lambda
\right\rangle
\overset{d}{\longrightarrow}
Z.
\end{equation}

\underline{Step 3.} We now show that the two-sided critical values are
uniformly asymptotically equivalent over the shrinking neighborhood of
$v_0$. Specifically, we claim that
\begin{equation}
\label{pthmregcase2eqn10}
\sup_{|h|\leq h_\epsilon/\sqrt N}
\left|
\hat c_N(1-\alpha;v_0+h)
-
c(1-\alpha;v_0)
\right|
\longrightarrow0
\end{equation}
on $E_N$.

To establish \eqref{pthmregcase2eqn10}, it suffices to show that
\begin{equation}
\label{pthmregcase2eqn11}
\sup_{|h|\leq h_\epsilon/\sqrt N}
d_H\left(
\mathcal P_U(\hat\Delta(v_0+h)),
\mathcal P_U(\Delta_0(v_0))
\right)
=o(1)
\end{equation}
on $E_N$.

The argument parallels Steps 3 and 4 of Case 1. Since
\eqref{pthmregeqn28} holds, there exist $\eta_+$ and $\eta_-$ and finite
constants $t_+$ and $t_-$ such that
\[
b(v_0+h_0)
\leq
A\eta_+
+t_+\Omega\mathbbm 1,
\qquad
\lVert\eta_+\rVert_\infty
\leq
2\lVert z^*\rVert_\infty,
\]
and
\[
b(v_0-h_0)
\leq
A\eta_-
+t_-\Omega\mathbbm 1,
\qquad
\lVert\eta_-\rVert_\infty
\leq
2\lVert z^*\rVert_\infty.
\]
Arguing as in the derivation of \eqref{pthmregeqn21}, these inequalities
imply that, uniformly over the approximate multiplier sets considered
below,
\begin{equation}
\label{pthmregcase2eqn12}
\left|
\langle e^*,\lambda\rangle
\right|
\leq C,
\end{equation}
where $e^*$ is defined by
\[
b(v)=b(v_0)+(v-v_0)e^*.
\]
Indeed,
\[
2h_0\langle e^*,\lambda\rangle
=
\langle b(v_0+h_0),\lambda\rangle
-
\langle b(v_0-h_0),\lambda\rangle,
\]
and each term on the right-hand side is uniformly bounded by the same
strong-duality argument used in \eqref{pthmregeqn21}.

Consequently, uniformly over
$|h|\leq h_\epsilon/\sqrt N$,
\begin{equation}
\label{pthmregcase2eqn13}
\left|
\langle b(v_0+h)-b(v_0),\lambda\rangle
\right|
\leq
C\frac{h_\epsilon}{\sqrt N}.
\end{equation}

Now let $\lambda\in\hat\Delta(v_0+h)$ for some
$|h|\leq h_\epsilon/\sqrt N$. Proceeding as in
\eqref{pthmregeqn17}--\eqref{pthmregeqn19}, we obtain, uniformly in $h$,
\[
\lVert\hat A^\top\lambda\rVert_\infty
\leq
C\frac{\kappa_N}{\sqrt N}.
\]
Using \eqref{pthmregcase2eqn13}, the analogue of
\eqref{pthmregeqn20} gives
\[
\langle b(v_0),\lambda\rangle
\geq
-C\frac{\kappa_N}{\sqrt N}.
\]
Moreover,
\[
\left(
\mathbbm 1^\top\Omega\lambda-1
\right)_+
\leq
C\frac{\kappa_N}{\sqrt N}.
\]
It therefore follows from Hoffman's theorem that
\begin{equation}
\label{pthmregcase2eqn14}
\sup_{|h|\leq h_\epsilon/\sqrt N}
\overset{\rightarrow}{d}_H
\left(
\mathcal P_U(\hat\Delta(v_0+h)),
\mathcal P_U(\Delta_0(v_0))
\right)
=o(1).
\end{equation}

For the reverse directed distance, fix
$\lambda\in\Delta_0(v_0)$ and define, as in Step 4 of Case 1,
\[
\Gamma_N(\lambda)
:=
\frac{\lambda}{1+C\alpha_N}.
\]
Using \eqref{pthmregcase2eqn13} and arguing as in the derivation of
\eqref{pthmregeqn23}, we obtain, uniformly over
$|h|\leq h_\epsilon/\sqrt N$,
\[
\Gamma_N(\lambda)\in\hat\Delta(v_0+h)
\]
for all sufficiently large $N$. Hence
\begin{equation}
\label{pthmregcase2eqn15}
\sup_{|h|\leq h_\epsilon/\sqrt N}
\overset{\rightarrow}{d}_H
\left(
\mathcal P_U(\Delta_0(v_0)),
\mathcal P_U(\hat\Delta(v_0+h))
\right)
=o(1).
\end{equation}
Equations \eqref{pthmregcase2eqn14} and
\eqref{pthmregcase2eqn15} establish
\eqref{pthmregcase2eqn11}.

Define
\[
\hat g_{N,h}(x)
:=
\sup_{\lambda\in\mathcal P_U(\hat\Delta(v_0+h))}
\langle x,\lambda\rangle,
\]
and
\[
g(x)
:=
\sup_{\lambda\in\mathcal P_U(\Delta_0(v_0))}
\langle x,\lambda\rangle.
\]
As in the derivation of \eqref{pthmregeqn25},
\eqref{pthmregcase2eqn11} implies that
$\hat g_{N,h}$ converges uniformly to $g$ on compact sets, uniformly over
$|h|\leq h_\epsilon/\sqrt N$. Assumption \ref{ass3} and the continuous
mapping theorem therefore imply that
\begin{equation}
\label{pthmregcase2eqn16}
\sup_{\substack{\mathbb W\in E_N\\
|h|\leq h_\epsilon/\sqrt N}}
d_{\mathrm{Pr}}\left(
\mathcal L\left(
\left.
\sup_{\lambda\in\hat\Delta(v_0+h)}
\left\langle
\zeta_b^*-\zeta_A^*\hat z,\lambda
\right\rangle
\,\right|\,\mathbb W
\right),
\mathcal L(Z)
\right)
\longrightarrow0.
\end{equation}
Assumption \ref{ass4} then implies
\eqref{pthmregcase2eqn10}.

\underline{Step 4.} We now complete the proof under
\eqref{pthmregeqn28}. Fix $\delta>0$. By
\eqref{pthmregcase2eqn10}, on $E_N$ and for all sufficiently large $N$,
\begin{equation}
\label{pthmregcase2eqn17}
\sup_{|h|\leq h_\epsilon/\sqrt N}
\hat c_N(1-\alpha;v_0+h)
\leq
c(1-\alpha;v_0)+\delta.
\end{equation}

Suppose first that
\[
\hat Z_{N,U}
>
c(1-\alpha;v_0)+2\delta.
\]
By \eqref{pthmregcase2eqn5}, on $E_N$ and for all sufficiently large $N$,
\[
\sqrt N\phi_U(\hat\mu,\hat a_N,v_0)
>
c(1-\alpha;v_0)+\delta.
\]
Since $\phi_U(\hat\mu,\hat a_N,v)$ is nondecreasing in $v$, we have
\[
\sqrt N\hat T_N(v_0+h)
\geq
\sqrt N\phi_U(\hat\mu,\hat a_N,v_0+h)
>
c(1-\alpha;v_0)+\delta
\]
for every $0\leq h<h_\epsilon/\sqrt N$. Combining this result with
\eqref{pthmregcase2eqn17} gives
\[
\left\{
\hat Z_{N,U}>
c(1-\alpha;v_0)+2\delta
\right\}
\cap E_N
\subseteq
\mathcal R_{N,\epsilon}^+
\]
for all sufficiently large $N$.

Similarly, using the fact that
$\phi_L(\hat\mu,\hat a_N,v)$ is nonincreasing in $v$,
\[
\left\{
\hat Z_{N,L}>
c(1-\alpha;v_0)+2\delta
\right\}
\cap E_N
\subseteq
\mathcal R_{N,\epsilon}^-
\]
for all sufficiently large $N$. Consequently,
\begin{equation}
\label{pthmregcase2eqn18}
\begin{aligned}
&
\left\{
\max\{\hat Z_{N,U},\hat Z_{N,L}\}
>
c(1-\alpha;v_0)+2\delta
\right\}
\cap E_N
\\
&\qquad\subseteq
\mathcal R_{N,\epsilon}^+
\cup
\mathcal R_{N,\epsilon}^-.
\end{aligned}
\end{equation}

It follows from \eqref{pthmregcase2eqn2} and
\eqref{pthmregcase2eqn18} that
\[
\begin{aligned}
\liminf_{N\to\infty}
Q\left(v_0\notin CI_{1-\alpha}\right)
&\geq
\liminf_{N\to\infty}
Q\left(
\max\{\hat Z_{N,U},\hat Z_{N,L}\}
>
c(1-\alpha;v_0)+2\delta
\right)
-\epsilon
\\
&\geq
Q\left(
Z>
c(1-\alpha;v_0)+2\delta
\right)
-\epsilon,
\end{aligned}
\]
where the last inequality follows from
\eqref{pthmregcase2eqn9} and the continuous mapping theorem.

Letting $\delta\downarrow0$ and using Assumption \ref{ass4}, we obtain
\[
\liminf_{N\to\infty}
Q\left(v_0\notin CI_{1-\alpha}\right)
\geq
\alpha-\epsilon.
\]
Since $\epsilon>0$ is arbitrary,
\[
\liminf_{N\to\infty}
Q\left(v_0\notin CI_{1-\alpha}\right)
\geq
\alpha.
\]
This establishes \eqref{pthmregcase2eqn1} under
\eqref{pthmregeqn28}.

\underline{Step 5.} Suppose now that \eqref{pthmregeqn29} holds. Define
\[
f_U(h)
:=
\phi_U\left(
\mu,2\lVert z^*\rVert_\infty,v_0+h
\right),
\qquad
h>0,
\]
and
\[
f_L(h)
:=
\phi_L\left(
\mu,2\lVert z^*\rVert_\infty,v_0-h
\right),
\qquad
h>0.
\]
By monotonicity, both $f_U$ and $f_L$ are nondecreasing functions of
$h$. Moreover, \eqref{pthmregeqn29} implies that
\[
\max\{f_U(h),f_L(h)\}
=\infty
\qquad
\forall\,h>0.
\]
It follows that at least one of the two functions is identically infinite
on $(0,\infty)$. 

Suppose without loss of generality that
\[
f_U(h)=\infty
\qquad
\forall\,h>0.
\]
By \eqref{lemInfEqn1} and \eqref{lemAuxeqn1}, on $E_N$ and for all
sufficiently large $N$,
\[
\hat T_N(v_0+h)=\infty
\qquad
\forall\,h>0.
\]
Hence
\[
\mathcal R_N^+
=
\left\{
\sqrt N\hat T_N(v_0)
>
\hat c_N(1-\alpha;v_0)
\right\}.
\]

At $v_0$, the system $b(v_0)\leq A\eta$, with $A$ and $b(v_0)$ given by
\eqref{eqnIP3}, has the unique solution $z^*$. Arguing as in Steps 2--4
above, but with $v$ fixed at $v_0$, gives
\[
\sqrt N\hat T_N(v_0)
\geq
\sup_{\lambda\in\Delta_0(v_0)}
\left\langle
\hat\zeta_b-\hat\zeta_Az^*,
\lambda
\right\rangle
-o(1),
\]
and
\[
\hat c_N(1-\alpha;v_0)
\longrightarrow
c(1-\alpha;v_0)
\]
on $E_N$. Therefore,
\[
\liminf_{N\to\infty}
Q\left(
\sqrt N\hat T_N(v_0)
>
\hat c_N(1-\alpha;v_0)
\right)
\geq
\alpha.
\]
It follows that
\[
\liminf_{N\to\infty}
Q\left(v_0\notin CI_{1-\alpha}\right)
\geq
\alpha.
\]

The argument is identical if
\[
f_L(h)=\infty
\qquad
\forall\,h>0.
\]
Thus \eqref{pthmregcase2eqn1} also holds under
\eqref{pthmregeqn29}.

Combining \eqref{pthmregcase2eqn1} with \eqref{eqnthm1}, which provides
the corresponding upper bound on the probability of exclusion, yields
\[
\lim_{N\to\infty}
Q\left(
v_0\in CI_{1-\alpha}
\right)
=
1-\alpha.
\]
This completes the proof of Case 2 and hence the proof of the theorem.
\end{proof}


\subsection{Proof of Theorem \ref{thm3}}
\label{Secthm3}

\begin{proof}
We show that the argument used to establish the power result in
Theorem \ref{thm1} continues to apply when the box constraints are
dropped, i.e., when $\hat{a}_N=\hat{e}_N=\infty$. We consider only the
first part of Theorem \ref{thm3}, as the arguments for the remaining
parts are analogous.

When $\hat{a}_N=\hat{e}_N=\infty$, the test statistic simplifies to
\[
\hat{T}_N(v)
=
\max\left\{
\hat{b}(v)^\top\lambda
\ \middle|\
\lambda\geq0,\ 
\hat{A}^\top\lambda=0,\ 
\mathbbm{1}^\top\hat{\Omega}\lambda\leq1
\right\},
\]
and the set $\hat{\Delta}(v)$ in \eqref{eqnEAM4} becomes
\[
\hat{\Delta}(v)
=
\left\{
\lambda\geq0
\ \middle|\
\lVert\hat{A}^\top\lambda\rVert_\infty
\leq\frac{\kappa_N}{\sqrt{N}},\
\mathbbm{1}^\top\hat{\Omega}\lambda\leq1,\
\langle\hat{b}(v),\lambda\rangle
\geq
\hat{T}_N(v)-\frac{\kappa_N}{\sqrt{N}}
\right\}.
\]

The argument used in the proof of Theorem \ref{thm1} continues to apply
with $\hat{a}_N=\hat{e}_N=\infty$, except for the argument leading to
\eqref{thmAVeqnBC}. There, the box constraints were used to obtain the
bound
\[
\lVert\eta_{\hat{\xi}}\rVert_\infty\leq\hat{a}_N.
\]
We now show that the assumption that the set of optimal solutions to
\eqref{eqnLP} is bounded provides the required bound on
$\eta_{\hat{\xi}}$.

Recall that $\eta_{\hat{\xi}}$ is any solution to
\[
\phi_N(\mu(\hat{\xi}),\infty,v_t)
\hat{\Omega}\mathbbm{1}
+
\hat{A}\eta_{\hat{\xi}}
\geq
\hat{b}(v_t).
\]
By the argument following \eqref{thmAVeqn36}, uniformly on $E_N$ and for
all sufficiently large $N$,
\[
\phi_N(\mu(\hat{\xi}),\infty,v_t)
\leq
C\frac{M_N}{\sqrt{N}}
\]
for some constant $C>0$. Moreover, by the definition of $E_N$,
uniformly on $E_N$,
\[
\lVert\hat{A}-A\rVert
\leq
\frac{M_N}{\sqrt{N}},
\qquad
\lVert\hat{b}(v_t)-b(v_t)\rVert
\leq
\frac{M_N}{\sqrt{N}}.
\]
By the definition of $v_t$, we also have
\[
\lVert b(v_t)-b(v_0)\rVert
\leq
C\frac{M_N}{\sqrt{N}}
\]
for some constant $C>0$. Hence, uniformly on $E_N$,
\[
\hat{A}\to A,
\qquad
\hat{b}(v_t)\to b(v_0),
\qquad
\phi_N(\mu(\hat{\xi}),\infty,v_t)\to0.
\]

We claim that the set of possible values of $\eta_{\hat{\xi}}$ is
uniformly bounded on $E_N$ for all sufficiently large $N$. Suppose, to
the contrary, that this is not the case. Then there exist a sequence of
sample realizations in $E_N$ and corresponding solutions
$\hat{\eta}_N$ such that
\[
\lVert\hat{\eta}_N\rVert\to\infty.
\]
Dividing
\[
\phi_N(\mu(\hat{\xi}),\infty,v_t)
\hat{\Omega}\mathbbm{1}
+
\hat{A}\hat{\eta}_N
\geq
\hat{b}(v_t)
\]
by $\lVert\hat{\eta}_N\rVert$ and passing to a convergent subsequence of
$\hat{\eta}_N/\lVert\hat{\eta}_N\rVert$, we obtain a vector
$\eta_0$ satisfying
\[
\lVert\eta_0\rVert=1
\qquad\text{and}\qquad
A\eta_0\geq0.
\]
In particular, $\eta_0\neq0$.

By the definition of $A$ in \eqref{eqnIP1}, the latter inequality implies
\[
M\eta_0\leq0
\qquad\text{and}\qquad
\rho^\top\eta_0\geq0.
\]
Let $x_0$ be any optimal solution to \eqref{eqnLP}. Then, for every
$t\geq0$,
\[
M(x_0+t\eta_0)
=
Mx_0+tM\eta_0
\leq
\beta,
\]
so $x_0+t\eta_0$ is feasible. Moreover,
\[
\rho^\top(x_0+t\eta_0)
=
\rho^\top x_0+t\rho^\top\eta_0
\geq
\rho^\top x_0.
\]
Since $x_0$ is optimal and the optimal value of \eqref{eqnLP} is finite,
we cannot have $\rho^\top\eta_0>0$. Hence
\[
\rho^\top\eta_0=0.
\]
It follows that
\[
\rho^\top(x_0+t\eta_0)=\rho^\top x_0
\qquad
\forall\,t\geq0.
\]
Thus $x_0+t\eta_0$ is an optimal solution to \eqref{eqnLP} for every
$t\geq0$. Since $\eta_0\neq0$, this contradicts the assumption that the
set of optimal solutions to \eqref{eqnLP} is bounded.

We therefore conclude that $\eta_{\hat{\xi}}$ is uniformly bounded on
$E_N$ for all sufficiently large $N$. This provides the bound that was
obtained from the box constraints in \eqref{thmAVeqnBC}. The remainder
of the proof is therefore identical to the argument following
\eqref{thmAVeqnBC}, and establishes \eqref{eqnthm2} when the box
constraints are dropped.

The proofs of the second and third parts of Theorem \ref{thm3} are
analogous.
\end{proof}

\subsection{Proof of Theorem~\ref{thm4}}
\label{Secthm4}

\begin{proof}

The proof closely parallels that of Theorem~\ref{thm1}, and is similar to the proof of Theorem 4.19 in \cite{GM25}. We provide the argument for the first part of the theorem;
the proofs of the remaining two parts are analogous. The main difference from Theorem~\ref{thm1} (besides the uniformity) is that the deterministic
restrictions directly constrain the primal variables to lie in the fixed
compact set $\mathcal{H}_D$. Consequently, there is no need to introduce the
data-dependent box constraint used in Theorem~\ref{thm1} to control the primal
solutions. Instead, we obtain a uniformly valid critical value by taking the
supremum over all $\eta\in\mathcal{H}_D$. 

We use the notation introduced in Section~\ref{Secthm1}. For each
$Q\in\mathcal{Q}$, let $E_N(Q)$ denote the event
\[
E_N(Q):=\left\{\mathbb{W}:\ 
\begin{array}{l}
\|\hat{\Omega}-\Omega(Q)\|\leq\alpha_N,\\[1mm]
d_{\mathrm{Pr}}\!\left(\mathcal{L}(\zeta^*\mid\mathbb{W}),\mathcal{N}(0,V(Q))\right)\leq\alpha_N,\\[1mm]
\sqrt{N}\|\hat{\mu}-\mu(Q)\|\leq M_N
\end{array}
\right\}.
\]
where $\alpha_N$ and $M_N$ are chosen such that
$\alpha_N\downarrow0$, $M_N\to\infty$, $M_N=o(\kappa_N)$,
$M_N\alpha_N=o(1)$,
\[
\sup_{Q\in\mathcal{Q}}
d_{\mathrm{Pr}}\!\left(
\mathcal{L}_Q\!\left(\sqrt{N}(\hat{\mu}-\mu(Q))\right),
\mathcal{N}(0,V(Q))
\right)
\leq\alpha_N,
\]
and
\[
\lim_{N\to\infty}\inf_{Q\in\mathcal{Q}}Q(E_N(Q))=1.
\]
Such sequences $\alpha_N$ and $M_N$ exist by
Assumptions~\ref{assU1}--\ref{assU3}. In what follows, $C>0$ denotes a
generic constant whose value may change from one line to the next. Recall that
\[
\hat{T}_N(v)=\phi_N(\hat{\mu},\infty,v).
\]

\medskip
\noindent
\textit{Step 1: Uniform validity.}
Arguing as in the derivation of \eqref{thmAVeqn7}, the saddle-point property
of the Lagrangian implies that, for any $Q\in {\cal Q}$ and any perturbation $\xi$,
\[
\sqrt{N}\,\phi_N(\mu_N(\xi),\infty,v_0)
\leq
\left\langle \xi_b-\xi_A\eta,\lambda_\xi\right\rangle,
\]
where $\eta$ is any solution to $b(v_0)\leq A\eta$ and
$\lambda_\xi\in\Delta_{0,N}(\xi,v_0)$, with \footnote{Here, for notational simplicity, we have suppressed the dependence of $\Delta_{0,N}(\xi,v_0)$ on $Q\in {\cal Q}$.}
\[
\Delta_{0,N}(\xi,v_0)
:=
\arg\max
\left\{
\left\langle b_N(\xi,v_0),\lambda\right\rangle
\;\middle|\;
\lambda\geq0,\;
A_N(\xi)^\top\lambda=0,\;
\mathbbm{1}^\top\Omega_N(\xi)\lambda\leq1
\right\}.
\]
By assumption, every such $\eta$ belongs to $\mathcal{H}_D$. Hence,
\begin{equation}
\label{pthm4eqn1}
\sqrt{N}\,\phi_N(\mu_N(\xi),\infty,v_0)
\leq
\sup_{\eta\in\mathcal{H}_D}
\sup_{\lambda\in\Delta_{0,N}(\xi,v_0)}
\left\langle \xi_b-\xi_A\eta,\lambda\right\rangle.
\end{equation}
Let ${\cal X}_N$ denote the set of pertubation $\xi$ defined by
\[
{\cal X}_N=\Bigl\{\xi \mid \|\xi_b\|\leq M_N, \ \|\xi_A\|\leq M_N, \ \|\xi_{\Omega}\|\leq \sqrt{N}\alpha_N\Bigr\}.
\]
Arguing as in Step~1 of the proof of Theorem 5.19 in \cite{GM25}, for all sufficiently large $N$, we have
\[
\sup_{Q\in {\cal Q}}\sup_{\mathbb{W}\in E_N(Q)}\sup_{\xi\in\mathcal{X}_N}
\overset{\rightarrow}{d}_H
\left(
\mathcal{P}_U\bigl(\Delta_{0,N}(\xi,v_0)\bigr),
\mathcal{P}_U\bigl(\hat{\Delta}(v_0)\bigr)
\right)
\leq C\alpha_N,
\]
and
\[
\sup_{\xi\in\mathcal{X}_N}
\sup_{\lambda\in\Delta_{0,N}(\xi,v_0)}
\|\mathcal{P}_U(\lambda)\|_1
=O(1).
\]
Combining these bounds with \eqref{pthm4eqn1}, we obtain, for all
$Q\in\mathcal{Q}$ and uniformly on $E_N(Q)$,
\begin{equation}
\label{pthm4eqn2}
\sup_{\xi\in\mathcal{X}_N}
\left\{
\sqrt{N}\,\phi_N(\mu_N(\xi),\infty,v_0)
-
\sup_{\eta\in\mathcal{H}_D}
\sup_{\lambda\in\hat{\Delta}(v_0)}
\left\langle \xi_b-\xi_A\eta,\lambda\right\rangle
\right\}
=o(1).
\end{equation}
The remainder of the argument establishing uniform validity of the resulting
confidence intervals follows Steps~2 and~3 of the proof of
Theorem~4.19 in \cite{GM25}, which establish the coupling and
anti-concentration arguments, respectively. The anti-concentration condition
required in Step~3 follows from Assumption~\ref{assU6}.

\medskip
\noindent
\textit{Step 2: Consistency against local alternatives.}
Fix $Q\in {\cal Q}$. As in Step~5 of the proof of Theorem~\ref{thm1}, it suffices to show that
\begin{equation}
\label{pthm4eqn3}
\lim_{t\to\infty}\lim_{N\to\infty}
Q\!\left(
\sqrt{N}\hat{T}_N\!\left(v_0+\frac{h}{\sqrt{N}}\right)
\leq
\hat{c}^D_N\!\left(1-\alpha;v_0+\frac{h}{\sqrt{N}}\right)
\text{ for some }h\geq t
\right)
=0.
\end{equation}
Since
\[
\hat{\Delta}(v)
\subseteq
\widetilde{\Delta}
:=
\left\{
\lambda\geq0
\;\middle|\;
\mathbbm{1}^\top\hat{\Omega}\lambda\leq1
\right\},
\]
we have
\[
\sup_{\eta\in\mathcal{H}_D}
\sup_{\lambda\in\hat{\Delta}(v)}
\left\langle\zeta_b^*-\zeta_A^*\eta,\lambda\right\rangle
\leq
\sup_{\eta\in\mathcal{H}_D}
\sup_{\lambda\in\widetilde{\Delta}}
\left\langle\zeta_b^*-\zeta_A^*\eta,\lambda\right\rangle.
\]
Let $\widetilde{c}^D_N(1-\alpha)$ denote the $(1-\alpha)$-quantile of
\[
\widetilde{v}^*
:=
\sup_{\eta\in\mathcal{H}_D}
\sup_{\lambda\in\widetilde{\Delta}}
\left\langle\zeta_b^*-\zeta_A^*\eta,\lambda\right\rangle.
\]
Arguing as in the derivation of \eqref{thmAVeqn30}, it is therefore enough to
show that
\begin{equation}
\label{pthm4eqn4}
\lim_{t\to\infty}\lim_{N\to\infty}
Q\!\left(
\left\{
\sqrt{N}\hat{T}_N\!\left(v_0+\frac{t}{\sqrt{N}}\right)
\leq
\widetilde{c}^D_N(1-\alpha)
\right\}
\cap E_N
\right)
=0.
\end{equation}

Define
\[
\widetilde{\Delta}_0
:=
\left\{
\lambda\geq0
\;\middle|\;
\mathbbm{1}^\top\Omega\lambda\leq1
\right\}.
\]
Arguing as in the derivation of \eqref{thmAVeqn34}, the continuous mapping
theorem and the assumed nondegeneracy of
\[
\sup_{\eta\in\mathcal{H}_D}
\sup_{\lambda\in\widetilde{\Delta}_0}
\left\langle\zeta_b-\zeta_A\eta,\lambda\right\rangle
\]
(which follows from Assumption \ref{assU6}) imply that its distribution function is continuous and strictly increasing at
its $(1-\alpha)$-quantile for $\alpha\in(0,1/2)$ (see Proposition B.1 in \cite{GM25}). Consequently,
\begin{equation}
\label{pthm4eqn5}
\widetilde{c}^D_N(1-\alpha)
\overset{p}{\longrightarrow}
\widetilde{c}^D(1-\alpha),
\end{equation}
where $\widetilde{c}^D(1-\alpha)$ denotes the $(1-\alpha)$-quantile of
\[
\sup_{\eta\in\mathcal{H}_D}
\sup_{\lambda\in\widetilde{\Delta}_0}
\left\langle\zeta_b-\zeta_A\eta,\lambda\right\rangle.
\]

Let
\[
\hat{\xi}
:=
\left(
\sqrt{N}(\hat{A}-A),
\sqrt{N}(\hat{b}-b),
\sqrt{N}(\hat{\Omega}-\Omega)
\right)
\]
denote the perturbation induced by the realized sample, and set
\[
v_t:=v_0+\frac{t}{\sqrt{N}}.
\]
We can decompose
\[
\sqrt{N}\hat{T}_N(v_t)=R_N(t)+D_N(t),
\]
where
\[
\begin{aligned}
R_N(t)
&:=
\sqrt{N}
\left\{
\phi_N(\mu_N(\hat{\xi}),\infty,v_t)
-
\phi(\mu,\infty,v_t)
\right\},\\
D_N(t)
&:=
\sqrt{N}\phi(\mu,\infty,v_t).
\end{aligned}
\]
By \eqref{lemLBeqn1},
\begin{equation}
\label{pthm4eqn6}
D_N(t)\geq Ct,
\qquad t\geq0.
\end{equation}

By Lemma~\ref{lemAux}, for all sufficiently large $N$, on the event
\[
\left\{
\sqrt{N}\hat{T}_N(v_t)
\leq
\widetilde{c}^D_N(1-\alpha)
\right\}
\cap E_N,
\]
the relevant population and perturbed problems have finite values, so that
$R_N(t)$ is finite. Moreover, by the saddle-point property of the Lagrangian,
arguing as in the derivation of \eqref{thmAVeqn36},
\begin{equation}
\label{pthm4eqn7}
\begin{aligned}
R_N(t)
\geq{}&
\sqrt{N}\bigl(\hat{b}(v_t)-b(v_t)\bigr)^\top\lambda_0
-
\eta_{\hat{\xi}}^\top
\bigl(\sqrt{N}(\hat{A}-A)\bigr)^\top\lambda_0\\
&\quad
-
\sqrt{N}s_{\hat{\xi}}\,
\mathbbm{1}^\top(\hat{\Omega}-\Omega)\lambda_0,
\end{aligned}
\end{equation}
where
\[
\lambda_0\in
\arg\max
\left\{
\langle b(v_t),\lambda\rangle
\;\middle|\;
\lambda\geq0,\;
A^\top\lambda=0,\;
\mathbbm{1}^\top\Omega\lambda\leq1
\right\},
\]
and
\[
(\eta_{\hat{\xi}},s_{\hat{\xi}})
\in
\arg\min
\left\{
s
\;\middle|\;
s\hat{\Omega}\mathbbm{1}+\hat{A}\eta
\geq\hat{b}(v_t),\;
s\geq0
\right\}.
\]
As in the argument following \eqref{thmAVeqn36}, uniformly on $E_N$ and over
all admissible $\lambda_0$,
\[
\sqrt{N}s_{\hat{\xi}}\,
\mathbbm{1}^\top(\hat{\Omega}-\Omega)\lambda_0
=o(1).
\]

In Step~5 of the proof of Theorem~\ref{thm1}, the box constraint was used to
ensure that $\eta_{\hat{\xi}}$ is bounded in probability. Here this follows
directly from the deterministic restrictions: since these restrictions are
unaffected by the perturbation $\hat{\xi}$, we necessarily have
$\eta_{\hat{\xi}}\in\mathcal{H}_D$. Hence, \eqref{pthm4eqn7} gives, uniformly
on $E_N$,
\begin{equation}
\label{pthm4eqn8}
R_N(t)
\geq
\inf_{\eta\in\mathcal{H}_D}
\left\{
\sqrt{N}\bigl(\hat{b}(v_t)-b(v_t)\bigr)^\top\lambda_0
-
\eta^\top
\bigl(\sqrt{N}(\hat{A}-A)\bigr)^\top\lambda_0
\right\}
+o_p(1).
\end{equation}
The random term on the right-hand side of \eqref{pthm4eqn8} is uniformly
bounded in probability over $t>0$. Indeed,
$\sqrt{N}\bigl(\hat{b}(v_t)-b(v_t)\bigr)$ does not depend on $t$,
$\mathcal{H}_D$ is compact, and $\mathcal{P}_U(\lambda_0)$ is uniformly
bounded (on $E_N$ and for all sufficiently large $N$) over all admissible $\lambda_0$. Combining
\eqref{pthm4eqn6} and \eqref{pthm4eqn8}, we obtain
\[
\sqrt{N}\hat{T}_N(v_t)
\geq
Ct-O_p(1).
\]
Together with \eqref{pthm4eqn5}, this implies \eqref{pthm4eqn4}. The remainder
of the argument proceeds as in the argument following
\eqref{thmAVeqnBC}, establishing \eqref{pthm4eqn3} and completing the proof.
\end{proof}

\section{Uniformity Conditions}
\label{SecAss}
\subsection{Uniformity conditions for Theorem~\ref{thm1}}
In this section, we state conditions on classes of DGPs (${\cal Q}$) under which the
conclusions in \eqref{eqnthm1} hold uniformly. The regularity assumption on the coefficient matrices (Assumption \ref{assU4}) is what makes the result in equation \eqref{pthm1eqn2} of Lemma \ref{lemNP} hold uniformly over ${\cal Q}$. The proof then proceeds by
allowing the events $E_N$ used in the proof of Theorem \ref{thm1} to depend on the DGP $Q\in\mathcal Q$ and then arguing as in \cite{GM25}.
Since the required modifications are straightforward, we omit the proof.

 As noted in \cite{GM25}, the anti-concentration conditions can be done away with if one is willing to modify the rejection rule so that the critical value is increased by an arbitrarily small fixed positive constant.

Below, let $A$ and $b$ be defined as in \eqref{eqnIP1}, \eqref{eqnIP2}, or
\eqref{eqnIP3}, depending on whether an upper, lower, or two-sided confidence
interval is considered. For each $Q\in\mathcal{Q}$, let $\Omega(Q)$ be defined,
as in Section~\ref{SecInference}, from the asymptotic covariance matrix of
\[
\sqrt{N}\left[
\begin{pmatrix}
\hat{b}(v)\\
\operatorname{Vec}(\hat{A})
\end{pmatrix}
-
\begin{pmatrix}
b(Q)(v)\\
\operatorname{Vec}(A(Q))
\end{pmatrix}
\right].
\]

\begin{assumption}
\label{assU0}
For each $Q\in {\cal Q}$, the LP with inputs $\mu(Q)$ has finite value, denoted $v_0(Q)$.
\end{assumption}

\begin{assumption}[Uniform asymptotic approximation]
\label{assU1}
 Let $V(Q)$ denote the asymptotic covariance
matrix of $\sqrt{N}(\hat{\mu}-\mu(Q))$.

Then
\[
\lim_{N\to\infty}
\sup_{Q\in\mathcal Q}
d_{\mathrm{Pr}}
\left(
\mathcal{L}_Q\left(
\sqrt{N}(\hat{\mu}-\mu(Q))
\right),
\mathcal{N}(0,V(Q))
\right)
=
0.
\]
\end{assumption}

\begin{assumption}[Uniform consistency of the covariance estimator]
\label{assU2}
Let $\Omega(Q)$ defined as above. Then
\begin{enumerate}[i)]
\item For every $\epsilon>0$,
\[
\lim_{N\to\infty}
\sup_{Q\in\mathcal Q}
Q\left(
\lVert\hat{V}-V(Q)\rVert>\epsilon
\right)
=
0.
\]
\item There exist constants
$0<\underline{\sigma}<\overline{\sigma}<\infty$
such that, uniformly over $Q\in\mathcal{Q}$, every nonzero diagonal element
of $V(Q)$ belongs to
$[\underline{\sigma},\overline{\sigma}]$.
\end{enumerate}
\end{assumption}

\begin{assumption}[Uniform bootstrap consistency]
\label{assU3}
There exists a random vector $\zeta_N^*$, whose conditional distribution
is determined by the realized sample $\mathbb W$, such that, for every
$\epsilon>0$,
\[
\lim_{N\to\infty}
\sup_{Q\in\mathcal Q}
Q\left(
d_{\mathrm{Pr}}
\left(
\mathcal{L}_Q(\zeta_N^*\mid\mathbb W),
\mathcal{N}(0,V(Q))
\right)
>
\epsilon
\right)
=
0.
\]
\end{assumption}

\begin{assumption}[Anti-concentration]
\label{assU5}
For each $Q\in\mathcal{Q}$, let
$\bigl(\zeta_{b_{\mathrm{aug}}}(Q),\zeta_{A_{\mathrm{aug}}}(Q)\bigr)$
be a random element whose distribution is the limiting distribution under $Q$
of
\[
\left(
\sqrt{N}\bigl(\hat{b}_{\mathrm{aug}}(v)-b_{\mathrm{aug}}(Q,v)\bigr),
\sqrt{N}\bigl(\hat{A}_{\mathrm{aug}}-A_{\mathrm{aug}}(Q)\bigr)
\right).
\]
There exists a constant $c>0$ such that
\[
\inf_{Q\in\mathcal{Q}}
\max_{\lambda\in\Delta_0(Q,v_0(Q))}
\operatorname{Var}
\left(
\left\langle
\zeta_{b_{\mathrm{aug}}}(Q)
-\zeta_{A_{\mathrm{aug}}}(Q)\eta^*(Q),
\lambda
\right\rangle
\right)
\geq c.
\]
Here $v_0(Q)$, $\Delta_0(Q,v_0(Q))$, and $\eta^*(Q)$ denote the
$Q$-indexed counterparts of the objects defined in
Assumption~\ref{ass4}.
\end{assumption}

\begin{assumption}[Uniform Bound on the Hoffman Constant]
\label{assU4}
 We assume that there exists a constant
$C<\infty$ such that
\[
\sup_{Q\in\mathcal Q}
\Lambda(A(Q))
\leq C,
\]
where $\Lambda(A)$ is defined as in \cite{GM25} and provides an upper bound
on the associated Hoffman constant.
\end{assumption}

{\bf Anti-Concentration Condition for Theorem~\ref{thm4}}

\begin{assumption}[Anti-concentration]
\label{assU6}
For each $Q\in\mathcal{Q}$, let $b_Q$ and $A(Q)$ be defined as in
\eqref{eqnIP1}, \eqref{eqnIP2}, or \eqref{eqnIP3}, and suppose that the
corresponding systems \eqref{secUeqn1} contain the common set of deterministic
constraints in \eqref{eqnDBC}. For each $Q\in\mathcal{Q}$, define
\[
\Delta_0(Q)
:=
\arg\max\left\{
\langle b_Q(v_0(Q)),\lambda \rangle:\ \lambda\geq 0, \ A(Q)^\top \lambda=0,\ \mathbbm{1}^\top\Omega(Q) \lambda \leq 1
\right\},
\]
where $v_0(Q)$ is defined as in Assumption \ref{assU0}.
Then there exists a constant $c>0$ such that
\[
\inf_{Q\in\mathcal{Q}} \sup_{\eta\in\mathcal{H}_D}
\sup_{\lambda\in\Delta_0(Q)}
\operatorname{Var}
\left(
\left\langle
\zeta_b(Q)-\zeta_A(Q)\eta,\lambda
\right\rangle
\right)
\geq c,
\]
where $\bigl(\zeta_b(Q),\zeta_A(Q)\bigr)$ are defined as in
Assumption~\ref{assU4}, applied to the unaugmented systems.
\end{assumption}

\section{Supporting and Auxiliary Lemmas}
\label{secAux}

The following lemma shows that the solution $\hat{z}$ of the QP~\eqref{eqnEAM6} is consistent for the minimum-norm element of the set ${\cal O}$ defined in~\eqref{eqnAux}. The lemma plays a central role in the analysis as it shows that the added box-constraints that we introduce in Section \ref{SecInference} intersect the set $\{\eta\mid b(v)\leq A \eta\}$ with probability approaching one, whenever the latter set is non-empty. It also gives a consistent estimator of the minimum-norm primal--dual solution pair, without requiring estimation of the value of the LP itself. The proof builds on arguments from the proof of Theorem~4.19 in \cite{GM25}. For completeness and the reader's convenience, we provide the full proof here. For a related result, see Theorem 2 in \cite{CSS}.
\begin{lemma}
\label{lemNP}
Let ${\cal O}$ be the set defined in equation~\eqref{eqnAux}, and let
\begin{equation}
\label{pthm1eqn1}
z^*
=
\argmin_{z\in{\cal O}}
\|z\|
\end{equation}
denote its minimum-norm element. Suppose that assumptions ~\ref{ass1} and ~\ref{ass2} hold. Then
\begin{equation}
\label{pthm1eqn2}
\lim_{N\to\infty}
Q\!\left(
\|\hat{z}-z^*\|_{\infty}
\leq
\frac{\kappa_{N}^{3/4}}{N^{1/4}}
\right)
=1
\end{equation}
and
\begin{equation}
\label{pthm1eqn21}
\lim_{N\to\infty}
Q\!\left(
\hat{f}_N
\leq
\frac{\kappa_{N}}{\sqrt{N}}
\right)
=1.
\end{equation}
Here, $\hat{z}$ and $\hat{f}_N$ are defined in equations~\eqref{eqnEAM6} and~\eqref{eqnEAM7}, respectively, and $\kappa_{N}$ satisfies
\[
\kappa_{N}\to\infty
\qquad\text{and}\qquad
\kappa_{N}=o(N^{1/4}).
\]
\end{lemma}
\begin{proof}[\bf{Proof of Lemma~\ref{lemNP}}]
Consider the event
\[
E_N
=
\left\{
\mathbb{W}
:
\|\hat{\mu}-\mu\|
\leq
\frac{M_N}{\sqrt{N}},
\quad
\|\hat{V}-V(Q)\|
\leq
\alpha_N
\right\},
\]
where $M_N\uparrow\infty$, $M_N=o(\kappa_N)$, and $\alpha_N\downarrow 0$.
By Assumptions~\ref{ass1} and~\ref{ass2}, if $\alpha_N$ decreases to zero sufficiently slowly, then
\[
\lim_{N\to\infty}Q(E_N)=1.
\]

We henceforth work on the event $E_N$. For all sufficiently large $N$, we have
\begin{equation}
\label{pthm1eqn3}
\hat{\pi}-\hat{\Pi}z^*
=
(\hat{\pi}-\pi)
+(\pi-\Pi z^*)
+(\Pi-\hat{\Pi})z^*
\le
\frac{\kappa_{N}}{\sqrt N}\,\hat S\mathbbm{1}.
\end{equation}
Here we use the fact that, on $E_N$, the diagonal entries of $\hat{S}$ associated with nondeterministic constraints are uniformly bounded away from zero for sufficiently large $N$.
Hence \eqref{pthm1eqn3} implies that $z^*$ is feasible for the QP \eqref{eqnEAM6}. Because $\hat z$ minimizes the norm over the
feasible region,
\begin{equation}
\label{pthm1eqn4}
\|\hat z\|\le \|z^*\|.
\end{equation}

Next, by strong duality for linear programming,
\[
\hat f_N
=
\min\Bigl\{
t:
\hat{\pi}-\hat{\Pi}z
\le
t\,\hat S\mathbbm{1},
\quad
t\ge 0
\Bigr\}.
\]
Combining this representation with \eqref{pthm1eqn3} yields
\begin{equation}
\label{pthm1eqn5}
0\le \hat f_N \le \frac{\kappa_{N}}{\sqrt N}
\end{equation}
for all sufficiently large $N$. Note in particular that the inclusion of $\hat{f}_N$ in the constraint defining the QP implies that the feasible region of
\eqref{eqnEAM6} is always nonempty and $\hat z$ is well defined.

We now consider the distance of $\hat z$ to the population feasible set
\[
\mathcal O=\{z:\pi-\Pi z\le 0\}.
\]
In what follows, $C>0$ denotes a generic constant whose value may vary from one line to the next. By Hoffman's theorem (\cite{AH}), there exists a constant $C>0$ (the so-called Hoffman constant), depending on $\pi$ and $\Pi$, such that
\begin{equation}
\label{pthm1eqn6}
d(\hat z,\mathcal O)
\le
C\bigl\|(\pi-\Pi\hat z)_+\bigr\|.
\end{equation}

Using the feasibility of $\hat z$ for \eqref{eqnEAM6},
\eqref{pthm1eqn5}, and the fact that
$\|\hat z\|\le \|z^*\|$, we obtain
\begin{align*}
(\pi-\Pi\hat z)_+
&=
\bigl(
\pi-\hat\pi
+\hat\pi-\hat\Pi\hat z
+\hat\Pi\hat z-\Pi\hat z
\bigr)_+ \\
&\le
(\pi-\hat\pi)_+
+
\frac{\kappa_{1N}}{\sqrt N}\,\hat S\mathbbm{1}
+
(\hat\Pi\hat z-\Pi\hat z)_+ \\
&\le
C\,\frac{\kappa_{N}}{\sqrt N}\,\hat S\mathbbm{1} .
\end{align*}
Substituting into \eqref{pthm1eqn6} gives
\[
\|\hat z-P_{\mathcal O}(\hat z)\|
\le
C\,\frac{\kappa_{N}}{\sqrt N},
\]
where $P_{\mathcal O}$ denotes the Euclidean projection onto
$\mathcal O$. Combining the latter inequality with \eqref{pthm1eqn4} gives 
\begin{equation}
\label{pthm1eqn7}
\|P_{\mathcal O}(\hat z)\|
\le
\|\hat z\|
+
C\frac{\kappa_{N}}{\sqrt N}
\le
\|z^*\|
+
C\frac{\kappa_{N}}{\sqrt N}.
\end{equation}

Finally, since $z^*=P_{\mathcal O}(0)$, the characterization of
Euclidean projections implies that
\[
\langle z^*,z\rangle \ge \|z^*\|^2
\qquad
\text{for every } z\in\mathcal O.
\]
Applying this inequality with
$z=P_{\mathcal O}(\hat z)$ and using \eqref{pthm1eqn7}, we obtain
\begin{align*}
\|z^*-P_{\mathcal O}(\hat z)\|^2
&=
\|z^*\|^2
+
\|P_{\mathcal O}(\hat z)\|^2
-
2\langle z^*,P_{\mathcal O}(\hat z)\rangle \\
&\le
\|P_{\mathcal O}(\hat z)\|^2-\|z^*\|^2 \\
&\le
C\frac{\kappa_{N}}{\sqrt N}.
\end{align*}
Hence
\[
\|z^*-P_{\mathcal O}(\hat z)\|
\le
C\Bigl(\frac{\kappa_{N}}{\sqrt N}\Bigr)^{1/2}
=
C\frac{\kappa_{N}^{1/2}}{N^{1/4}}.
\]
Combining this bound with the inequality that precedes \eqref{pthm1eqn7} yields
\[
\|\hat z-z^*\|
\le
\|\hat z-P_{\mathcal O}(\hat z)\|
+
\|P_{\mathcal O}(\hat z)-z^*\|
\le
C\frac{\kappa_{N}^{1/2}}{N^{1/4}}.
\]
Since $\kappa_{N}\to\infty$, the right-hand side is
$o(\kappa_{N}^{3/4}/N^{1/4})$, and therefore
\[
\|\hat z-z^*\|_{\infty}
\le
\frac{\kappa_{N}^{3/4}}{N^{1/4}}
\]
for all sufficiently large $N$.

Together with \eqref{pthm1eqn5}, this proves the lemma.
\end{proof}



\begin{lemma}
\label{lemAux}
Let $A$ and $b(v)$ be as in either \eqref{eqnIP1}, \eqref{eqnIP2}, or \eqref{eqnIP3}. Let
\[
\mu=(A,b,\Omega)
\quad\text{and}\quad
\hat{\mu}=(\hat{A},\hat{b},\hat{\Omega})
\]
be defined in Section~\ref{SecInference}. Define
\[
\phi(\mu,a,v)
:=
\max\Bigl\{
b(v)^\top\lambda
-a\|\omega\|_1
\;\Big|\;
\lambda\geq0,\;
A^\top\lambda=\omega,\;
\mathbbm{1}^\top\Omega\lambda\leq1
\Bigr\},
\]
where $a>0$ and $v\in\mathbb{R}$. Let $\phi(\hat{\mu},a,v)$ be defined analogously with $(A,b,\Omega)$ replaced by $(\hat{A},\hat{b},\hat{\Omega})$. Then, for every $a>0$ and $v\in\mathbb{R}$,
\begin{equation}
\label{lemAuxeqn1}
\phi(\mu,a,v)=\infty
\quad\Longrightarrow\quad
Q\!\left(\phi(\hat{\mu},a,v)=\infty\right)=1.
\end{equation}
Moreover, on the event that the indices of the nonzero diagonal entries of $\hat{\Omega}$ and $\Omega$ coincide,
\begin{equation}
\label{lemAuxeqn2}
\phi(\hat{\mu},a,v)=\infty
\quad\Longrightarrow\quad
\phi(\mu,a,v)=\infty.
\end{equation}
\end{lemma}

\begin{proof}[\bf{Proof of Lemma~\ref{lemAux}}]
Suppose that $\phi(\mu,a,v)=\infty$. Then there exists a sequence $\{\lambda_n\}$ such that
\[
b(v)^\top\lambda_n
-a\|A^\top\lambda_n\|_1
\longrightarrow\infty,
\]
with $\lambda_n\geq0$ and $\mathbbm{1}^\top\Omega\lambda_n\leq1$ for all $n$. Dividing by $\|\lambda_n\|$ and passing to a convergent subsequence, we may assume that $\lambda_n/\|\lambda_n \longrightarrow \lambda^*$,
where $\lambda^*\geq0$ and $\|\lambda^*\|=1$.

We necessarily have
\[
b(v)^\top \lambda^*
-a\|A^\top \lambda^*\|_1\geq 0,
\]
where $\lambda^*\geq0$ and
\[
\mathbbm{1}^\top\Omega\lambda^*=0.
\]
Since $\Omega$ is diagonal with nonnegative entries, the latter condition implies that
$\lambda^*$ assigns zero weight to all stochastic constraints. We claim that, without loss of generality, we may assume that
\begin{equation}
\label{lemAeqn1}
b(v)^\top \lambda^*
-a\|A^\top \lambda^*\|_1>0.
\end{equation}
Otherwise,
\[
\max\left\{
b(v)^\top\lambda-a\|A^\top\lambda\|_1
\;\middle|\;
\lambda\geq0,\;
\mathbbm{1}^\top\Omega\lambda=0
\right\}=0.
\]
By strong duality for linear programs, there exist $t^*$ and $\eta^*$ such that
\[
b(v)\leq A\eta^*+t^*\Omega\mathbbm{1},
\qquad
-a\mathbbm{1}\leq \eta^*\leq a\mathbbm{1}.
\]
Thus, for any $\lambda\geq0$ satisfying
$\mathbbm{1}^\top\Omega\lambda\leq1$,
\[
\lambda^\top b(v)
\leq
\lambda^\top A\eta^*
+t^*\mathbbm{1}^\top\Omega\lambda
\leq
\lambda^\top A\eta^*+t^*.
\]
Since $\|\eta^*\|_\infty\leq a$, we obtain
\[
\lambda^\top b(v)-a\|A^\top\lambda\|_1
\leq
\lambda^\top A\eta^*
-a\|A^\top\lambda\|_1+t^*
\leq t^*.
\]
Hence,
\[
\max\left\{
b(v)^\top\lambda-a\|A^\top\lambda\|_1
\;\middle|\;
\lambda\geq0,\;
\mathbbm{1}^\top\Omega\lambda\leq1
\right\}
\leq t^*,
\]
contradicting $\phi(\mu,a,v)=\infty$. Therefore, \eqref{lemAeqn1} holds.

Consequently, when $\phi(\mu,a,v)=\infty$, there exists a nonzero vector
$\lambda^*\geq0$ satisfying \eqref{lemAeqn1} and assigning positive weight only to the deterministic constraints of the system $b(v)\leq A\eta$. Since these constraints are unchanged under estimation, we have
\[
\hat{b}(v)^\top\lambda^*
-a\|\hat{A}^\top\lambda^*\|_1
=
b(v)^\top\lambda^*
-a\|A^\top\lambda^*\|_1
>0,
\]
and, because $\lambda^*$ places no weight on stochastic constraints,
\[
\mathbbm{1}^\top\hat{\Omega}\lambda^*=0.
\]
Therefore, $t\lambda^*$ is feasible for the perturbed LP for every $t>0$, and
\[
\phi(\hat{\mu},a,v)
\geq
\lim_{t\to\infty}
\left(
\hat{b}(v)^\top t\lambda^*
-a\|\hat{A}^\top t\lambda^*\|_1
\right)
=\infty.
\]
This proves that
\[
Q\!\left(\phi(\hat{\mu},a,v)=\infty\right)=1,
\]
and establishes \eqref{lemAuxeqn1}.

Conversely, suppose that $\phi(\hat{\mu},a,v)=\infty$ and that the indices of the nonzero diagonal entries of $\hat{\Omega}$ and $\Omega$ coincide. Applying the same argument to the perturbed problem yields a nonzero vector $\lambda^*\geq0$ satisfying
\[
\hat{b}(v)^\top\lambda^*
-a\|\hat{A}^\top\lambda^*\|_1>0
\]
and assigning weight only to deterministic constraints. Since the deterministic constraints are identical in the perturbed and population problems, we also have
\[
\mathbbm{1}^\top\Omega\lambda^*=0,
\]
and therefore
\[
b(v)^\top\lambda^*
-a\|A^\top\lambda^*\|_1>0.
\]
It follows that $\phi(\mu,a,v)=\infty$, proving \eqref{lemAuxeqn2}.
\end{proof}


\begin{lemma}
\label{lemInf}
Let $\phi$ be defined as in Lemma \ref{lemAux}, and suppose that $A$ and $b(v)$ are as in either \eqref{eqnIP1} or \eqref{eqnIP3}, and let $\eta^*$ be the minimum-norm solution of
\[
b(v_0)\leq A\eta.
\]
Then, for any $a>\lVert\eta^*\rVert_\infty$,
\begin{equation}
\label{lemInfEqn1}
\phi(\mu,a,v_0+h)=+\infty
\quad \forall\,h>0
\quad\Longrightarrow\quad
\phi(\mu,a',v_0+h)=+\infty
\quad
\forall\,a'>0,\ \forall\,h>0.
\end{equation}

Likewise, suppose that $A$ and $b(v)$ are as in either \eqref{eqnIP2} or \eqref{eqnIP3}, and let $\eta^*$ denote the minimum-norm solution of
\[
b(v_0)\leq A\eta.
\]
Then, for any $a>\lVert\eta^*\rVert_\infty$,
\begin{equation}
\label{lemInfEqn2}
\phi(\mu,a,v_0-h)=+\infty
\quad \forall\,h>0
\quad\Longrightarrow\quad
\phi(\mu,a',v_0-h)=+\infty
\quad
\forall\,a'>0,\ \forall\,h>0.
\end{equation}
\end{lemma}

\begin{proof}[\bf{Proof of Lemma~\ref{lemInf}}]
We prove the first claim. The proof of the second claim is analogous.

Let
\[
D:=\{i:\Omega_{ii}=0\}
\]
denote the indices corresponding to the deterministic constraints. By the
dual representation of $\phi$, we have
\[
\phi(\mu,a,v)
=
\min\left\{
t:
b(v)\leq A\eta+t\Omega\mathbbm{1},\
\|\eta\|_\infty\leq a,\
t\geq0
\right\}.
\]
It follows that
\begin{equation}
\label{lemInfeqn1}
\phi(\mu,a,v)<\infty
\quad\Longleftrightarrow\quad
\exists\,\eta\ \text{such that}\
\|\eta\|_\infty\leq a
\quad\text{and}\quad
b_D(v)\leq A_D\eta.
\end{equation}
Indeed, the necessity is immediate. Conversely, if
$b_D(v)\leq A_D\eta$, then, since $\Omega_{ii}>0$ for every
$i\notin D$, one can choose $t<\infty$ sufficiently large so that
\[
b(v)\leq A\eta+t\Omega\mathbbm{1}.
\]

Suppose now that
\[
\phi(\mu,a,v_0+h)=+\infty
\qquad\forall\,h>0.
\]
We claim that, in fact, the system
\begin{equation}
\label{lemInfeqn2}
b_D(v_0+h)\leq A_D\eta
\end{equation}
has no solution for any $h>0$.

Suppose, to the contrary, that for some $h_0>0$ there exists
$\eta_0$ satisfying
\[
b_D(v_0+h_0)\leq A_D\eta_0.
\]
Since $\eta^*$ satisfies $b(v_0)\leq A\eta^*$, we also have
\[
b_D(v_0)\leq A_D\eta^*.
\]
For $\tau\in(0,1)$, define
\[
\eta_\tau:=(1-\tau)\eta^*+\tau\eta_0.
\]
Since $b(v)$ is affine in $v$ for the systems under consideration,
\[
b_D(v_0+\tau h_0)
=
(1-\tau)b_D(v_0)
+
\tau b_D(v_0+h_0).
\]
Therefore,
\[
\begin{aligned}
b_D(v_0+\tau h_0)
&\leq
(1-\tau)A_D\eta^*
+
\tau A_D\eta_0
\\
&=
A_D\eta_\tau.
\end{aligned}
\]
Moreover, since $\|\eta^*\|_\infty<a$, we have
\[
\|\eta_\tau\|_\infty<a
\]
for all sufficiently small $\tau>0$. Hence, by
\eqref{lemInfeqn1},
\[
\phi(\mu,a,v_0+\tau h_0)<\infty
\]
for all sufficiently small $\tau>0$, contradicting the assumption that
\[
\phi(\mu,a,v_0+h)=+\infty
\qquad\forall\,h>0.
\]
This proves the claim in \eqref{lemInfeqn2}.

Since \eqref{lemInfeqn2} has no solution for any $h>0$, it follows
immediately from \eqref{lemInfeqn1} that, for every $a'>0$,
\[
\phi(\mu,a',v_0+h)=+\infty
\qquad\forall\,h>0,
\]
which establishes the first claim.
\end{proof}


\begin{lemma}
\label{lemLB}
Let $\phi$ be as defined in Lemma~\ref{lemAux}. There exists a constant
$C>0$, depending only on the fixed  matrices $A$ and $\Omega$, such that for every
$a\geq \|z^*\|_{\infty}$, where $z^*$ denotes the minimum-norm element of
$\mathcal{O}$,\footnote{For \eqref{lemLBeqn1} and \eqref{lemLBeqn2}, it is
sufficient to assume $a\geq\|x^*\|_{\infty}$ and
$a\geq\|y^*\|_{\infty}$, respectively.} the following bounds hold:
\begin{align}
\label{lemLBeqn1}
(A,b)\text{ as in }\eqref{eqnIP1}
&\Longrightarrow
\phi(\mu,a,v_0+h)\geq Ch,
&& h\geq0,
\\
\label{lemLBeqn2}
(A,b)\text{ as in }\eqref{eqnIP2}
&\Longrightarrow
\phi(\mu,a,v_0-h)\geq Ch,
&& h\geq0,
\\
\label{lemLBeqn3}
(A,b)\text{ as in }\eqref{eqnIP3}
&\Longrightarrow
\phi(\mu,a,v_0+h)\geq C|h|,
&& h\in\mathbb{R}.
\end{align}

Moreover, there exists a constant $C'>0$, also depending only on the
matrices $A$ and $\Omega$, such that whenever $a\geq\|z^*\|_{\infty}$ and
$\phi(\mu,a,v_0+h)<\infty$, we have
\begin{align}
\label{lemLBeqn4}
(A,b)\text{ as in }\eqref{eqnIP1}
&\Longrightarrow
\phi(\mu,a,v_0+h)\leq C'h,
&& h\geq0,
\\
\label{lemLBeqn5}
(A,b)\text{ as in }\eqref{eqnIP2}
&\Longrightarrow
\phi(\mu,a,v_0-h)\leq C'h,
&& h\geq0,
\\
\label{lemLBeqn6}
(A,b)\text{ as in }\eqref{eqnIP3}
&\Longrightarrow
\phi(\mu,a,v_0+h)\leq C'|h|,
&& h\in\mathbb{R}.
\end{align}
\end{lemma}

\begin{proof}[\bf{Proof of Lemma~\ref{lemLB}}]
We prove \eqref{lemLBeqn1}, \eqref{lemLBeqn3}, and \eqref{lemLBeqn4}. 
The proof of \eqref{lemLBeqn2} is analogous to that of \eqref{lemLBeqn1}, 
while the proofs of \eqref{lemLBeqn5} and \eqref{lemLBeqn6} follow similarly 
from the argument for \eqref{lemLBeqn4} and are therefore omitted.

We first prove \eqref{lemLBeqn1}. Let
\[
\mathcal D:=\{(\lambda,u_1,u_2)\geq 0:A^\top\lambda=u_1-u_2,\ \mathbbm1^\top\Omega\lambda\leq1\}
\]
denote the feasible region of the LP associated with $\phi$ (the feasible region of the expression of $\phi$ that precedes equation \eqref{thmAVeqn4}). By Farkas' lemma,
$\phi(\mu, a,v_0+h)>0$ for all $h>0$. If this quantity is infinite, the desired bound follows immediately, for any $C>0$. Otherwise, for any $0<h'<h$, $\phi(\mu, a,v_0+h')$ is finite and there exists $(\lambda_{h'},u_{1h'},u_{2h'})\in\operatorname{Extr}(\mathcal D)$\footnote{Note that the set of extreme points is non-empty, due to the non-negativity constraints $(\lambda,u_1,u_2)\geq 0$.} such that
\[
\phi(\mu, a,v_0+h')
=
b(v_0+h')^\top\lambda_{h'}
-a\|A^\top\lambda_{h'}\|_1>0.
\]
Let $e^*$ denote the canonical basis vector corresponding to the coordinate of $b(v)$ containing $v$ (the vector with all zeros except a one in the last coordinate, given the expression in \eqref{eqnIP1}). Then
\[
b(v_0+h')^\top\lambda_{h'}
-a\|A^\top\lambda_{h'}\|_1
=
b(v_0)^\top\lambda_{h'}
-a\|A^\top\lambda_{h'}\|_1
+h'\langle e^*,\lambda_{h'}\rangle.
\]
Since $\phi(\mu, a,v_0)=0$ and the feasible region does not depend on $v$,
\[
b(v_0)^\top\lambda_{h'}
-a\|A^\top\lambda_{h'}\|_1\leq0,
\]
and hence $\langle e^*,\lambda_{h'}\rangle>0$. Therefore,
\[
\phi(\mu, a,h)\geq b(v_0)^\top\lambda_{h'} -a\|A^\top\lambda_{h'}\|_1 +h\langle e^*,\lambda_{h'}\rangle
\geq
(h-h')\langle e^*,\lambda_{h'}\rangle .
\]
Noting that the set $\operatorname{Extr}(\mathcal D)$ is finite, and letting $C>0$ be defined by
\[
C:=\min_{(\lambda,u_{1},u_{2})\in\operatorname{Extr}(\mathcal D):
\langle e^*,\lambda\rangle>0}
\langle e^*,\lambda\rangle>0, 
\]
we get
\[
\phi(\mu, a,h)\geq C(h-h').
\]
Letting $h'\downarrow0$ gives
\[
\phi(\mu,a,v_0+h)\geq Ch
\]
as claimed.

We now prove \eqref{lemLBeqn3}. Let $\phi_U$, $\phi_L$, and $\phi_{TS}$ denote the functions $\phi$ from Lemma~\ref{lemAux} corresponding to the formulations in \eqref{eqnIP1}, \eqref{eqnIP2}, and \eqref{eqnIP3}, respectively. In view of \eqref{lemLBeqn1} and \eqref{lemLBeqn2}, it suffices to show that, for every $a\geq0$ and $v\in\mathbb{R}$,
\begin{equation}
\label{plemLBeqn1}
\phi_{TS}(\mu,a,v)
=
\max\{\phi_L(\mu,a,v),\,\phi_U(\mu,a,v)\}.
\end{equation}
Let $(A_U,b_U)$, $(A_L,b_L)$, and $(A_{TS},b_{TS})$ denote the pairs $(A,b)$ corresponding to \eqref{eqnIP1}, \eqref{eqnIP2}, and \eqref{eqnIP3}, respectively. Then
\[
A_{TS}
=
\begin{pmatrix}
A_U & 0\\
0 & A_L
\end{pmatrix},
\qquad
b_{TS}
=
\begin{pmatrix}
b_U\\
b_L
\end{pmatrix}.
\]
Let $\Omega_U$ and $\Omega_L$ denote the matrices $\Omega$ associated with
\eqref{eqnIP1} and \eqref{eqnIP2}, respectively. A straightforward computation then gives
\begin{equation}
\label{plemLBeqn2}
\begin{aligned}
\phi_{TS}(\mu,a,v)
=
\max\Bigl\{
&b_U(v)^\top\lambda_U
-a\|A_U^\top\lambda_U\|_1
+b_L(v)^\top\lambda_L
-a\|A_L^\top\lambda_L\|_1
\\
&\Bigm|
(\lambda_U,\lambda_L)\geq0,\;
\|\Omega_U\lambda_U\|_1+
\|\Omega_L\lambda_L\|_1\leq1
\Bigr\}.
\end{aligned}
\end{equation}
From the last expression, if $\lambda_U$ is feasible for the LP defining
$\phi_U$, then $(\lambda_U,0)$ is feasible for the LP defining
$\phi_{TS}$ and achieves the same objective value. The same argument applies
to any $\lambda_L$ feasible for the LP defining $\phi_L$. Therefore,
\[
\phi_{TS}(\mu,a,v)
\geq
\max\{\phi_L(\mu,a,v),\,\phi_U(\mu,a,v)\}.
\]
Moreover, if either $\phi_U(\mu,a,v)=\infty$ or
$\phi_L(\mu,a,v)=\infty$, then \eqref{plemLBeqn1} follows immediately.

Suppose now that $\phi_L(\mu,a,v)$ and $\phi_U(\mu,a,v)$ are finite. For any
$(\lambda_U,\lambda_L)$ feasible for the LP in \eqref{plemLBeqn2}, arguing as
in the proof of \eqref{lemAeqn1}, we have that
\[
\|\Omega_U\lambda_U\|_1=0
\quad\Longrightarrow\quad
b_U(v)^\top\lambda_U-a\|A_U^\top\lambda_U\|_1\leq 0,
\]
since otherwise $\phi_U(\mu,a,v)=\infty$. Therefore,
\[
\begin{aligned}
&b_U(v)^\top\lambda_U-a\|A_U^\top\lambda_U\|_1
+b_L(v)^\top\lambda_L-a\|A_L^\top\lambda_L\|_1
\\
&\qquad\leq
b_L(v)^\top\lambda_L-a\|A_L^\top\lambda_L\|_1
\leq \phi_L(\mu,a,v).
\end{aligned}
\]
By symmetry, if $\|\Omega_L\lambda_L\|_1=0$, the objective value is bounded
above by $\phi_U(\mu,a,v)$. It remains to bound the objective value when
\[
\|\Omega_U\lambda_U\|_1>0
\quad\text{and}\quad
\|\Omega_L\lambda_L\|_1>0.
\]
In this case, let
\[
t_U:=\|\Omega_U\lambda_U\|_1,
\qquad
t_L:=\|\Omega_L\lambda_L\|_1,
\]
and define
\[
\tilde{\lambda}_U:=\frac{\lambda_U}{t_U},
\qquad
\tilde{\lambda}_L:=\frac{\lambda_L}{t_L}.
\]
Since $(\lambda_U,\lambda_L)$ is feasible, we have $t_U+t_L\leq1$. Moreover,
\[
\tilde{\lambda}_U\geq0,
\qquad
\|\Omega_U\tilde{\lambda}_U\|_1=1,
\]
and therefore
\[
b_U(v)^\top\tilde{\lambda}_U
-a\|A_U^\top\tilde{\lambda}_U\|_1
\leq
\phi_U(\mu,a,v).
\]
Similarly,
\[
b_L(v)^\top\tilde{\lambda}_L
-a\|A_L^\top\tilde{\lambda}_L\|_1
\leq
\phi_L(\mu,a,v).
\]
The objective value of the LP in \eqref{plemLBeqn2} at
$(\lambda_U,\lambda_L)$ is therefore
\[
\begin{aligned}
&b_U(v)^\top\lambda_U-a\|A_U^\top\lambda_U\|_1
+b_L(v)^\top\lambda_L-a\|A_L^\top\lambda_L\|_1
\\
&\quad=
t_U\Bigl(
b_U(v)^\top\tilde{\lambda}_U
-a\|A_U^\top\tilde{\lambda}_U\|_1
\Bigr)
+
t_L\Bigl(
b_L(v)^\top\tilde{\lambda}_L
-a\|A_L^\top\tilde{\lambda}_L\|_1
\Bigr)
\\
&\quad\leq
t_U\phi_U(\mu,a,v)+t_L\phi_L(\mu,a,v)
\\
&\quad\leq
\max\{\phi_L(\mu,a,v),\,\phi_U(\mu,a,v)\},
\end{aligned}
\]
where the last inequality follows from $t_U+t_L\leq1$.

We now prove \eqref{lemLBeqn4}. Suppose that
$a\geq \|z^*\|_{\infty}$. Then $\phi(\mu,a,v_0)=0$, and by Farkas'
lemma,
\[
\phi(\mu,a,v_0+h)>0
\]
(possibly with value $+\infty$) for every $h>0$. Suppose that
$\phi(\mu,a,v_0+h)<\infty$. Since the LP defining $\phi$ attains its
maximum, there exists $(\lambda_h,u_{1h},u_{2h})\in\operatorname{Extr}(\mathcal D)$ such that
\begin{align*}
\phi(\mu,a,v_0+h)
&=
b(v_0+h)^\top\lambda_h
-a\|A^\top\lambda_h\|_1\\
&=
b(v_0)^\top\lambda_h
-a\|A^\top\lambda_h\|_1
+h\langle e^*,\lambda_h\rangle .
\end{align*}
Since $\phi(\mu,a,v_0)=0$, we must have
\[
b(v_0)^\top\lambda_h-a\|A^\top\lambda_h\|_1\leq 0,
\]
and therefore
\[
\phi(\mu,a,v_0+h)
\leq h\langle e^*,\lambda_h\rangle.
\]
Moreover, because the left-hand side is strictly positive, we necessarily have $\langle e^*,\lambda_h\rangle>0$.
Hence,
\[
\phi(\mu,a,v_0+h)\leq C'h,
\]
where
\[
C'
:=
\max_{\substack{
(\lambda,u_1,u_2)\in\operatorname{Extr}(\mathcal D)\\
\langle e^*,\lambda\rangle>0
}}
\langle e^*,\lambda\rangle .
\]

\end{proof}

The following lemma is based on parts (a) and (b) of Theorem 1 in \cite{SR2} and provides an alternative characterization of the MFCQ, and is used in the statement and proof of Theorem \ref{thm3}

\begin{lemma}[Dual Characterization of MFCQ]
\label{lemMFCQ}

Consider the linear program
\[
\min_{x\in\mathbb{R}^d}
\left\{
c^\top x :
A_E x=b_E,\;
A_I x\leq b_I
\right\},
\]
and suppose that it is feasible and has a finite optimal value. Let
\[
\mathcal{P}
:=
\left\{
x\in\mathbb{R}^d :
A_E x=b_E,\;
A_I x\leq b_I
\right\}.
\]
Then $\mathcal{P}$ satisfies MFCQ, i.e., the rows of $A_E$ are linearly
independent and there exists $x^*\in\mathbb{R}^d$ such that
\[
A_E x^*=b_E
\qquad\text{and}\qquad
A_I x^*<b_I,
\]
if and only if the set of optimal solutions to the dual problem,
\[
\Delta_0
:=
\arg\max_{\lambda_E,\lambda_I}
\left\{
b_E^\top\lambda_E+b_I^\top\lambda_I :
A_E^\top\lambda_E+A_I^\top\lambda_I=c,\;
\lambda_I\leq0
\right\},
\]
is bounded.
\end{lemma}

\begin{proof}[\bf{Proof of Lemma \ref{lemMFCQ}}]

\underline{Necessity}  Suppose, for the sake of contradiction, that $\mathcal{P}$ is regular and $\Delta_0$ is unbounded. Since the MFCQ holds, there exists a point $x_0$ satisfying $A_E x_0 = b_E$ and $A_I x_0 < b_I$. The unboundedness of $\Delta_0$ implies the existence of a nonzero vector $\lambda = (\lambda_E', \lambda_I')' $ such that  

\begin{equation}
\label{5eqn1}
\lambda_I \leq 0, \quad\quad A_E^T \lambda_E + A_I^T \lambda_I = 0, \quad\quad b_I^T \lambda_I + b_E^T \lambda_E = 0.
\end{equation}  

Since the MFCQ ensures that $A_E$ has full row rank, we cannot have $\lambda_I = 0$ in \eqref{5eqn1}. Consequently, the choice of $\lambda$ in \eqref{5eqn1} leads to  

\[
\langle \lambda_E, A_E x_0 \rangle + \langle \lambda_I, A_I x_0 \rangle > \langle \lambda_E, b_E \rangle + \langle \lambda_I, b_I \rangle.
\]  

This results in $0 > 0$, a contradiction.

\underline{Sufficiency}  Suppose, for the sake of contradiction, that $\Delta_0$ is bounded but the MFCQ does not hold. The failure of MFCQ implies the existence of a sequence of perturbations $b_n \to b$ (where only the $b$-vector is perturbed) such that  

\begin{equation}
\label{5eqn2}
\mathcal{P}_n = \{ x \in \mathbb{R}^d \mid A_E x = b_{n,E}, \ A_I x \leq b_{n,I} \} = \emptyset.
\end{equation}  

By assumption, since the value of the unperturbed program is finite, the set  
\[
\{\lambda \mid \lambda_I \leq 0, \ A_E^T \lambda_E + A_I^T \lambda_I = c \}
\]  
is nonempty. Moreover, \eqref{5eqn2} implies that the corresponding perturbed dual programs are unbounded:  

\[
\max_{\lambda} \Bigl\{ \lambda_E^T b_{n,E} + \lambda_I^T b_{n,I} \mid \lambda_I \leq 0, \ A_E^T \lambda_E + A_I^T \lambda_I = c \Bigr\} = +\infty, \quad \forall n \geq 1.
\]  

This implies the existence of an unbounded sequence $\{\lambda_n\}$ such that  

\[
\lambda_n^T b_n \to +\infty, \quad\quad \lambda_{n,I} \leq 0, \quad\quad A_E^T \lambda_{n,E} + A_I^T \lambda_{n,I} = c.
\]  

Dividing through by $\|\lambda_n\|$ and considering a limit $\lambda^* \neq 0$ of a convergent subsequence of $\{\lambda_n / \|\lambda_n\|\}_{n \geq 1}$, we obtain  

\[
\langle \lambda^*, b \rangle \geq 0, \quad\quad \lambda_I^* \leq 0, \quad\quad A_E^T \lambda_E^* + A_I^T \lambda_I^* = 0.
\]  

However, if $\lambda \in \Delta_0$, then for any $t > 0$, the vector $\lambda + t \lambda^*$ also belongs to $\Delta_0$, contradicting the assumption that $\Delta_0$ is bounded. This contradiction completes the proof.  
\end{proof}

\begin{lemma}
\label{lemDelta}
Let $\Delta_0(v_0)$ be defined as in the expression that precedes Assumption \ref{ass4}. Then $\Delta_0(v_0)$ contains a nonzero element.
\end{lemma}

\begin{proof}[\bf{Proof of Lemma \ref{lemDelta}}]
We give the argument for the upper one-sided formulation
\eqref{eqnIP1}; the lower one-sided case is analogous, and the two-sided
case follows from the decomposition in \eqref{plemLBeqn1}.

Let $a:=2\lVert x^*\rVert_\infty$. Consider the set ${\cal D}$ used in the proof of Lemma \ref{lemLB} given by
\[
{\cal D}
:=
\left\{
(\lambda,u_1,u_2)\geq0:
A^\top\lambda=u_1-u_2,\ 
\mathbbm{1}^\top\Omega\lambda\leq1
\right\}.
\]
Since $v_0$ is the value of the LP,
\[
\phi(\mu,a,v_0)=0,
\qquad
\phi(\mu,a,v_0+h)>0
\quad\text{for every }h>0.
\]
For each $h>0$ sufficiently small for which the latter value is finite (we consider the complementary case where $\phi(\mu,a,v_0+h)=\infty$ for all $h>0$ further below),
there must exist an extreme point
$(\lambda_h,u_{1h},u_{2h})\in\operatorname{Extr}({\cal D})$ such that
\[
b(v_0)^\top\lambda_h
-a\lVert A^\top\lambda_h\rVert_1
+h\langle e^*,\lambda_h\rangle
>0.
\]
On the other hand, $\phi(\mu,a,v_0)=0$ implies
\[
b(v_0)^\top\lambda_h
-a\lVert A^\top\lambda_h\rVert_1
\leq0.
\]
Hence
\[
\langle e^*,\lambda_h\rangle>0,
\]
so in particular $\lambda_h\neq0$.

Now let $h_n\downarrow0$. Since $\operatorname{Extr}(D)$ is finite,
there exists a subsequence along which
$(\lambda_{h_n},u_{1h_n},u_{2h_n})$ is constant, say equal to
$(\lambda_0,u_{10},u_{20})$. Along this subsequence,
\[
b(v_0)^\top\lambda_0
-a\lVert A^\top\lambda_0\rVert_1
>
-h_n\langle e^*,\lambda_0\rangle.
\]
Letting $n\to\infty$ and using $\phi(\mu,a,v_0)=0$ gives
\[
b(v_0)^\top\lambda_0
-a\lVert A^\top\lambda_0\rVert_1
=0.
\]
Thus $\lambda_0\neq0$ is optimal at $v_0$. Lifting
$(\lambda_0,u_{10},u_{20})$ back to the augmented formulation yields a
nonzero element of $\Delta_0(v_0)$.

If instead
\[
\phi(\mu,a,v_0+h)=\infty
\]
for every sufficiently small $h>0$, the argument is slightly different. Arguing as in the derivation of \eqref{lemAeqn1}, for each such $h>0$ there exists a nonzero $\lambda_h\geq0$, placing weight only on the deterministic constraints,
such that
\[
\mathbbm{1}^\top\Omega\lambda_h=0\quad\text{and}\quad \|\lambda_h\|=1
\]
and
\[
b(v_0+h)^\top\lambda_h
-a\lVert A^\top\lambda_h\rVert_1
>0.
\]
 Letting $h_n\downarrow0$, and 
passing to a subsequence if necessary, we have
\[
\lambda_{h_n}\longrightarrow\lambda_0
\]
for some $\lambda_0\neq0$, such that
\[
\|\lambda_0\|=1,\qquad \lambda_0\geq0,
\qquad
\mathbbm{1}^\top\Omega\lambda_0=0.
\]

Since
\[
b(v_0+h)
=
b(v_0)+he^*,
\]
we have
\[
b(v_0)^\top\lambda_{h_n}
-a\lVert A^\top\lambda_{h_n}\rVert_1
>
-h_n\langle e^*,\lambda_{h_n}\rangle.
\]
Taking the limits in the latter inequality, it follows that
\[
b(v_0)^\top\lambda_0
-a\lVert A^\top\lambda_0\rVert_1
\geq0.
\]
On the other hand, since $\phi(\mu,a,v_0)=0$, we necessarily have
\[
b(v_0)^\top\lambda_0
-a\lVert A^\top\lambda_0\rVert_1
=0,
\]
and $\lambda_0\neq0$ is optimal
at $v_0$, and the corresponding point in the augmented representation belongs to $\Delta_0(v_0)$. This shows that the latter set contains a nonzero element.

The lower one-sided case follows identically. For the two-sided
formulation, the decomposition in \eqref{plemLBeqn1} implies that an
optimal nonzero multiplier from either one-sided problem, with the other
block set equal to zero, is an element of $\Delta_0(v_0)$.
\end{proof}


\section{Details on the examples}
\label{SecDetEx}

In this section we give additional details for the examples in Section \ref{SecExemp}.

\subsection{Details for Example~\ref{ExKKLS}}\label{SecDetEx_KKLS}

Example~\ref{ExKKLS} has two values of the profitability shock. Here we allow $w_t\in\mathcal{W}=\{w_1,\dots,w_W\}$ with $w_1<\dots<w_W$, so that there are $2W$ states $z=(k,w)$, ordered with the $W$ states with $k=0$ first; accordingly $oo,s,vp,fc,ec\in\mathbb{R}^W$, $\pi_a\in\mathbb{R}^{2W}$ and $\pi=(\pi_0\tr,\pi_1\tr)\tr\in\mathbb{R}^{4W}$, with $\pi_0$ and $\pi_1$ as in \eqref{eqnKKLS}. We write $I_n$ for the $n\times n$ identity matrix and $\mathbbm{1}_n$ for the $n$-vector of ones, and drop the subscript when the dimension is clear.

\subsubsection{Theoretical background}
We first recall the representation of payoffs and firm values in dynamic discrete choice models, and then show that $\theta_V$ is an affine function of $\pi$ for certain types of counterfactual policies.

In every period $t=0,1,2,\dots$ a firm chooses an action $a_t\in\mathcal A$ after observing a state $z_t\in\mathcal Z$ and idiosyncratic shocks $\eps_t=(\eps_{at})_{a\in\mathcal{A}}$. The \textit{flow payoff} $\pi_a(z)$ is the per-period return of action $a$ in state $z$, and the total per-period utility is $\pi_a(z_t)+\eps_{at}$. Given a discount factor $\delta\in[0,1)$ and transition distributions $F(z_{t+1}\mid z_t,a_t)$, the firm solves
\begin{equation*}
V(z_t,\eps_t)=\max_{a\in\mathcal A}\left\{\pi_a(z_t)+\eps_{at}
+\delta\,\ev\left[V(z_{t+1},\eps_{t+1})\mid a,z_t\right]\right\}.
\end{equation*}
The shock vectors $\eps_t$ are i.i.d. across firms and periods, independent of $z_t$, and have joint distribution $G$. Let $V(z)=\int V(z,\eps)\,dG(\eps)$ be the \textit{ex ante} value function, $v_a(z)=\pi_a(z)+\delta\,\ev[V(z_{t+1})\mid a,z]$ the \textit{conditional} value function, $p_a(z)=\Pr\{v_a(z)+\eps_a\ge v_j(z)+\eps_j\ \forall j\in\mathcal{A}\}$ the conditional choice probability (CCP), and $p(z)=(p_a(z))_{a\in\mathcal{A}}$.

\citet{AM} show in their Lemma~1 that for every $a$ there is a function $\psi_a(\cdot)$, fully determined by $G$, such that $V(z)=v_a(z)+\psi_a(p(z))$ for all $z$. When $G$ is the type 1 extreme value distribution, $\psi_a(p(z))=\gamma_E-\ln p_a(z)$, where $\gamma_E$ is Euler's constant. Stack $\pi_a(z)$, $V(z)$, $p_a(z)$ and $\psi_a(p(z))$ over states into the vectors $\pi_a$, $V$, $p_a$ and $\psi_a(p)$, and let $F_a$ be the transition matrix of the state under action $a$. Then, for a reference action $J$,
\begin{equation}\label{eq:V_J}
V=(I-\delta F_J)^{-1}\left(\pi_J+\psi_J(p)\right),
\end{equation}
so that the ex ante value function is affine in the flow payoff of the reference action once the CCPs are known. \citet{KSS}, in their eq.~(3)--(5), show that for any other action $a\ne J$, the payoff $\pi_a$ is an affine function of $\pi_J$:
\begin{equation}\label{eq:payoffs}
 \pi_a=M_a(F)\pi_J+b_a(p,F),
\end{equation}
where $M_a(F)=(I-\delta F_a)(I-\delta F_J)^{-1}$ and $b_a(p,F)=M_a(F)\psi_J(p)-\psi_a(p)$. This equation restricts the set of possible payoffs, but does not point-identify them.

A counterfactual policy replaces the primitives $(\pi,F)$ by $(\tilde\pi,\tilde F)$, which generate a new value function $\tilde V$ and new CCPs $\tilde p$. \citet{KKLS} restrict attention to \emph{affine payoff counterfactuals}, $\tilde\pi=H\pi+g$, for a known matrix $H$ and a known vector $g$. Since $\pi$ is not identified, $\tilde p$ and $\tilde V$ are in general not identified either.

Let $f$ and $\tilde f$ be the stationary distributions, assumed unique, of the baseline and counterfactual Markov chains of the state, with transition matrices $\sum_a\diag(p_a)F_a$ and $\sum_a\diag(\tilde p_a)\tilde F_a$. One of the objects of interest in \cite{KKLS} is \textit{the long-run average change in the firm value}, $\theta_V=\tilde f\tr\tilde V-f\tr V$. In the setting of Example~\ref{ExKKLS}, let $F_w$ be the $W\times W$ transition matrix of $w$ and $f_w$ its stationary distribution, also assumed unique. Since $k_{t+1}=a_t$, we have $F_0=\left(\begin{smallmatrix}F_w&0\\F_w&0\end{smallmatrix}\right)$ and $F_1=\left(\begin{smallmatrix}0&F_w\\0&F_w\end{smallmatrix}\right)$. We take $J=0$ as the reference action.
\begin{proposition}\label{prop:linear_thetaV}
In the setting of Example~\ref{ExKKLS}, consider a counterfactual that either
\begin{enumerate}
\item[(a)] leaves the payoff of the reference action and the transitions unchanged, i.e., $\tilde\pi_0=\pi_0$ and $\tilde F=F$, and shifts the payoff of action $1$ by a prespecified vector $g_1\in\mathbb{R}^{2W}$, $\tilde\pi_1=\pi_1+g_1$; or
\item[(b)] eliminates action $1$ and with it the incumbency states: $\tilde{\mathcal A}=\{0\}$ and $\tilde{\mathcal Z}=\{0\}\times\mathcal W$, leaving the exogenous process $F_w$ and the payoff of action $0$ on the remaining states unchanged, so that $\tilde F_0=F_w$ and $\tilde\pi_0=oo$.
\end{enumerate}
Then $\theta_V=\gamma\tr\pi+c_0$ for every $\pi$ satisfying \eqref{eq:payoffs}, where $\gamma\in\mathbb{R}^{4W}$ and $c_0\in\mathbb{R}$ are identified and are functions of $(p,F)$ and the policy. In particular, $\theta_V$ is an affine function of $\pi$ on the identified set $\Pi^I$.
\end{proposition}
\begin{proof}
In case (a) the counterfactual is affine with $H=I$, so that the diagonal blocks of $H$ corresponding to the two actions are $H_{00}=H_{11}=I$, and $\tilde F=F$, so $M_1(\tilde F)=M_1(F)$, the condition $H_{11}M_1(F)-M_1(\tilde F)H_{00}=0$ holds trivially, and $\tilde p$ is identified by Corollary~1(ii) of \cite{KSS}. In case (b) only one action remains, so $\tilde p_0\equiv1$. In both cases $\tilde p$, and hence $\tilde f$, are functions of $(p,F)$ and the policy only; in case (b), $\tilde f=f_w$.

Next, write $\tilde\pi_0=H_{00}\pi_0$, with $H_{00}=I_{2W}$ in case (a) and $H_{00}=(I_W\;\;0)$ in case (b), and let $S_0=(I_{2W}\;\;0)$ select $\pi_0$ from $\pi$, i.e., $\pi_0=S_0\pi$. Applying \eqref{eq:V_J} with $J=0$ to the baseline and to the counterfactual model gives $V=(I-\delta F_0)^{-1}(S_0\pi+\psi_0(p))$ and $\tilde V=(I-\delta\tilde F_0)^{-1}(H_{00}S_0\pi+\psi_0(\tilde p))$, where $\tilde F_0=F_0$ in case (a) and $\tilde F_0=F_w$ in case (b). Hence $\theta_V=\tilde f\tr\tilde V-f\tr V=\gamma\tr\pi+c_0$, where
\[\gamma=S_0\tr\bigl[H_{00}\tr(I-\delta\tilde F_0\tr)^{-1}\tilde f-(I-\delta F_0\tr)^{-1}f\bigr], \quad c_0=\tilde f\tr(I-\delta\tilde F_0)^{-1}\psi_0(\tilde p)-f\tr(I-\delta F_0)^{-1}\psi_0(p).\]
Both $\gamma$ and $c_0$ depend only on $(p,F)$ and the policy.
\end{proof}

\begin{remark} Case (a) covers any prespecified addition to the payoff of operating. A lump-sum subsidy $c$ to fixed costs corresponds to $g_1=c\,\mathbbm{1}_{2W}$. A subsidy that raises variable profits by a known vector $\Delta\in\mathbb{R}^W$ transforms
\[
\pi_1=\begin{pmatrix}vp-fc-ec\\ vp-fc\end{pmatrix}
\quad\text{into}\quad
\tilde\pi_1=\begin{pmatrix}vp+\Delta-fc-ec\\ vp+\Delta-fc\end{pmatrix}
=\pi_1+\begin{pmatrix}\Delta\\ \Delta\end{pmatrix},
\]
so $g_1=(\Delta\tr,\Delta\tr)\tr$. For a subsidy proportional to revenues, $\Delta$ is a known function of variable profits (which are known under Restriction~1 below) and the demand elasticity.

Case (b) is a policy that removes the option to be active, such as an export ban or the closure of a market.
\end{remark}
\subsubsection{Identified set for $\pi$}\label{SecDetEx_KKLS_restr}
In the setting of Example~\ref{ExKKLS}, equation \eqref{eq:payoffs} with $J=0$ and $a=1$ can be written as
\begin{equation}\label{eq:payoffs_matrix}
 M_{\pi}\pi=b_1(p,F), \text{ where }M_{\pi}=\bigl(-M_1(F)\;\;I_{2W}\bigr)=\begin{pmatrix}-(I+\delta F_w) & \delta F_w & I_W & 0\\ -\delta F_w & -(I-\delta F_w) & 0 & I_W\end{pmatrix}.
\end{equation}
Equation \eqref{eq:payoffs_matrix} alone leaves $\pi_0$ unrestricted, so the bounds on $\theta_V$ are infinite whenever $\gamma\neq0$. \citet{KKLS} consider three additional sets of restrictions on the payoff components. Each is a system of linear inequalities in $\pi$, which we write as $R_j\pi\le r_j$:
\begin{restriction*}
\begin{enumerate}
\item[1.] The payoff of the outside option is normalized to zero, the fixed and entry costs are non-negative, and variable profits are known: $oo=0$; $fc\ge 0$; $ec\ge 0$; $vp$ known. Equivalently,
\[
\underbrace{\begin{pmatrix} I & 0 & 0 & 0\\-I & 0 & 0 & 0\\0 & 0 & 0 & I\\ 0 & 0 & I & -I\end{pmatrix}}_{R_1}\pi\le
\underbrace{\begin{pmatrix} 0\\0\\vp\\ 0\end{pmatrix}}_{r_1}.
\]
\item[2.] The payoff of staying active, $vp-fc$, is non-decreasing in $w$, and the entry cost is bounded below by the current payoff of staying active and above by the expected discounted value of this payoff: $vp-fc\le ec\le \mathbbm{1}_W\,f_w\tr(vp-fc)/(1-\delta)$. Equivalently,
\[
\underbrace{\begin{pmatrix} 0 & 0 & 0 & D\\[2pt]
0 & 0 & I & 0\\[2pt]
0 & 0 & -I & I-\tfrac{1}{1-\delta}\mathbbm{1}_W f_w\tr\end{pmatrix}}_{R_2}\pi\le \underbrace{\begin{pmatrix} 0\\0\\0\end{pmatrix}}_{r_2}.
\]
\item[3.] The scrap value $s$ does not depend on the exogenous state $w$. Equivalently,
\[
\underbrace{\begin{pmatrix} 0 & D & 0 & 0\\0 & -D & 0 & 0\end{pmatrix}}_{R_3}\pi\le\underbrace{\begin{pmatrix} 0\\0\end{pmatrix}}_{r_3}.
\]
\end{enumerate}
\end{restriction*}
Here $I$ and $0$ stand for identity and zero matrices of appropriate dimensions, and $D=\left(\begin{smallmatrix}
1 & -1 &    \dots    & 0       \\
  & \ddots & \ddots &    \\
  0& \dots       & 1      & -1
\end{smallmatrix}\right)$ is the $(W-1)\times W$ first-difference matrix.

Stacking $M_{\pi}$, $-M_{\pi}$, and the matrices for Restriction 1 ($R_1$), Restrictions 1--2 ($R_1$ and $R_2$), or Restrictions 1--3 ($R_1$, $R_2$, and $R_3$) into the matrix $R$, and the vectors $b_1(p,F)$, $-b_1(p,F)$, and $r_1$, $(r_1\tr,r_2\tr)\tr$, or $(r_1\tr,r_2\tr,r_3\tr)\tr$, respectively, into the vector $r$, we obtain the identified set for the payoffs, $\Pi^I=\{\pi:R\pi\le r\}$.

It can be shown that the set $\Pi^I$ is bounded under Restriction~1, so that the bounds on $\theta_V$ are finite. 
\begin{remark}\label{rem:affine_to_LP}
To write $\max_{R\pi\le r}\{c_0+\gamma\tr\pi\}$ in the canonical form \eqref{eqnLP}, let $x=(x_0,\pi\tr)\tr$ and set
\begin{equation}
\label{eq:affine_to_LP}
M=
\begin{pmatrix}
0&R\\
1&0\tr\\
-1&0\tr
\end{pmatrix},
\qquad
\beta=
\begin{pmatrix}
r\\
1\\
-1
\end{pmatrix},
\qquad
\rho=
\begin{pmatrix}
c_0\\
\gamma
\end{pmatrix}.
\end{equation}
The last two constraints impose $x_0=1$, so
\[
\max_{R\pi\le r}\{c_0+\gamma\tr\pi\}
=
\max_{Mx\le\beta}\langle\rho,x\rangle.
\]
\end{remark}

\begin{remark} Let $f_1=(f(1,w))_{w\in\mathcal{W}}$ and $\tilde f_1=(\tilde f(1,w))_{w\in\mathcal{W}}$ collect the stationary probabilities of the incumbency states. In both cases of Proposition~\ref{prop:linear_thetaV},
\[
\gamma=(-q\tr,\;q\tr,\;0\tr,\;0\tr)\tr,\qquad\text{so that}\qquad \gamma\tr\pi=q\tr(s-oo),
\]
where $q=\tilde f_1-f_1$ in case (a) and $q=-f_1$ in case (b). Under Restriction~1, therefore, $\theta_V=c_0+q\tr s$, that is, $\theta_V=c_0+\sum_{w}[\tilde f(1,w)-f(1,w)]\,s_w$ in case (a) and $\theta_V=c_0-\sum_{w}f(1,w)\,s_w$ in case (b).

Hence $\theta_V$ is point identified if and only if $q\tr s$ is constant on $\Pi^I$. Since Restrictions~1--3 bound the level of $s$ only through inequalities, this is generally not the case: $\theta_V$ is partially identified unless $q$ is orthogonal to every direction in which $s$ varies over $\Pi^I$. Note that in case (a), $\tilde V-V$ is point identified \citep[Proposition~9]{KSS}, but $\theta_V=\tilde f\tr(\tilde V-V)+(\tilde f-f)\tr V$ is not, because $V$ depends on $\pi$.
\end{remark}


\section{Details on the Monte Carlo simulations}
\label{SecDetMC}

\subsection{Setting in Section~\ref{SecMC_AV}}\label{SecDetMC_AV}
\subsubsection{Details about the proposed procedure}\label{SecDetMC_AV_BM}

We implement $\hat T_N$ in \eqref{eqnEAM2} and $\hat\Delta$ in \eqref{eqnEAM5} using an alternative normalizing matrix that performs better in finite samples. Let $\eta^*$ denote the relevant minimum-norm solution: the primal solution for the upper CI, the dual solution for the lower CI, and the primal-dual solution for the two-sided CI\@. If $(\zeta_b,\zeta_A)$ denotes the Gaussian limit of $(\sqrt{N}(\hat b-b),\sqrt{N}(\hat A-A))$, define the diagonal matrix $\widetilde\Omega$ by $\widetilde\Omega_{jj}=\operatorname{sd}([\zeta_b-\zeta_A\eta^*]_j)$. We then set $\Omega^{\mathrm{new}}_{jj}=\max\{\Omega_{jj},\widetilde\Omega_{jj}\}$, where $\Omega$ is defined in \eqref{eq:Omega} and the maximum is taken entrywise.

To estimate $\widetilde\Omega$, let $(\xi_b^*,\xi_A^*)$ be a bootstrap draw from the estimated distribution of $(\zeta_b,\zeta_A)$ and let $\hat\eta$ denote the corresponding estimator of $\eta^*$: $\hat x$, $\hat y$, or $\hat z=(\hat x^\top,\hat y^\top)^\top$ for the upper, lower, or two-sided CI, respectively, with $\hat z$ defined in \eqref{eqnEAM6}. Define $\widehat{\widetilde\Omega}_{jj}=\operatorname{sd}^*([\xi_b^*-\xi_A^*\hat\eta]_j)$ and $\widehat\Omega^{\mathrm{new}}_{jj}=\max\{\widehat\Omega_{jj},\widehat{\widetilde\Omega}_{jj}\}$, where $\operatorname{sd}^*$ denotes the bootstrap standard deviation conditional on the sample. Finally, $\widehat\Omega^{\mathrm{new}}_{\mathrm{aug}}$ is obtained by appending zero diagonal entries for the box-constraint rows, which are treated as deterministic. Since $\hat\eta\xrightarrow{p}\eta^*$ and the bootstrap consistently estimates the distribution of $(\zeta_b,\zeta_A)$, $\widehat{\widetilde\Omega}\xrightarrow{p}\widetilde\Omega$ and hence $\widehat\Omega^{\mathrm{new}}\xrightarrow{p}\Omega^{\mathrm{new}}$. Thus, $\widehat\Omega^{\mathrm{new}}_{\mathrm{aug}}$ satisfies the same requirements as the normalizing matrix used in our theoretical results.

\begin{figure}[htbp]
    \centering
    \begin{subfigure}[t]{0.495\textwidth}
        \centering
        \includegraphics[width=\textwidth]{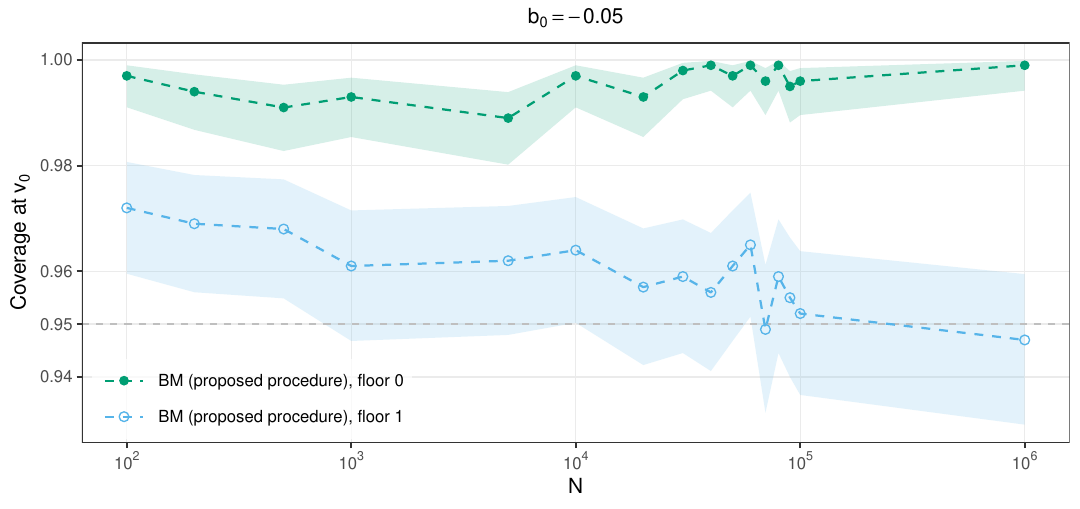}
        \caption{Upper CI}
        \label{fig:AV_floor_upper}
    \end{subfigure}
    \hfill
    \begin{subfigure}[t]{0.495\textwidth}
        \centering
        \includegraphics[width=\textwidth]{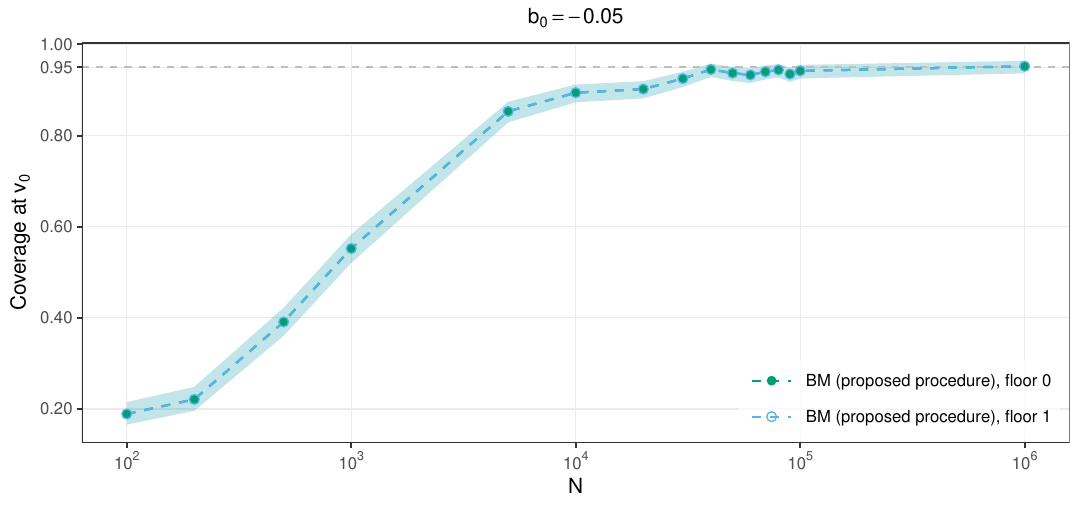}
          \caption{Lower CI}
        \label{fig:AV_floor_lower}
    \end{subfigure}
    \caption{\small Illustration of Remark~\ref{rem:floor}. The effect of box size choice on the coverage of the upper and lower confidence intervals for the design in Section~\ref{SecMC_AV}, $b_0=-0.05$. $1,000$ simulations were run. Nominal coverage level $1-\alpha=0.95$. Floor $0$ and floor $1$ correspond to \eqref{eq:box} and \eqref{eq:box_floor}  respectively. Shaded bands are pointwise $95\%$ \cite{Wilson} intervals for the simulated coverage, reflecting Monte Carlo error.}
    \label{fig:AV_floor}
\end{figure}

Next, Figure~\ref{fig:AV_floor} illustrates the claim of Remark~\ref{rem:floor}, i.e., the effect of the modification to the box size proposed in \eqref{eq:box_floor}. Figure~\ref{fig:AV_floor_upper} shows that when the unique solution to the corresponding problem (here, the primal one) is zero, the box size given in \eqref{eq:box} leads to the procedure being conservative, while using $\hat{a}_N$ defined as in \eqref{eq:box_floor} restores nominal coverage asymptotically. When the unique solution is not zero (e.g., for the lower one-sided system in this design), the two choices \eqref{eq:box} and \eqref{eq:box_floor} yield essentially the same results, as Figure~\ref{fig:AV_floor_lower} illustrates.

\subsubsection{Other details of Section~\ref{SecMC_AV_AV}}\label{SecDetMC_AV_AV}
The following remark derives the asymptotic coverage of \citeauthor{AV}'s confidence intervals as a function of $\bar{v}$.
\begin{remark}\label{remAV}
In the design of Section~\ref{SecMC_AV} with $b_{0}<0$, for any $\bar v>0$ the nominal $(1-\alpha)$ one-sided confidence interval of \cite{AV} has asymptotic coverage $\Phi\big(z_{1-\alpha}/\sqrt{1+r}\big)$ and the two-sided interval has asymptotic coverage $2\Phi\big(z_{1-\alpha/2}/\sqrt{1+r}\big)-1$, where $\Phi$ is the standard normal cdf and $r\ge0$ depends on $\bar{v}$ and is defined in \eqref{eq:r} below. Here $r=0$ if and only if Assumption~B3 of \cite{AV} holds, so coverage is nominal under B3 and strictly below nominal when B3 is
violated. We now derive this result.

\citet{AV} considers an LP of the form $B(\theta)=\min_{x}p\tr x$ s.t. $M^{V}x\ge c$. It is obtained from \eqref{eqnLP} by setting $p=-\rho$, $M^{V}=-M$ and $c=-\beta$, so that $B(\theta)=-v(\theta)$; below $M$ denotes $M^{V}$. Following \citeauthor{AV}'s notation, let $\lambda^{*}\in\Lambda(\theta_{0})$ be the KKT vector, $A$ the set of constraints binding at the optimum, and $S_{A}=\{v:M_{A}\tr v=p\}$. In this design $A=\{1,2\}$ and $M_{A}$ is square and invertible, so $S_{A}=\{v^{*}\}$ with $v^{*}=\lambda^{*}_{A}$, and \citeauthor{AV}'s collection $\mathcal A$ of admissible binding sets is $\{A\}$. Assumption~B3,\footnote{Unless stated otherwise, numbered equations, assumptions, lemmas and theorems cited in this section are those of \cite{AV}.} $\max_{A\in\mathcal A}\min_{v\in S_{A}}\|v\|\le\bar v$, therefore reads
$\|v^{*}\|\le\bar v$.

Let $\hat{x}$ be the fold-1 vertex and $\hat A$ its binding set, so that $\hat M^{(1)}_{\hat A}\hat x=\hat c^{(1)}_{\hat A}$. Let $\check v$ denote the constrained least-squares vector of eq.~(10), computed on fold~1:
\begin{equation*}
  \check{v}\in\arg\min_{\|v\|\le\bar v}\left\|\hat M^{(1)}_{\hat A}{}\tr v-p\right\|^{2}.
\end{equation*}
The problem is convex, so $\check v$ is characterized by its first-order conditions: with $\hat\tau\ge0$ the multiplier on the Euclidean norm constraint, $(\hat M^{(1)}_{\hat A}\hat M^{(1)}_{\hat A}{}\tr+\hat\tau I)\check{v}=\hat M^{(1)}_{\hat A}p$ and $\hat\tau(\|\check v\|-\bar v)=0$. Let $\check v(\tau)$ and $\tau$ denote the solution and multiplier of the same system with $\hat M^{(1)}_{\hat A}$ replaced by $M_{A}$. Since $M_{A}M_{A}\tr$ is positive definite, $\check v(\tau)$ is unique and continuous in $M_{A}$; as $P(\hat A=A)\to1$ (proof of Theorem~2.2) and $\hat M^{(1)}_{A}\to_{p}M_{A}$, it follows that $\check v\to_{p}\check v(\tau)$ and $\hat\tau\to_{p}\tau$. Explicitly, $\check v(\tau)=(M_{A}M_{A}\tr+\tau I)^{-1}M_{A}M_{A}\tr v^{*}$. If $\|v^{*}\|\le\bar v$, then $\tau=0$ and $\check v(0)=v^{*}$. If $\|v^{*}\|>\bar v$, the norm constraint binds and $\tau>0$ is the unique root of $\|\check v(\tau)\|=\bar v$, unique because $\|\check v(\tau)\|$ is strictly decreasing in $\tau$; this $\check v(\tau)$ is not in $S_{A}$.

Two projections of $\check v$ must be kept apart. In the proof of Theorem~2.3 (see eq.~(117)), \citeauthor{AV} defines $\tilde{v}_{n}$ as the projection of $\check v$ onto $S(\hat A,M)\equiv\arg\min_{\|v\|\le\bar v}\|p-M_{\hat A}\tr v\|^{2}$. We denote by $\tilde{v}_{n}$ the projection onto $S_{A}$ instead. Recall that $S_{A}=\{v^{*}\}$. On $\{\hat A=A\}$, $S(\hat A,M)=S(A,M)=\{\check v(\tau)\}$, a singleton because the objective is strictly convex. On this event \citeauthor{AV}'s $\tilde{v}_{n}$ is therefore $\check{v}(\tau)$, while ours is always $v^{*}$. The two coincide if and only if $\tau=0$, that is, if and only if B3 holds.

Denote the fold-$f$ estimation error on the binding rows evaluated at $\hat x$ by  $\Delta^{(f)}_{A}(\hat{x})\equiv\hat c^{(f)}_{A}-c_{A}-\left(\hat M^{(f)}_{A}-M_{A}\right)\hat{x}$, for $f=1,2$. On $\{\hat A=A\}$ the vertex condition gives $M_{A}\hat x-c_{A}=\Delta^{(1)}_{A}(\hat{x})$. For
any $\tilde v_n\in S_{A}$ Lemma~B.1 gives $p\tr\hat x=\tilde v\tr M_{A}\hat x$ and $B(\theta_{0})=\tilde v\tr c_{A}$. Substituting both into $\breve B_{n}=\check v\tr(\hat c^{(2)}_{\hat A}-\hat M^{(2)}_{\hat A}\hat x)+p\tr\hat x$
yields on $\{\hat A=A\}$,
\begin{equation*}
  \breve B_{n}-B(\theta_{0})
  =\check v\tr\Delta^{(2)}_{A}(\hat{x})+(\tilde v_{n}-\check v)\tr\Delta^{(1)}_{A}(\hat{x}).
\end{equation*}
As noted above, when B3 fails, \citeauthor{AV}'s $\tilde v_{n}$ is not in $S_{A}$, so the decomposition above does not hold for it. It does hold for our choice $\tilde v_{n}\in S_{A}$, but then $\tilde v_{n}-\check{v}\xrightarrow{p} v^{*}-\check{v}(\tau)\ne0$.

In this design $x^{\ast}=0$, so $\sqrt{n_{f}}\Delta^{(f)}_{A}(x^{\ast})=\sqrt{n_{f}}(\hat\nu^{(f)},\hat\zeta^{(f)})\tr$ is asymptotically $\mathcal{N}(0,\Sigma_{A})$ with $\Sigma_{A}=(1/12)I$, since both $\hat\nu^{(f)}$ and $\hat\zeta^{(f)}$ have asymptotic variance $1/12$, and the two estimation errors are asymptotically independent. Let $n_1=\gamma n$, so $n_2=(1-\gamma)n$. Since $\hat x-x^{*}=O_{p}(n^{-1/2})$ and $\hat M^{(f)}_{A}-M_{A}=O_{p}(n^{-1/2})$,
$\sqrt{n_{2}}\big(\breve B_{n}-B(\theta_{0})\big)$ is asymptotically normal with mean zero and variance
\begin{equation*}
  \check{v}(\tau)\tr\Sigma_{A}\check{v}(\tau)
  +\frac{1-\gamma}{\gamma}\left(v^{*}-\check{v}(\tau)\right)\tr\Sigma_{A}\left(v^{*}-\check{v}(\tau)\right).
\end{equation*}
The first term is $\sigma^{2}(A,x^{*},\check v(\tau),\Sigma)$ of Lemma~B.3, for which \citeauthor{AV}'s $\hat\sigma^{2}_{n}$ is consistent. Hence the asymptotic variance of $\sqrt{n_{2}}(\breve B_{n}-B(\theta_{0}))$ is $\sigma^{2}(A,x^{*},\check v(\tau),\Sigma)(1+r)$ with
\begin{equation}
  r=\frac{1-\gamma}{\gamma}\cdot
  \frac{\|v^{*}-\check v(\tau)\|^{2}}{\|\check v(\tau)\|^{2}}
  =\frac{1-\gamma}{\gamma}\cdot
  \frac{\|\tau(M_{A}M_{A}\tr+\tau I)^{-1}v^{*}\|^{2}}{\bar v^{\,2}},
  \label{eq:r}
\end{equation}
where the second equality uses $v^{*}-\check v(\tau)=\tau(M_{A}M_{A}\tr+\tau I)^{-1}v^{*}$ and, when the norm constraint binds,
$\|\check v(\tau)\|=\bar v$; when it does not bind, $\tau=0$ and both expressions are zero. This gives the coverage formulas stated at the beginning.

For example, when $b_{0}=-0.02$, we have $A=\{1,2\}$, $M_{A}=\bigl(\begin{smallmatrix}-0.98&1\\1&-1\end{smallmatrix}\bigr)$, $v^{*}=(50,50)\tr$ and $\|v^{*}\|=70.71$. With $\bar v=34.45$ the norm constraint binds, $\tau\approx1.06\times10^{-4}$ and $\check{v}(\tau)=(24.23,24.49)\tr$. With equal folds ($\gamma=1/2$), $r\approx1.108$, so the asymptotic coverage of the nominal $95\%$ one-sided interval is $0.871$ and that of the two-sided interval is $0.823$.
\end{remark}

\subsubsection{Mapping into the canonical form of \cite{CSS}}\label{SecDetMC_AV_CSS}
\citet{CSS} test the null that the system
\begin{equation*}
  B\,\mu \;+\; \left(D + B\,\Pi\right)\delta \;\le\; \bar{d}(v)
\end{equation*}
admits a solution $\delta$, where $\mu$ is the estimated intercept, $\Pi$ the estimated slope, $(B,D)$ known matrices and
$\bar{d}(v)$ the known $v$-dependent right-hand side. 

For the upper one-sided CI, the correspondence is
\begin{equation*}
 \delta = x, \quad \mu = \begin{bmatrix} -\beta \\ 0 \end{bmatrix}, \quad
  \Pi = \begin{bmatrix} M \\ -\rho\tr \end{bmatrix}, \qquad
  B = I_{d+1}, \quad
  D = 0_{(d+1)\times p}, \quad
  \bar{d}(v) = \begin{bmatrix} 0_{d} \\ -v \end{bmatrix} .
\end{equation*}

For the lower one-sided CI, the correspondence is 
\begin{equation*}
  \delta = y, \quad \mu = \begin{bmatrix} \rho \\ 0 \end{bmatrix}, \quad
  \Pi = \begin{bmatrix} -M\tr \\ \beta\tr \end{bmatrix}, \quad
  B = \begin{bmatrix}
        -I_{p} & 0 \\
         I_{p} & 0 \\
         0     & 1 \\
         0     & 0
      \end{bmatrix}, \quad
  D = \begin{bmatrix} 0_{(2p+1)\times d} \\ -I_{d} \end{bmatrix}, \quad
  \bar{d}(v) = \begin{bmatrix} 0_{p} \\ 0_{p} \\ v \\ 0_{d} \end{bmatrix} .
\end{equation*}

Finally, for the two-sided CI 
\begin{equation}
\begin{aligned}
 & \delta = \begin{bmatrix} x \\ y \end{bmatrix}, \quad \mu = \begin{bmatrix} -\beta \\ \rho \\ 0 \\ 0 \end{bmatrix}, \quad
  \Pi = \begin{bmatrix}
          M       & 0       \\
          0       & -M\tr   \\
          -\rho\tr & 0      \\
          0       & \beta\tr
        \end{bmatrix}, \quad
  B = \begin{bmatrix}
        I_{d} & 0      & 0 & 0 \\
        0     & 0      & 1 & 0 \\
        0     & -I_{p} & 0 & 0 \\
        0     &  I_{p} & 0 & 0 \\
        0     & 0      & 0 & 1 \\
        0     & 0      & 0 & 0
      \end{bmatrix}, \\
 & D = \begin{bmatrix}
        0_{(d+2p+2)\times p} & 0_{(d+2p+2)\times d} \\
        0_{d \times p} & -I_{d}
      \end{bmatrix}, \quad \bar{d}(v) = \begin{bmatrix} 0_{d}\tr\\ -v \\ 0_{p}\tr \\ 0_{p}\tr \\ v \\ 0_{d}\tr
      \end{bmatrix}
\end{aligned}\label{eq:css_ts}
\end{equation}

Because the estimated box constraints are treated as deterministic conditional on the sample, the box-augmented systems are obtained from those above as follows, where $d_\delta:=\dim(\delta)$ and $d_\mu:=\dim(\mu)$:
\begin{equation*}
  B \;\mapsto\; \begin{bmatrix} B \\ 0_{2d_{\delta} \times d_{\mu}} \end{bmatrix},
  \qquad
  D \;\mapsto\; \begin{bmatrix} D \\ -I_{d_{\delta}} \\ I_{d_{\delta}} \end{bmatrix},
  \qquad
  \bar{d}(v) \;\mapsto\; \begin{bmatrix} \bar{d}(v) \\ \hat{a}_N\mathbf{1}_{2d_{\delta}} \end{bmatrix}.
\end{equation*}

\begin{figure}[htbp]
    \centering
    \begin{subfigure}[t]{0.495\textwidth}
        \centering
        \includegraphics[width=\textwidth]{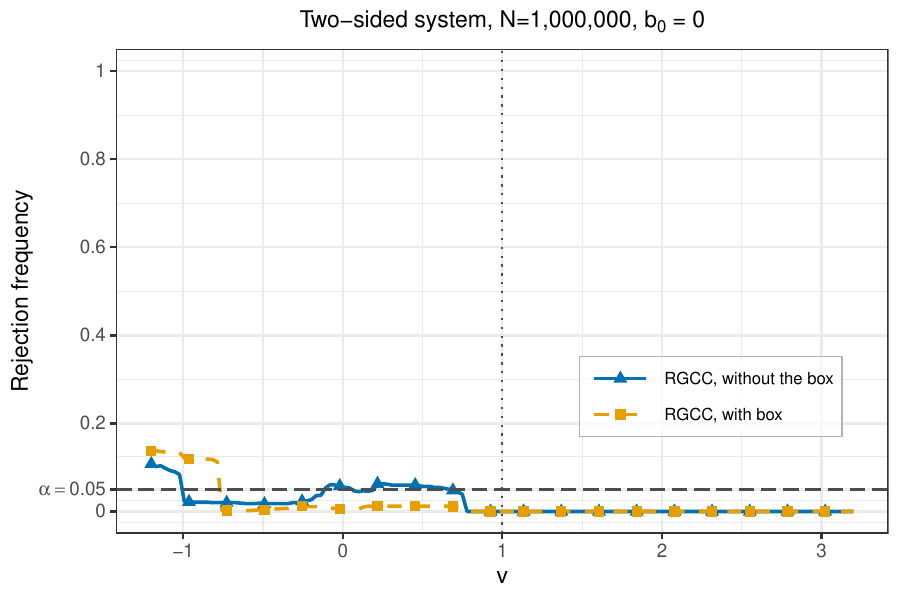}
        \caption{Power curve for $b_0=0$}
        \label{fig:CSS_twosided_b00}
    \end{subfigure}
    \hfill
    \begin{subfigure}[t]{0.495\textwidth}
        \centering
        \includegraphics[width=\textwidth]{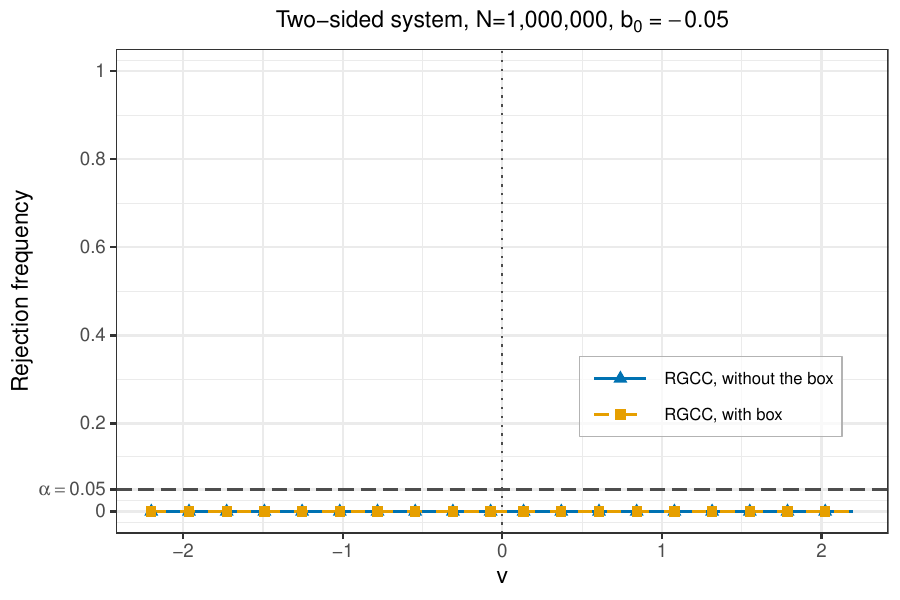}
          \caption{Power curve for $b_0=-0.05$.}
        \label{fig:CSS_twosided_b0m005}
    \end{subfigure}
    \caption{\small Illustration of Remark~\ref{rem:css_ts_degen} using the simulations from Section~\ref{SecMC_AV}. The RGCC test of \cite{CSS}, applied directly to the system in \eqref{eq:css_ts}, controls size with and without the box but has trivial power.}
    \label{fig:AV_CSS}
\end{figure}

The following observation applies to the two-sided system:
\begin{remark}[Degeneracy of the two-sided mapping]\label{rem:css_ts_degen}
Let $(\hat\mu, \hat\Pi)$ be any estimators of the reduced form
of \cite{CSS} applied to the system from Equation~\eqref{eq:css_ts}, obtained by plugging estimators 
$(\hat M, \hat\beta, \hat\rho)$. Fix any
$\delta^* = (x^{*}{}\tr, y^{*}{}\tr)\tr \in \mathbb{R}^{p+d}$ and let
$a = (y^{*}{}\tr, x^{*}{}\tr, 1, 1)\tr$. Then
\begin{equation*}
a\tr(\hat\mu + \hat\Pi\delta^*)
= y^{*}{}\tr(-\hat\beta + \hat Mx^*)
+ x^{*}{}\tr(\hat\rho - \hat M\tr y^*)
- \hat\rho\tr x^* + \hat\beta\tr y^* = 0
\end{equation*}
identically in $(\hat M, \hat\beta, \hat\rho)$. Consequently
$\Sigma_\infty a = 0$ with $a \neq 0$, so Assumption 1(iii) of
\cite{CSS} fails at every $\delta^*$. Note that the failure is
unaffected by the box augmentation. Figure~\ref{fig:AV_CSS} illustrates the performance of the RGCC test when applied to the system in \eqref{eq:css_ts} with and without box augmentation for both values of $b_0$. The test controls size, but has trivial power. Moreover, a stronger result can be established: the same degeneracy arises for any admissible representation of the two-sided primal--dual system in the canonical form of \citet{CSS}, and is therefore not an artifact of the particular choice of $(B,D,\mu,\Pi,\bar d)$ in \eqref{eq:css_ts}.
\end{remark}

 \subsubsection{Mapping into the canonical form of \cite{BPSSMT}}\label{SecDetMC_AV_BPSSMT}
\citet{BPSSMT} consider the following system:
\begin{equation*}
  A_{0}\,x_{0} \;+\; A_{1}\,x_{1} \;=\; \tilde{\beta},
  \qquad x_{0} \text{ free}, \qquad x_{1} \ge 0.
\end{equation*}

For the upper one-sided CI, the correspondence is as follows:
 $x_{0} = x$, $x_{1} = (s_{1}, s_{2})$
 
\begin{equation*}
  x_0=x, \quad x_1 =\begin{bmatrix}s_1\\s_2\end{bmatrix}, \quad A_{0} = \begin{bmatrix} M \\ \rho\tr \end{bmatrix}, \qquad
  A_{1} = \begin{bmatrix} I_{d} & 0 \\ 0 & -1 \end{bmatrix}, \qquad
  \tilde{\beta} = \begin{bmatrix} \beta \\ v \end{bmatrix},
\end{equation*}
where $s_1$ and $s_2$ are the slack variables that transform inequalities to equalities, i.e., $Mx + s_{1} = \beta$ and $\rho\tr x - s_{2} = v$.  With the box augmentation,
$x_{1} = (s_{1}, s_{2}, u_{1}, u_{2})$ and
\begin{equation*}
 x_0=x, \quad x_{1}=\begin{bmatrix}s_1\\s_2\\u_1\\u_2\end{bmatrix}, \quad A_{0} = \begin{bmatrix} M \\ \rho\tr \\ I_{p} \\ I_{p} \end{bmatrix},
  \quad
  A_{1} = \begin{bmatrix}
            I_{d} & 0  & 0      & 0     \\
            0     & -1 & 0      & 0     \\
            0     & 0  & -I_{p} & 0     \\
            0     & 0  & 0      & I_{p}
          \end{bmatrix},
  \quad
  \tilde{\beta} = \begin{bmatrix} \beta \\ v \\ -\hat{a}_N\mathbf{1}_{p} \\ \hat{a}_N\mathbf{1}_{p} \end{bmatrix}.
\end{equation*}
 
Next, for the lower one-sided CI, $A_0$ is the zero matrix, and the rest of the correspondence is  
\begin{equation*}
 x_{1}=\begin{bmatrix}y\\s_3\end{bmatrix},\quad A_{1} = \begin{bmatrix} M\tr & 0 \\ \beta\tr & 1 \end{bmatrix}, \qquad
  \tilde{\beta} = \begin{bmatrix} \rho \\ v \end{bmatrix},
\end{equation*}
where $s_3$ is defined from $\beta\tr y + s_{3} = v$. With the box,
$x_{1} = (y, s_{3}, u_{3})$ and
\begin{equation*}
  x_{1}=\begin{bmatrix}y\\s_3\\u_3\end{bmatrix},\quad A_{1} = \begin{bmatrix}
            M\tr     & 0 & 0     \\
            \beta\tr & 1 & 0     \\
            I_{d}    & 0 & I_{d}
          \end{bmatrix},
  \quad
  \tilde{\beta} = \begin{bmatrix} \rho \\ v \\ \hat{a}_{N}\mathbf{1}_{d} \end{bmatrix}.
\end{equation*}

\begin{remark}[Zero-variance target in the lower system]\label{rem:BPSSMT_lower}
The screening procedure of \cite{BPSSMT} targets $-\tilde\beta$, which is deterministic for this system. Hence, the method's performance is fully determined by the quality of the certification direction found on the first fold.
\end{remark}
 
Finally, for the two-sided system:
 \begin{equation*}
  x_{0} = x, \quad x_{1}=\begin{bmatrix}y\\s_1\\s_2\\s_3\end{bmatrix},\quad A_{0} = \begin{bmatrix} M \\ \rho\tr \\ 0_{p \times p} \\ 0_{1 \times p} \end{bmatrix},
  \quad
  A_{1}^{\text{no-box}} = \begin{bmatrix}
            0        & I_{d} & 0  & 0 \\
            0        & 0     & -1 & 0 \\
            M\tr     & 0     & 0  & 0 \\
            \beta\tr & 0     & 0  & 1
          \end{bmatrix},
  \quad
  \tilde{\beta} = \begin{bmatrix} \beta \\ v \\ \rho \\ v \end{bmatrix}.
\end{equation*}
With the box, 
\begin{equation*}
  x_{1}=\begin{bmatrix}y\\s_1\\s_2\\s_3\\u_1\\u_2\\u_3\end{bmatrix},\quad A_{0} = \begin{bmatrix} M \\ \rho\tr \\ 0 \\ 0 \\ I_{p} \\ I_{p} \\ 0_{d \times p} \end{bmatrix},
  \qquad
  A_{1} = \begin{bmatrix}
            A_{1}^{\text{no box}} & 0      & 0     & 0     \\
            0_{p \times n_{1}}      & -I_{p} & 0     & 0     \\
            0_{p \times n_{1}}      & 0      & I_{p} & 0     \\
            I_{d} & 0 & 0 & I_{d}
          \end{bmatrix},
  \qquad
  \tilde{\beta} = \begin{bmatrix} \beta \\ v \\ \rho \\ v \\ -\hat{a}_{N}\mathbf{1}_{p} \\ \hat{a}_{N}\mathbf{1}_{p} \\ \hat{a}_{N}\mathbf{1}_{d} \end{bmatrix}.
\end{equation*}
 

\subsection{Setting in Section~\ref{SecMC_KKLS}}\label{SecDetMC_KKLS}
Following the replication code of \cite{KKLS} available \href{https://zenodo.org/records/18376448}{here}, we make the following decisions. We consider the case with exogenous state $w_t=(w_{1t},w_{2t},w_{3t})$ consisting of three independent shocks, each taking two values: $\{w_{1l},w_{1h}\}=\{-0.354,+0.354\}$, $\{w_{2l},w_{2h}\}=\{-0.395,+0.395\}$, and $\{w_{3l},w_{3h}\}=\{-0.433,+0.433\}$. Hence, there are $X=16$ states in total. The (residual) inverse demand function is linear: $P_t = \bar w + w_{1t} + w_{2t} - \eta Q_t.$ Here $w_{1t}$ and $w_{2t}$ are demand shocks, and marginal cost is constant in quantity and equal to the supply shock: $mc_t = w_{3t}$. Variable profits are then $vp_t = \left(\bar w + w_{1t} + w_{2t} - mc_t\right)^2/(4\eta)$. The idiosyncratic shocks are type-I extreme value. The other parameter values are $s = 4.5$, $ec = 5$, $fc = 0.5$, $\delta = 0.9$, $\eta=4$, $\bar{w}=6.8$. CCPs are estimated by a grouped binomial logit. We implement nonparametric bootstrap clustered at the firm level ($N$ firms resampled with replacement, $T$ fixed).

We consider a lower confidence interval on $v_0=\min_{Mx\le\beta}\theta_V$. The details on $x$, $\rho$, $M$, and $\beta$ under the different sets of restrictions considered can be found in Section~\ref{SecDetEx_KKLS_restr}.
\end{appendix}


\end{document}